\documentclass[traditabstract]{aa} 
\usepackage{graphicx}
\usepackage{txfonts}
\usepackage{longtable}
\usepackage{multirow}
\usepackage{lscape}
\usepackage{xcolor}
\usepackage{natbib,twoopt}
\usepackage{bm}
\usepackage[breaklinks=true]{hyperref} 
\bibpunct{(}{)}{;}{a}{}{,}             
\makeatletter

\def\ms{\hbox{\,m\,s$^{-1}$}}         
\def\cms{\hbox{\,cm\,s$^{-1}$}}       
\def\m2s2{\hbox{\,m$^{2}$\,s$^{-2}$}} 
\def\Msun{\hbox{$\mathrm{M}_{\odot}$}}           
\def\Rsun{\hbox{$\mathrm{R}_{\odot}$}}

\def\Me{\hbox{$\mathrm{M}_{\oplus}$}\,}             
\def\Re{\hbox{$\mathrm{R}_{\oplus}$}\,}
\def\mp{M_{\rm p}}
\def\rp{R_{\rm p}}
\DeclareMathOperator{\e}{e}

\newcommand{\pchord}{\texttt{PolyChord}} 
\newcommand{\mps}{\ensuremath{\textrm{m}~\textrm{s}^{-1}}}

\newcommand{\lrhk}{\ensuremath{\log R'_{\rm HK}}}
\newcommand{\bis}{\ensuremath{{\rm BIS}}}

\usepackage{titlesec}

\titleformat{\paragraph}
{\normalfont\normalsize}{\theparagraph}{1em}{}
\titlespacing*{\paragraph}
{0pt}{3.25ex plus 1ex minus .2ex}{1.5ex plus .2ex}

\begin{document}

\title{In-depth characterization of the Kepler-10 three-planet system with \\ 
HARPS-N radial velocities and \emph{Kepler} transit timing variations} 
\titlerunning{In-depth characterization of the Kepler-10 three-planet system}
\authorrunning{Bonomo et al.}

\author{A.~S.~Bonomo \inst{1} 
\and L.~Borsato \inst{2}
\and V.~M.~Rajpaul \inst{3, 4}
\and L.~Zeng \inst{5}
\and M.~Damasso \inst{1}
\and N.~C.~Hara \inst{6,7}
\and M.~Cretignier \inst{6,8}
\and A.~Leleu \inst{6}
\and N.~Unger \inst{6}
\and X.~Dumusque \inst{6}
\and F.~Lienhard \inst{4,9}
\and A.~Mortier \inst{10}
\and L.~Naponiello \inst{1}
\and L.~Malavolta \inst{2,11}
\and A.~Sozzetti \inst{1}
\and D.~W.~Latham\inst{12}
\and K.~Rice\inst{13,14}
\and R.~Bongiolatti \inst{15}
\and L.~Buchhave \inst{16}
\and A.~C.~Cameron\inst{17}
\and A.~F.~Fiorenzano \inst{18}
\and A.~Ghedina\inst{18}
\and R.~D.~Haywood\inst{19}\fnmsep\thanks{STFC Ernest Rutherford Fellow}
\and G.~Lacedelli\inst{11, 20}
\and A.~Massa \inst{21}
\and F.~Pepe\inst{6}
\and E.~Poretti \inst{18,22}
\and S.~Udry\inst{6} 
}

\institute{
INAF - Osservatorio Astrofisico di Torino, via Osservatorio 20, 10025 Pino Torinese, Italy  
\and INAF - Osservatorio Astronomico di Padova, Vicolo dell’Osservatorio 5, 35122 Padova, Italy
\and Academy for the Mathematical Sciences, c/o Isaac Newton Institute for Mathematical Sciences, 20 Clarkson Road, Cambridge CB3 0EH, UK
\and Astrophysics Group, Cavendish Laboratory, University of Cambridge, JJ Thomson Avenue, Cambridge CB3 0HE, UK
\and Department of Earth and Planetary Sciences, Harvard University, 20 Oxford Street, Cambridge, 02138, MA, USA
\and D\'epartement d'astronomie de l'Universit\'e de Gen\`eve, Chemin Pegasi 51, 1290 Versoix, Switzerland
\and Universit\'e Aix Marseille, CNRS, CNES, LAM, Marseille, France
\and Department of Physics, University of Oxford, Oxford OX13RH, UK
\and Department of Physics, ETH Zurich, Wolfgang-Pauli-Strasse 2, CH-8093 Zurich, Switzerland
\and School of Physics \& Astronomy, University of Birmingham, Edgbaston, Birmingham, B15 2TT, UK
\and Dipartimento di Fisica e Astronomia ``Galileo Galilei'', Universit\`a degli Studi di Padova, Vicolo dell'Osservatorio 3, 35122 Padova, Italy
\and Center for Astrophysics | Harvard \& Smithsonian, 60 Garden Street, Cambridge, MA 02138, USA
\and SUPA, Institute for Astronomy, University of Edinburgh, Blackford Hill, Edinburgh EH9 3HJ, UK
\and Centre for Exoplanet Science, University of Edinburgh, Edinburgh EH9 3FD, UK
\and Dipartimento di Fisica, Universit\`a degli Studi di Milano, Via Celoria 16, I-20133 Milano, Italy
\and DTU Space, Technical University of Denmark, Elektrovej 328, DK-2800 Kgs. Lyngby, Denmark
\and Centre for Exoplanet Science, SUPA School of Physics and Astronomy, University of St Andrews, St Andrews, Fife KY16 9SS, UK
\and Fundaci\'on Galileo Galilei - INAF, Rambla Jos\'e Ana Fernandez P\'erez 7, E-38712 Bre\~{n}a Baja, Tenerife, Spain
\and Department of Astrophysics, University of Exeter, Stocker Rd, Exeter EX4 4QL, UK
\and Instituto de Astrof\'isica de Canarias (IAC), 38205 La Laguna, Tenerife, Spain
\and Dipartimento di Fisica, Universit\`a degli Studi di Torino, via Pietro Giuria 1, 10125 Torino, Italy
\and INAF - Osservatorio Astronomico di Brera, Via E. Bianchi 46, 23807 Merate, Italy
}

\date{Received 18 November 2024 / Accepted 10 February 2025}

\offprints{\email{aldo.bonomo@inaf.it}}

\abstract{The old G3V star Kepler-10 is known to host two transiting planets, the ultra-short-period super-Earth Kepler-10\,b ($P=0.837$~d; $\rp=1.47~\Re$) and the long-period sub-Neptune Kepler-10\,c ($P=45.294$~d; $\rp=2.35~\Re$), and a non-transiting planet that causes variations in the Kepler-10\,c transit times. Measurements of the mass of Kepler-10\,c in the literature have shown disagreement, depending on the radial-velocity dataset and/or the modeling technique used. Here we report on the analysis of almost 300 high-precision radial velocities gathered with the HARPS-N spectrograph at the Telescopio Nazionale Galileo over $\sim11$~years, and extracted with the YARARA-v2 tool, which corrects for possible systematics and/or low-level activity variations at the spectrum level. To model these radial velocities, we used three different noise models and various numerical techniques,
which all converged to the solution: $M_{\rm p, b}=3.24 \pm 0.32$~\Me (10$\sigma$) and $\rho_{\rm p, b}=5.54 \pm 0.64~\rm g\;cm^{-3}$ for planet b; $M_{\rm p, c}=11.29 \pm 1.24$~\Me (9$\sigma$) and $\rho_{\rm p, c}=4.75 \pm 0.53~\rm g\;cm^{-3}$ for planet c; and $M_{\rm p, d}\sin{i}=12.00 \pm 2.15$~\Me (6$\sigma$) and $P=151.06 \pm 0.48$~d for the non-transiting planet Kepler-10\,d. 
This solution is further supported by the analysis of the Kepler-10\,c transit timing variations and their simultaneous modeling with the HARPS-N radial velocities. 
While Kepler-10\,b is consistent with a rocky composition and a small or no iron core, Kepler-10\,c may be a water world that formed beyond the water snowline and subsequently migrated inward. 
}

\keywords{planetary systems – planets and satellites: composition – planets and satellites: detection – planets and satellites: fundamental parameters – techniques: radial velocities}

 
\maketitle

%

 \section{Introduction}
The determination of accurate and precise masses and bulk densities of transiting small planets with radial-velocity (RV) follow-up 
allows us to infer their composition, under reasonable assumptions of planet interior differentiation (e.g., \citealt{2008PASP..120..983Z}) or mixing (e.g., \citealt{Dorn2017}).
The ever-increasing number of well-characterized small planets has revealed a 
surprising variety in their densities, and hence their compositions. 
Generally, high bulk densities correspond to rocky planets with either terrestrial (e.g., \citealt{2015ApJ...800..135D}) or iron-rich (e.g., \citealt{ 2019NatAs...3..416B}) interiors. 
Intermediate densities can be reproduced by massive icy shells (water worlds) and/or thin H/He  (hydrogen/helium) atmospheres shrouding the planet rocky interiors (e.g., \citealt{2018MNRAS.481.1839M}), with a strong degeneracy between the icy and gaseous mass fractions (e.g., \citealt{Rogers_2010, 2014PNAS..11112622S}); 
low densities definitively require thicker H/He atmospheres (e.g., \citealt{2011Natur.470...53L}).

The fundamental goal in estimating planet compositions lies in understanding the main mechanisms that give rise 
to the observed compositional properties and the related well-known valley in the 
distribution of planetary radii around solar-type stars at $R_{\rm trans}\sim1.7-2.0$~\Re, which likely separates prevalently rocky planets
($R_{\rm p} \lesssim R_{\rm trans}$) from volatile-rich ones ($R_{\rm p} \gtrsim R_{\rm trans}$; \citealt{2017AJ....154..109F, 2018MNRAS.479.4786V}). 
Such mechanisms include planet formation at different regions of the protoplanetary disks --~dry or ice-rich regions 
inside or beyond the water iceline, respectively \citep{2017MNRAS.472..245L, 2019PNAS..116.9723Z} -- and/or at different timescales, 
that is, in gas-rich or gas-poor environments. 
Several post-formation processes that may afterwards modify the planet composition have been identified, 
among which 
(i) atmospheric photoevaporation driven by intense stellar XUV radiation, mainly in the first $\sim100$~Myr 
after formation (e.g., \citealt{2013ApJ...775..105O, 2017ApJ...847...29O, 2014ApJ...792....1L}); 
(ii) core-powered mass loss, that is, atmospheric escape due to the luminosity of the cooling rocky core, acting on longer timescales \citep[$\sim 1$ Gyr;][]{2018MNRAS.476..759G, 2019MNRAS.487...24G}; and
(iii) erosion due to giant impacts (e.g., \citealt{2015ApJ...812..164L, Reinhardt2022}). 
The observed diversity in planet compositions and the radius valley are probably caused by more than one mechanism. 
Nonetheless, if one of the abovementioned (formation or post-formation) mechanisms dominates over the others, 
it may produce a distinct slope in the position of the radius valley as a function of orbital period, 
which indicates a possible variation of the transition between rocky and non-rocky planets 
with the level of incident flux received by the planet (e.g., \citealt{2018MNRAS.479.5303L, 2019ApJ...875...29M}). 
This possible dependence of the rocky/non-rocky transition with orbital period must then be compared with the inferred compositions from bulk densities,
which are expected to be rocky or volatile-rich below and above the radius valley, respectively, with the changing level of incident flux. 
For that purpose, more precise measurements of the bulk densities of small planets with relatively long orbital periods (hence lower insolation), $P\gtrsim30$~d, are needed, 
as only a few of them have been characterized well to date. 

Determining the mass and bulk density of small planets with long orbital periods can be much more difficult 
than for shorter-period, inner planets, for three main reasons. Firstly, the semiamplitude of the (Doppler) RV signal induced by the planet decreases as $P^{-1/3}$;
secondly, wider temporal baselines of RV measurements are required to properly sample long-period signals; 
and thirdly, these signals may be more affected by correlated noise due to stellar magnetic activity variations
and/or the presence of additional planets in multiplanet systems. A clear example of this difficulty is the planetary system K2-3, 
which consists of three small planets -- K2-3\,b, c, and d with radii $\rp=2.3$, 1.8, and 1.6~\Re ~-- 
which orbit an early M dwarf with periods $P=10.05$, 24.65, and 44.56~d, the latter in the habitable zone \citep{2015ApJ...804...10C}. 
Even with more than 300 HARPS-N (High Accuracy Radial velocity Planet Searcher in North hemisphere) and HARPS high-precision radial velocities, 
it was not possible to determine the masses/densities of the two outer planets with a precision better than 
$3\sigma$ (\citealt{2018A&A...615A..69D}; see also \citealt{2019AJ....157...97K}).

Kepler-10 is a ``historical'' planetary system containing two transiting planets: Kepler-10\,b, 
the first rocky planet discovered by the \emph{Kepler} space mission \citep{2011ApJ...729...27B}, 
with a radius of $R_{\rm b}=1.47 \pm 0.03$~\Re
and an ultra-short orbital period (USP) of $P_{\rm b}=0.837$~d; and Kepler-10\,c, 
which has both a larger radius $R_{\rm b}=2.35_{-0.04}^{+0.09}$~\Re and 
a much longer orbital period $P_{\rm c}=45.294$~d (\citealt{2011ApJS..197....5F, 2014ApJ...789..154D}, hereafter D14).
The planets orbit a 10.5-Gyr old GV star, which is thus expected to be magnetically quiet 
(average CaII activity index $\log{R^{'}_{\rm HK}}\simeq -4.97$), and is located at a distance of $186$~pc. 

While there has been consensus on the mass and density of the inner rocky planet Kepler-10\,b, 
determinations of the mass and density of Kepler-10\,c in the literature have been largely discrepant, 
illustrating that the characterization of small planets with long orbital periods may be considerably problematic, even at low levels of stellar magnetic activity.
For instance, with 148 high-precision RVs collected with the HARPS-N spectrograph at the Telescopio Nazionale Galileo 
\citep{2012SPIE.8446E..1VC, 2014SPIE.9147E..8CC}, 
D14 found a RV semiamplitude of $K_{\rm c}=3.26 \pm 0.36$~\ms, which implies a planetary mass and density of 
$M_{\rm c}=17.2 \pm 1.9$~\Me and $\rho_{\rm c}=7.1 \pm 1.0~\rm g\,cm^{-3}$. 
By using more sparse 72 RV measurements acquired with 
the HIRES (HIgh Resolution Echelle Spectrometer) spectrograph at the Keck telescope \citep{2010ApJ...721.1467H}, 
\citet[hereafter W16]{2016ApJ...819...83W} found $K_{\rm c} = 1.09 \pm 0.58$~\ms, 
which corresponds to $M_{\rm c}=5.7_{-2.9}^{+3.2}$~\Me  and $\rho_{\rm c}=2.4_{-1.2}^{+1.4}~\rm g\,cm^{-3}$ 
(one may actually even wonder whether, with a statistical significance of 1.9~$\sigma$, 
the signal of Kepler-10\,c is truly detected in the HIRES RV data). 
By combining the 148 HARPS-N and the 72 HIRES RVs, W16 obtained $K_{\rm c} = 2.67 \pm 0.34$~\ms, 
and hence $M_{\rm c}=13.98 \pm 1.79$~\Me  and $\rho_{\rm c}=5.94 \pm 0.75~\rm g\,cm^{-3}$. 
However, this combination of HARPS-N and HIRES data did not solve the issue of the evident mass discrepancy (see Fig.~5, bottom right panel in W16), but led to an average mass, which was found to be closer to the value found by D14
because the number of HARPS-N RVs was approximately twice that gathered with HIRES.

\begin{table*}[t!]
\small
\centering
\caption{Radial-velocity semiamplitudes ($K$) and corresponding masses ($\mp$) of the Kepler-10 planets from the literature.}
\begin{tabular}{l c c c c c} 
\hline
\rule{0pt}{2.6ex} Reference & \citet{2011ApJ...729...27B} & \citet{2014ApJ...789..154D} & \citet{2016ApJ...819...83W} & \citet{2016ApJ...819...83W} & \citet{2017MNRAS.471L.125R} \\ [0.3ex] 
\hline
\rule{0pt}{2.6ex} Instruments & HIRES & HARPS-N & HIRES & HARPS-N+HIRES & HARPS-N+HIRES \\ [0.3ex]
\hline
\multicolumn{6}{c}{\rule{0pt}{2.6ex} Kepler-10\,b} \\ [0.3ex]
\hline
$K$ [\ms] & $3.3^{+0.8}_{-1.0}$ & $2.38 \pm 0.35$ & $3.31 \pm 0.59$ & $2.67 \pm 0.30$ & $2.32^{+0.21}_{-0.18}$ \\ [0.3ex]
$\mp$ [\Me] & $4.56^{+1,17}_{-1.29}$ & $3.33 \pm 0.49$ & $4.61 \pm 0.83$ & $3.72 \pm 0.42$ & $3.24 \pm 0.28$ \\
\hline
\multicolumn{6}{c}{\rule{0pt}{2.6ex} Kepler-10\,c} \\ [0.3ex]
\hline
$K$ [\ms] & - & $3.26 \pm 0.26$ & $1.09 \pm 0.58$ & $2.67 \pm 0.34$ & $1.41^{+0.25}_{-0.23}$ \\ [0.3ex]
$\mp$ [\Me] & $<20$ & $17.2 \pm 1.9$ & $5.7^{+3.2}_{-2.9}$ & $13.98 \pm 1.79$ & $7.37^{+1.32}_{-1.19}$ \\
\hline
\multicolumn{6}{c}{ \rule{0pt}{2.6ex} Kepler-10\,d} \\ [0.3ex]
\hline
$P$ [days] & - & - & - & 101.36$^1$ & $102 \pm 1$ \\ [0.3ex]
$K$ [\ms]  & - & - & - & - & $0.85^{+0.24}_{-0.14}$ \\ [0.3ex]
$\mp$ [\Me] & - & - & - & 6.84$^1$ & $5.90^{+1.70}_{-1.01}$  \\ 
\hline
\end{tabular}
\label{table_planet_masses_literature}
\begin{flushleft}
\footnotemark[1]{Uncertainties were not provided.}
\end{flushleft}
\end{table*}

In addition to comparing the mass determinations obtained with HARPS-N and HIRES, 
W16 confirmed the variations in the Kepler-10\,c transit times with an amplitude of $\sim5$~min, 
which were previously discovered by \citet{2015ApJ...813...14K}. Such transit timing variations (TTVs)
indicate the presence of an additional planet in the system, Kepler-10\,d, which dynamically perturbs Kepler-10\,c. 
In an attempt to take the constraints of TTVs into account, W16 performed a preliminary combined fit of TTVs and RVs, though (i) considering for simplicity a limited number of possible orbital configurations, near 2:1 or 3:2 resonance, for Kepler-10\,d, and (ii) fixing the so-called TTV ``super-period'' at the peak of the periodogram of the TTVs at 475.22~d (see Sect.~6 in W16).
By doing so, W16 found as their best solution that Kepler-10\,d is in apparent 2:1 resonance with 
Kepler-10\,c, and has a period of $101.36$~d and a mass of 6.84~\Me.

To try and reconcile the discrepancy in the RV semiamplitude, 
and hence planetary mass/density of Kepler-10\,c, with the HIRES and HARPS-N data,
\citet{2017MNRAS.471L.125R} employed a framework based on Gaussian process (GP) regression 
with a quasi-periodic kernel to account for possible correlated noise (e.g., \citealt{2014MNRAS.443.2517H, 2015ApJ...808..127G}), 
by simultaneously modeling the RVs and some activity indicators, 
such as the $\log{R^{'}_{\rm HK}}$ and bisector activity indices \citep{Rajpaul2015}.
As a result, they found $K_{\rm c}=1.27^{+0.42}_{-0.35}$~\ms and $K_{\rm c}=1.64^{+0.42}_{-0.34}$~\ms from 
the HIRES and HARPS-N data, respectively, and $K_{\rm c}=1.41^{+0.25}_{-0.23}$~\ms when combining 
both datasets. The latter $K_{\rm c}$ would imply $M_{\rm c}=7.37^{+1.32}_{-1.19}$~\Me 
and $\rho_{\rm c}=3.14^{+0.63}_{-0.55}~\rm g\,cm^{-3}$, 
which are considerably lower than those measured by both D14 and W16.
From model comparison using Bayesian evidence, \citet{2017MNRAS.471L.125R} also found that 
the three-planet model was favored over the two-planet one, and reported both the orbital period and mass of Kepler-10\,d, 
namely $P_{\rm d}=102\pm1$~d and $M_{\rm d}=5.90^{+1.70}_{-1.01}$~\Me, 
in agreement with the findings of W16. 
Moreover, they found a higher evidence for the GP model, and derived a period of $55.5 \pm 0.8$~d for the quasi-periodic variations, though 
it is not clear whether that period is related to the stellar rotation \citep{2017MNRAS.471L.125R}. 

Table~\ref{table_planet_masses_literature} summarizes the RV semiamplitudes and corresponding masses of the Kepler-10 planets as retrieved in the different aforementioned works from the literature. Since it is clear from all these works that Kepler-10 is a complex planetary system, we continued to monitor it 
in 2017, 2019, and 2020, by practically doubling the number of HARPS-N RVs within the HARPS-N/GTO program.
After extracting most of the HARPS-N RVs with upgraded spectra reduction softwares (Sect.~\ref{radial_velocities}), 
we analyzed them with several tools (Sects.~\ref{RV_periodograms} and \ref{RV_modeling}), also in combination with the Kepler-10\,c TTVs (Sect.~\ref{sec:analysis_dynamical}).
In this way we improved the characterization of the Kepler-10 planetary system (Sect.~\ref{planet_parameters}) 
with regard to both its orbital architecture and the determination of
precise masses, and thus possible compositions, of the Kepler-10\,b and Kepler-10\,c transiting planets. 
This in turn allowed us to put additional constraints on the formation and evolution of the Kepler-10
planetary system (Sect.~\ref{discussion_conclusions}).

\section{Data}

\subsection{Kepler photometry} 
Even though the \emph{Kepler} light curve was analyzed in several previous works 
(\citealt{2011ApJ...729...27B, 2014ApJ...781...67F}; D14; W16; \citealt{2019ApJ...883...79D, 2022A&A...658A.132S}), 
we performed a new analysis of the \emph{Kepler} data to 
(i) re-compute the twenty-four transit times of Kepler-10\,c in short (58~s) cadence, and the first two transit epochs  observed in long-cadence mode (29.4~min), which are not provided by W16;
and (ii) re-derive the Kepler-10\,c transit parameters.

\subsection{Radial velocities}
\label{radial_velocities}
\subsubsection{HARPS-N radial velocities}
In total, we gathered 308 HARPS-N spectra, 55 of which before the failure of the red side of the charge-coupled device (CCD) 
in late September 2012 \citep{2014A&A...572A...2B}, and 253 after the replacement of the CCD. 
The former 55 spectra were reduced with the original HARPS-N data reduction software (DRS) version 3.7, and the latter 253 with the ESPRESSO (Echelle SPectrograph for Rocky Exoplanets and Stable Spectroscopic Observations) DRS version 2.3.5 adapted to HARPS-N \citep{2021plat.confE.106D}. The radial velocities were extracted from all the spectra by cross-correlating them with a G2V stellar template (e.g., \citealt{2002A&A...388..632P}).
The new DRS pipeline allows for better long-term RV accuracy, due to a careful selection of non-saturated thorium and argon lines that are used to calibrate the spectrograph wavelength solution. 
In the DRS-3.7 Kepler-10 RV data, significant variations in the flux of the original thorium-argon calibration lamp induced an artificial quadratic long-term trend of $\sim5$~\ms\ until the replacement of the thorium-argon lamp at the beginning of June 2020, which further produced an RV offset at the level of a \ms.
Those systematics disappear or are at least strongly mitigated to a level below the \ms\, when using the new pipeline. In addition, the new DRS uses a novel algorithm to determine the wavelength solution that is more stable from calibration to calibration. The night-to-night RV rms offset due to wavelength solution calibration is reduced from $\sim$80 to $\sim$50\cms \citep{2021plat.confE.106D}. 
However, no significant improvements from the new DRS are expected for the first 55 HARPS-N spectra taken with the old CCD, given both the relatively small amount of data and the short timespan coverage. For this reason, we used the original HARPS-N DRS (version 3.7) for those spectra.

Given the high complexity of the Kepler-10 system, as shown by the discrepancy in the Kepler-10\,c mass determination 
from previous works, we also made use of the YARARA-v2 tool
\citep{cretignier2021, 2022A&A...659A..68C} to correct for telluric lines and subtle instrumental effects 
that cannot be accounted for by the new DRS pipeline, as well as for possible stellar activity variations, despite the low stellar activity level. In doing so, we had to discard 17 out of 253 HARPS-N spectra ($\sim7$\%) because they did not pass the signal-to-noise ratio (S/N) and vetting criteria of YARARA-v2. 

\begin{figure}[t!]
\centering
\vspace{-1.2cm}
\includegraphics[width=9.5cm, angle=180]{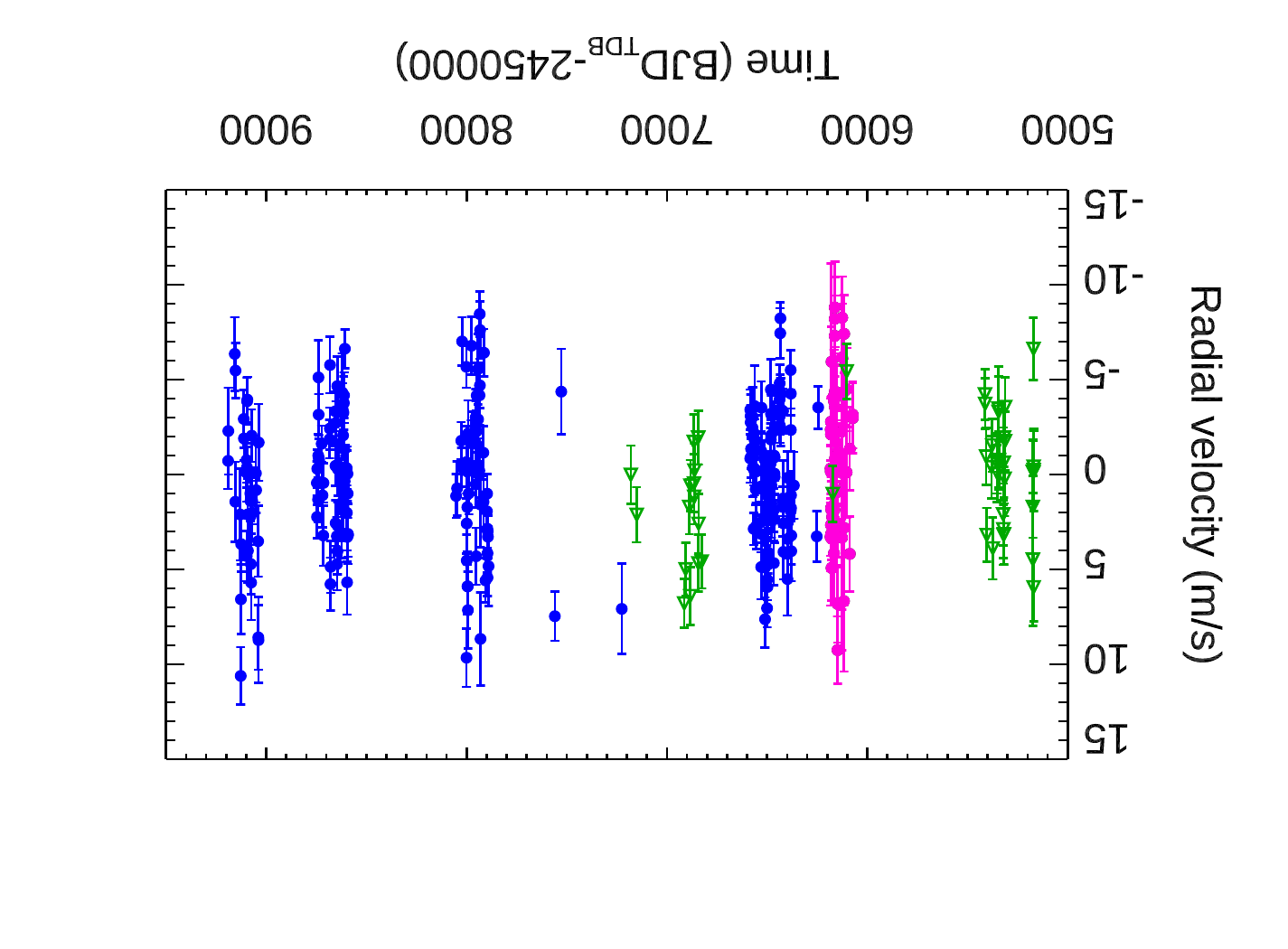}
\caption{Kepler-10 radial velocities. Filled blue and magenta circles show the HARPS-N data collected with the new and old CCD, respectively, 
and empty green triangles display the HIRES measurements. 
Radial-velocity zero points as determined with the DE-MCMC analysis (Sect.~\ref{DE-MCMC_analysis} and Table~\ref{table_harpsn-hires_param}) 
were subtracted from each dataset. }
\label{figure_RVvsTIME_HARPSN-HIRES}
\end{figure}

\begin{figure*}[t!]
\centering
\includegraphics[width=18.5cm]{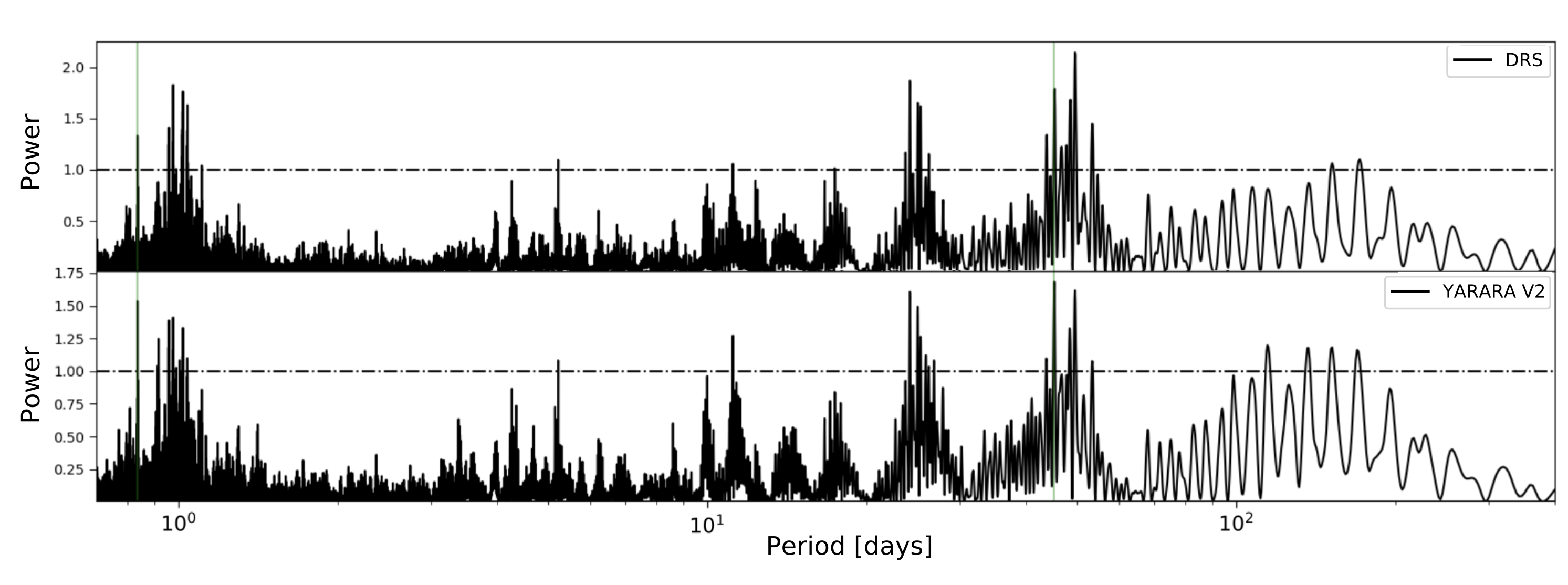}
\caption{Generalized Lomb-Scargle periodograms of the 236 HARPS-N radial velocities as reduced with
the new DRS (top panel) and the new DRS+YARARA-v2 (bottom panel). The power of the periodograms was normalized by the 1\% false alarm probability (FAP) level. 
Note the increase in power of the peaks at the periods of Kepler-10\,b and Kepler-10\,c (vertical green lines) 
with the YARARA-v2 reduction.}
\label{figure_DRSvsYARARA}
\end{figure*}

Both the 55 DRS-3.7 and 236 YARARA-v2 HARPS-N RVs used in this work are shown in Fig.~\ref{figure_RVvsTIME_HARPSN-HIRES}. They were released by  \citet{2023A&A...677A..33B} along with the stellar activity indicators full width at half maximum (FWHM), contrast and bisector span of the cross-correlation function (CCF), and the CaII H\&K S-index and $\log{R^{'}_{\rm HK}}$\footnote{We point out that the CCF contrast and FWHM activity indicators are affected by strong variations, which are highly correlated, due to changes of the HARPS-N focus. However, this did not result in any systematic RV variation, as the product $\rm Contrast \cdot FWHM$ is conserved.}. 
The improvement achieved with YARARA-v2 can be seen in Fig.~\ref{figure_DRSvsYARARA},
which shows the increase in power of the peaks at the expected orbital periods of both Kepler-10\,b and Kepler-10\,c 
(from \emph{Kepler} photometry) in the generalized Lomb-Scargle (GLS, \citealt{2009A&A...496..577Z}) periodogram of the 236 HARPS-N RVs  reduced with the new DRS+YARARA-v2 (bottom panel) compared to the new DRS (top panel). 
For this reason, we used the former for the 236 RVs collected from 2013 to 2021 (no remarkable improvement is instead expected for the 55 RVs gathered earlier for the same aforementioned reasons that we did not apply the new DRS for their extraction).
We note that we checked for possible outliers in the HARPS-N RVs by using Chauvenet's criterion (e.g., \citealt{2023A&A...677A..33B}), and found none.

\subsubsection{HIRES radial velocities}
The HIRES Kepler-10 spectra were recorded using different deckers that define the length of the effective slit.  
The first 17 HIRES observations that spanned approximately 38 days were made with the B5-decker (0.861 x 3.5~$\rm arcsec^2$)\footnote{\url{https://www2.keck.hawaii.edu/inst/hires/manual2.pdf}}. The longer C2-decker (0.861 x 14~$\rm arcsec^2$), which allows for better subtraction of night sky emission and scattered moonlight as compared to the B5-decker, was used for most of the subsequent observations.

We investigated the quality of the B5 RV set by fitting a two-Keplerian (i.e., two-planet) model to subsets of the data and analyzing the best-fit parameters. The importance of the consistency of the extracted semiamplitude in time for subsets of data is discussed in \citet{hara2022b}. For this analysis, we used PyORBIT \citep{2016ascl.soft12008M} by fixing the periods and transit times of planets b and c, and assuming circular orbits. The only free parameters were the semiamplitudes of the two planets and the stellar RV jitter. Since planet b has a very short orbital period, we expect to get consistent semiamplitudes for this planet across the data as long as we include a sufficient number of RVs.
Fitting this model to the first 17 B5-decker observations in W16, we measured a remarkably large semiamplitude of 4.8\ms for planet b followed by a drop to 1.6\ms for the closest two-month chunk of C2-decker RVs.
We also compared the B5-decker RVs with RV chunks of similar duration within the HARPS-N data with and without the YARARA-v2 correction. For this, we randomly selected two-month windows of data with at least 15 observations, to match the characteristics of the B5 RV set, and fitted the two-Keplerian model described above. The median number of included measurements was equal to 20, which is comparable to the number of B5-decker observations. The extracted semiamplitude of planet b reached around 4\ms for only a very few subsets of the data; none reached 4.8\ms as extracted for the B5-decker RVs, and we found no semiamplitude jumps as large as the one between the B5 and the C2-decker RVs.
Therefore, the B5 data appear to be contaminated. To avoid the few noisy B5 measurements, which complicate a meaningful extraction of orbital information, and keep the dataset uniform, we considered only the C2 RVs.

As for the HARPS-N data, we searched for possible outliers in the HIRES C2-decker RVs with Chauvenet's criterion, and found four measurements
at the epochs 2455314.006, 2455344.978, 2456908.977, and 2456909.885 $\rm BJD_{UTC}$, which were thus discarded.

\section{Data analysis} 
\label{data_analysis} 

\subsection{Transit timing variations and parameters of Kepler-10\,c}
\label{Transit_fitting_Kepler10c}
\begin{figure}[t!]
\centering
\includegraphics[width=0.90\columnwidth]
{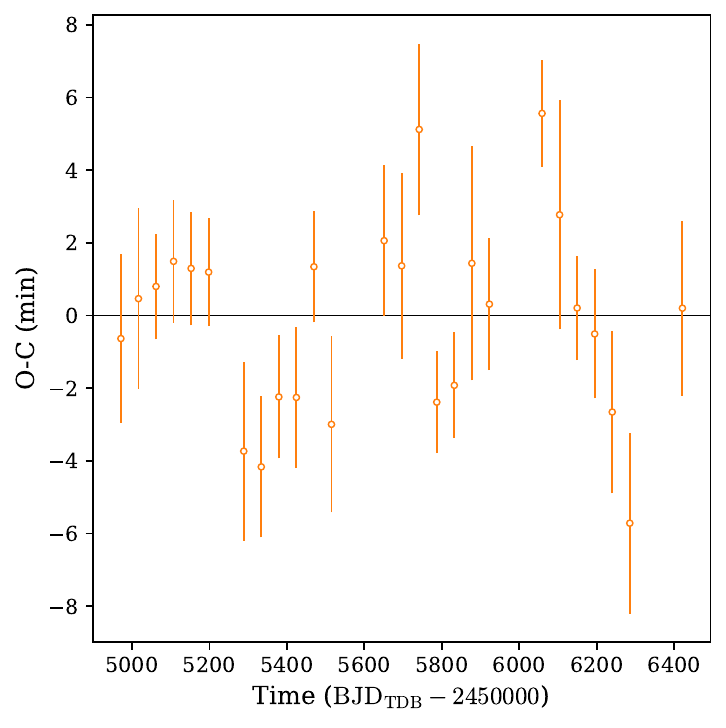}
\caption{Kepler-10\,c observed-calculated (O-C) diagram showing the TTV pattern.
The calculated times ($T_{c,\mathrm{lin}}$) are computed from a linear ephemeris: 
$T_{c,\mathrm{lin}} = T_\mathrm{ref} + N \times P = 2454971.678363 \pm  0.000659 + N \times 45.294278 \pm  0.000039$,
where $N$ is an integer number that identifies each transit time with respect to the reference time $T_\mathrm{ref}$.
}
\label{fig:K10c_OC}
\end{figure}

To compute the Kepler-10\,c transit timing variations, 
we first fitted each transit independently by (i) using the same differential evolution Markov chain Monte Carlo (DE-MCMC; \citealt{TerBraak2006, 2013PASP..125...83E}) Bayesian framework as in D14; 
(ii) considering intervals of the light curve centered at the predicted times  of the Kepler-10\,c transits from the linear ephemeris in D14, with a temporal window of twice the transit duration;
(iii) adopting a two-planet transit model in case of superposition of the Kepler-10\,b and Kepler-10\,c transits, by fixing the parameters of the Kepler-10\,b transits to the solution of D14; (iv) imposing a Gaussian prior on the stellar density, as derived from previous asteroseismic analyses of the \emph{Kepler} light curve; and (v) fixing both the orbital period of planet c and the coefficients of the quadratic limb-darkening law to those found by D14 (no significant changes were noted when adopting instead Gaussian priors with the values and uncertainties given in D14). 
For the first two transits observed in long-cadence mode (29.4 min), we oversampled the model to 1-min samples \citep{2010MNRAS.408.1758K}. 
The Kepler-10\,c transit times are given in Table~\ref{table_TTVs}; their variations with respect to a linear ephemeris are shown in Fig.~\ref{fig:K10c_OC}, and are fully consistent with those reported by both \citet{2015ApJ...813...14K} and W16.

We performed a subsequent, simultaneous DE-MCMC modeling of all the Kepler-10\,c transits as in D14, after adjusting the observed-calculated (O-C) offsets from the previously computed transit times. The determined transit parameters are in excellent agreement with those from previous works, and are reported in Table~\ref{Table_Kepler-10_system_parameters}.

\begin{table}
\small
\centering
\caption{Transit times of Kepler-10\,c.}
\begin{tabular}{c c c} 
\hline
Transit & Transit Times  &  Uncertainty  \\
        & [$\rm BJD_{TDB}-2454900$]    &    [days]  \\
\hline
0   &      71.6779   &   0.0016   \\
1   &     116.9730   &   0.0017   \\
2   &     162.2675   &   0.0010   \\
3   &     207.5622   &   0.0012   \\
4   &     252.8564   &   0.0011   \\
5   &     298.1506   &   0.0010   \\
6   &     388.7357   &   0.0017   \\
7   &     434.0297   &   0.0013   \\
8   &     479.3253   &   0.0012   \\
9   &     524.6196   &   0.0013   \\
10  &     569.9164   &   0.0011   \\
11  &     615.2076   &   0.0017   \\
12  &     751.0940   &   0.0014   \\
13  &     796.3878   &   0.0018   \\
14  &     841.6846   &   0.0016   \\
15  &     886.9737   &   0.0010   \\
16  &     932.2683   &   0.0010   \\
17  &     977.5649   &   0.0022   \\
18  &    1022.8584   &   0.0013   \\
19  &    1158.7449   &   0.0010   \\
20  &    1204.0372   &   0.0022   \\
21  &    1249.3297   &   0.0010   \\
22  &    1294.6235   &   0.0012   \\
23  &    1339.9163   &   0.0015   \\
24  &    1385.2085   &   0.0017   \\
25  &    1521.0954   &   0.0017   \\
\hline
\end{tabular}
\label{table_TTVs}
\end{table}

\subsection{Analysis of the transit timing variation signal}
\label{sec:analysis_TTV}

We analyzed the TTV signal presented in Fig. \ref{fig:K10c_OC} to see what constraints we can obtain on the presence of non-transiting planets, and to make a comparison with the signal found in the RV analysis.
A periodogram of the TTVs shown in Fig. \ref{fig:K10c_OC} yields a peak at $P_{\rm TTV}=466.2_{-16.2}^{+17.2}$ days, which encompasses with $1\sigma$ the value given by W16. TTVs with timescales of tens to a few hundred times the orbital period of the planets are typically attributed to proximity to a mean motion resonance (MMR) with an additional planet in the system, defined by $P_{\rm out}/P_{\rm in} =(k+q)/k$, with $k$ and $q$ integers. Either the pair is inside the MMR, in which case the period of the TTVs scales as $P~(M_{\rm p}/M_\star)^{-2/3}$ \citep[e.g.,][]{NeVo2016}, or the pair of planets is relatively close to the MMR, in which case the main TTV periodicity depends only on the orbital periods of the planets \citep[e.g.,][]{Lithwick2012}. Furthermore, short-term TTVs can be attributed to synodic "chopping" effects \cite[e.g.,][]{Deck2015}.  

First, we assumed that the perturbing planet is inside a MMR with Kepler-10\,c. Given the Kepler-10\,c orbital period of 45.2943~d, we estimated the mass that a perturbing planet would need to generate TTVs with a period of $\sim 466$ days using the formulas from \cite{NeVo2016}. We found that the perturbing planet should have a mass ranging between about one Saturn mass (for a 7:6 resonance, either inside or outside Kepler-10\,c) and about 10 Jupiter masses (for a 2:1 resonance either inside or outside Kepler-10\,c). In addition to the fact that such a massive object should have easily been found in the RV data, these configurations are also at the edge of instability \citep{DePaHo2013}. 

We then assumed that the perturbing planet is not inside, but relatively close to, a MMR of the form 
$P_{\rm out}/P_{\rm in} =(k+q)/k$, with $q=1$ or $2$. The main TTV frequency now depends only on the orbital period of Kepler-10\,c, which we fixed to 45.2943~d, and the perturbing planet, with a period called the ``super period'' $1/(k/P_{\rm in}-(k+q)/P_{\rm out})$ \citep[e.g.,][]{Lithwick2012}. For each  resonance considered, the perturbing planet can be either the inner or the outer planet. Then, for each of these configurations, the perturbing planet can be on either side of the MMR, since we only measure the absolute value of the super period. Supposing that $P_{\rm TTV}$ is such a super period, we can derive the orbital period of the companion, depending on the MMR it is close to. These potential companion periods are summarized in Table \ref{tab:super}.

\renewcommand{\arraystretch}{1.33}

\begin{table}
\small
\centering
\caption{Possible orbital periods of the perturbing planet d in the case of proximity to mean motion resonance.}
\begin{tabular}{c c c c} 
\hline
k+q & k  &  Period  & Period \\
 &  & (interior to MMR)  & (exterior to MMR) \\
\hline
\multicolumn{4}{c}{$P_{\rm d} < P_{\rm c}$}\\
\hline
$ 2 $&$ 1 $&$ 21.598_{-0.036}^{+0.036} $&$ 23.803_{-0.043}^{+0.044} $ \\
$ 3 $&$ 2 $&$ 29.248_{-0.033}^{+0.033} $&$ 31.206_{-0.037}^{+0.038} $ \\
$ 4 $&$ 3 $&$ 33.164_{-0.028}^{+0.028} $&$ 34.816_{-0.031}^{+0.031} $ \\
$ 5 $&$ 4 $&$ 35.544_{-0.024}^{+0.024} $&$ 36.953_{-0.026}^{+0.026} $ \\
$ 6 $&$ 5 $&$ 37.143_{-0.021}^{+0.021} $&$ 38.366_{-0.022}^{+0.023} $ \\
$ 7 $&$ 6 $&$ 38.291_{-0.019}^{+0.019} $&$ 39.369_{-0.020}^{+0.020} $ \\
$ 3 $&$ 1 $&$ 14.624_{-0.017}^{+0.016} $&$ 15.603_{-0.019}^{+0.019} $ \\
$ 5 $&$ 3 $&$ 26.658_{-0.018}^{+0.018} $&$ 27.714_{-0.019}^{+0.020} $ \\
$ 7 $&$ 5 $&$ 31.909_{-0.016}^{+0.016} $&$ 32.808_{-0.016}^{+0.017} $ \\
\hline
\multicolumn{4}{c}{$P_{\rm d} > P_{\rm c}$}\\
\hline
$ 2 $&$ 1 $&$ 82.57_{-0.26}^{+0.26} $&$ 100.33_{-0.38}^{+0.39} $ \\
$ 3 $&$ 2 $&$ 64.79_{-0.11}^{+0.11} $&$ 71.41_{-0.13}^{+0.13} $ \\
$ 4 $&$ 3 $&$ 58.497_{-0.066}^{+0.065} $&$ 62.412_{-0.074}^{+0.075} $ \\
$ 5 $&$ 4 $&$ 55.274_{-0.047}^{+0.047} $&$ 58.026_{-0.051}^{+0.052} $ \\
$ 6 $&$ 5 $&$ 53.316_{-0.037}^{+0.036} $&$ 55.429_{-0.039}^{+0.040} $ \\
$ 7 $&$ 6 $&$ 52.000_{-0.030}^{+0.029} $&$ 53.712_{-0.031}^{+0.032} $ \\
$ 3 $&$ 1 $&$ 123.85_{-0.39}^{+0.39} $&$ 150.50_{-0.57}^{+0.59} $ \\
$ 5 $&$ 3 $&$ 73.121_{-0.083}^{+0.082} $&$ 78.015_{-0.093}^{+0.094} $ \\
$ 7 $&$ 5 $&$ 62.202_{-0.043}^{+0.042} $&$ 64.667_{-0.045}^{+0.046} $ \\
\hline
\end{tabular}
\tablefoot{The period of planet c was fixed at 45.2943~d for this analysis.}
\label{tab:super}
\end{table}

Interestingly, the signal at $151.06\pm0.48$~d found in the HARPS-N RVs (see Sects.~\ref{RV_periodograms}, \ref{RV_modeling}, and Table~\ref{table_harpsn_param}) is $1\sigma$ consistent with the orbital period of $150.50_{-0.57}^{+0.59}$~d required to generate the observed TTV period if the companion is relatively close to the 3:1~MMR with Kepler-10\,c on an outer orbit. Given the relatively low uncertainty on both these periods, the probability that this is due to chance is quite low. The TTV signal therefore agrees with an orbital period of $\sim 151$~d for Kepler-10\,d.
In addition, the abovementioned 3:1 resonance is of second order ($q=2$), and so the amplitude of the TTVs induced by the proximity to this resonance vanishes for zero eccentricities \citep{HaLi2016}. 
This explains the relatively low TTV amplitude observed, but also implies that other TTV harmonics than the proximity to the 3:1~MMR might have comparable amplitudes.

\begin{figure}
\centering
\includegraphics[width=0.49\columnwidth]{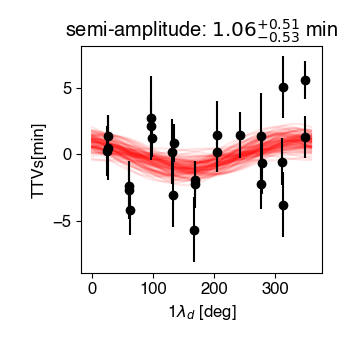}
\includegraphics[width=0.49\columnwidth]{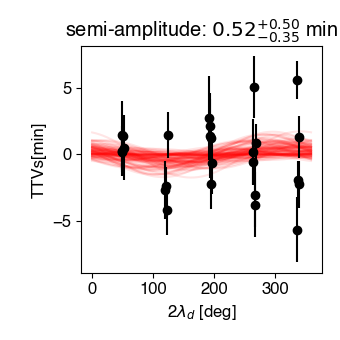}\\
\includegraphics[width=0.49\columnwidth]{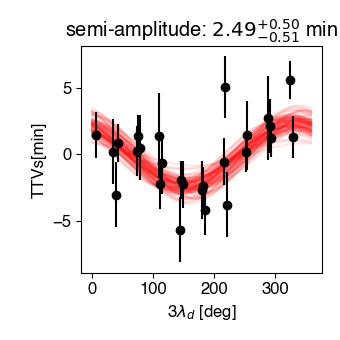}
\includegraphics[width=0.49\columnwidth]{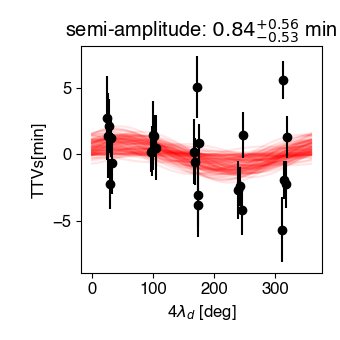}
\caption{TTVs of Kepler-10\,c folded with respect to harmonics of the mean longitude of Kepler-10\,d ($\lambda_d$) at the times of transit.}
\label{fig:TTVs_harmonics}
\end{figure}

We therefore verified if the TTV signal contains additional information on the interactions between planets c and d. Generalizing the results of \citet{Deck2015}, we know that the TTV signal expands as a function of the harmonics of the mean longitude of the perturbing planet ($\lambda_{\rm d}$). We show the TTVs folded with respect to these harmonics in Fig.~\ref{fig:TTVs_harmonics}. The amplitude of each harmonic $j\lambda_{\rm d}$ is the sum of the contribution of the synodic (i.e., conjunction) harmonic $j(\lambda_{\rm c}-\lambda_{\rm d})$ and the effect of the MMRs of the form $P_{\rm d}/P_{\rm c} = j/m$, with $m$ an integer. As can be seen in Fig.~\ref{fig:TTVs_harmonics}, only the third harmonic shows a significant TTV amplitude. Since it is this harmonic that contains the effect of the 3:1 MMR, we therefore conclude that the proximity to this resonance is likely dominating the TTV signal.

From this analysis, we can draw three conclusions: (i) the main periodicity observed in the TTVs of planet c can be explained by the signal at 151~d also seen in the HARPS-N RVs; (ii) since the relative proximity to this resonance generates TTVs for eccentric planets only, we expect Kepler-10\,c and d to have non-zero eccentricities; and (iii) even low S/N TTVs can help confirm periodic signals found in RVs.

\subsection{Periodogram analyses of radial velocities.}
\label{RV_periodograms}
To search for periodic signals attributable to planets other than Kepler-10\,b and Kepler-10\,c in the HARPS-N RVs, in view of the fact that at least one more planet is expected from the TTVs of Kepler-10\,c, 
we performed two periodogram analyses: the classical GLS periodogram and the 
false inclusion probability (FIP) periodogram \citep{hara2024}.

\subsubsection{Generalized Lomb-Scargle periodograms and iterative planet fitting}
The peaks at the periods of Kepler-10\,b and Kepler-10\,c are clearly visible in the GLS periodogram of the HARPS-N RVs (Figs.~\ref{figure_GLS_periodograms} and \ref{figure_DRSvsYARARA}). 
To search for additional signals, we first modeled the RV signals of Kepler-10\,b and Kepler-10\,c as in Sect.~\ref{DE-MCMC_analysis}, 
removed them from the original RV time series, and ran again the GLS periodogram on the residuals of the two-planet model. 
This procedure leads to a better fit of the planet signals than the sinusoidal fit from the GLS parameters, 
allowing in particular a slightly eccentric Keplerian fit for Kepler-10\,c.
The periodogram of the residuals showed a third signal with a period of $150.8\pm0.7$~d and a theoretical false alarm probability (FAP) 
of $3.4\cdot 10^{-7}$ (see Fig.~\ref{figure_GLS_periodograms}). We then performed a three-planet modeling, searched for other signals with the GLS periodogram 
in the best-fit residuals, and found a fourth signal with a period of $83.08\pm0.27$~d and a FAP of $3.6\cdot 10^{-4}$ (Fig.~\ref{figure_GLS_periodograms}); the power at $\sim 26$~d seen in Fig.~\ref{figure_DRSvsYARARA} instead disappeared, being somehow related to both the signals of the three planets and the data sampling. However, the inclusion of this fourth signal was not favored by Bayesian model comparisons (Sect.~\ref{RV_modeling}), and thus it was not considered in the final modeling. 
None of these signals was found in the usable stellar magnetic activity indicators, such as the S-index and the bisector of the CCF.

\subsubsection{False inclusion probability}

Periodograms are prone to aliasing. As a consequence, a combination of planetary signals and noise can create peaks at periods that do not correspond to the planet orbital period or the stellar rotation period or one of its harmonics \citep{hara2017,nava2020}. To assess precisely which signals are present, it is better practice to search for several planets and an appropriate noise model simultaneously. In this section, we outline an analysis which is described in more detail in Appendix \ref{app:fip_asp}.

\begin{figure*}[h!]
    \centering    \includegraphics[width=\linewidth]{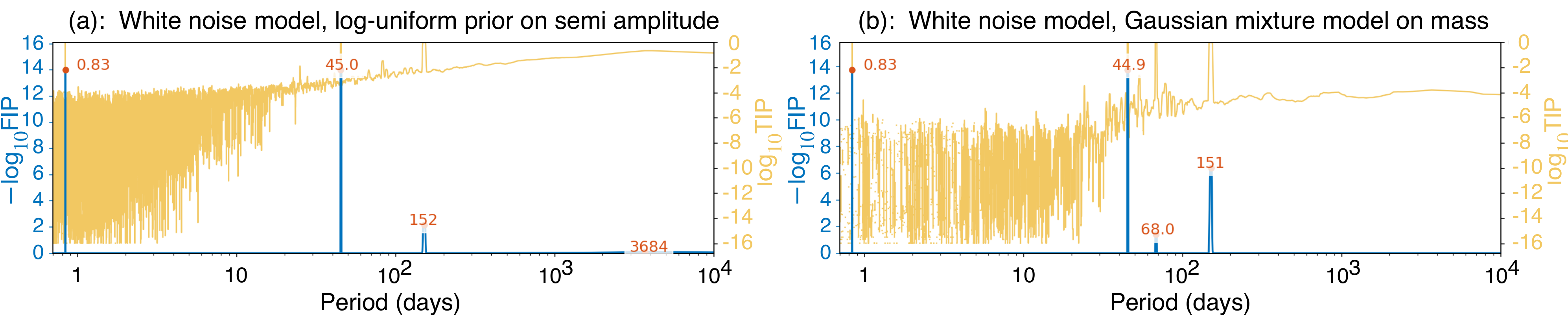}
    \caption{False inclusion probability periodograms of the Kepler-10 HARPS-N data processed with YARARA-v2 for different priors on semiamplitude. (a) was obtained with a log-uniform prior on semiamplitude, (b) with a Gaussian mixture model on mass. In yellow we represent the true inclusion probability (TIP), the probability of having a planet in the range $[\omega - \Delta \omega,  \omega + \Delta \omega]$ as a function of $\omega$. In blue, we represent $-\log_{10} FIP$, where FIP = 1 - TIP is the false inclusion probability.}
    \label{fig:fip_perio}
\end{figure*}

We first applied the Bayesian methodology described in \cite{hara2022a}, whose end goal is to compute directly the posterior probability of the presence of a planet. 
By $T_{\mathrm{obs}}$ we denote the total observation timespan, and define $\Delta \omega = 1/T_{\mathrm{obs}}$. 
We considered a grid of tightly spaced frequencies $\omega_k$ spanning from 0 to 1.5 cycles per day; for each $\omega_k$ in the grid,  we computed the true inclusion probability (TIP), defined as the posterior probability that there is a planet with an orbital frequency in $[\omega_k - \Delta \omega,  \omega_k + \Delta \omega]$. 
As shown in \cite{hara2024}, this detection criterion is optimal in the sense that, provided the likelihood and prior models are correct, it maximizes true detections for a given tolerance to false ones.

The computation of the TIP was done for a certain choice of priors and likelihood functions. We employed uniform priors on angles, a Beta prior on eccentricity as in \cite{kipping2014}, and a log-uniform prior on the period. As shown in \cite{hara2022a}, the prior on the semiamplitude and noise models can lead to drastic differences in statistical significance. We used either a log-uniform prior on semiamplitude, or a mixture Gaussian model on the mass, and a white noise model. The results are shown in Fig.~\ref{fig:fip_perio}. 

In all cases, we obtained very significant signals at 0.83 and 44.9~d with a false inclusion probability (FIP, equal to 1 - TIP), that is, a probability of not being present, of less than $10^{-12}$. Depending on the priors, the FIP of the 150-152~d signal varied from $7\cdot 10^{-2}$ to $5\cdot10^{-6}$. We took a further step and computed the TIP marginalized on the two noise models, white and red (see \ref{app:fip} for details). 
To further investigate the origin of the 151~d signal and try to evaluate whether it was strictly periodic or not, we applied the methods of \cite{hara2022b}. The analysis presented in  detail in Appendix~\ref{app:fip_asp} shows that the 151~d signal is compatible with a purely sinusoidal signal.

\subsection{Modeling of radial velocities}
\label{RV_modeling}
After revealing planet-induced Doppler signals with periodogram analyses of the HARPS-N RV dataset of Kepler-10, 
we fitted them with a multi-Keplerian model by 
(i) varying the number of planets in the system from two to four; 
(ii) employing three different Bayesian modeling approaches (see Sects.~\ref{DE-MCMC_analysis}-\ref{Polychord_analysis})
in order to evaluate the robustness of the orbital solution, given the previous discrepancies in the literature;  and
(iii) using a Gaussian likelihood function, which accounts for additional uncorrelated (white) noise (e.g., \citealt{2005ApJ...631.1198G}) 
and/or correlated (red) noise through GP regression with quasi-periodic (QP) and/or 
squared exponential (SE) covariance functions (e.g., \citealt{2015ApJ...808..127G, Rajpaul2015}).

The Keplerian signal of each planet was modeled with five parameters: 
the orbital period $P$; 
the inferior conjunction time $T_{\rm c}$, which is equivalent to the mid-transit time for the transiting planets; 
$\sqrt{e}\cos(\omega)$ and $\sqrt{e}\sin(\omega)$, where $e$ and $\omega$ are the orbital eccentricity and the argument of periastron;
and the RV semiamplitude $K$. 
A zero point $\gamma$ and a jitter term $\sigma_{\rm j}$ were also fitted for each dataset. 
The jitter terms were summed in quadrature to the formal RV uncertainties to 
take additional uncorrelated noise of unknown origin (stellar, instrumental, etc.) into account. 

To check whether possible correlated noise is also present in the data and might be modeled along with the planetary signals, 
despite the very low stellar magnetic activity level, we further considered Gaussian Processes with 
the following hyperparameters: the amplitude $h$ of the GP, the exponential decay term $\lambda$, 
the periodic GP term $P_{\rm rot}$, and the inverse harmonic complexity term $w$. 
The first two hyperparameters are in common between the QP and SE 
covariance functions (kernels).

To be consistent among the different RV analyses, we used the same priors on the model parameters, which are summarized in Table~\ref{table_priors_RV_parameters}, specifically 
\begin{itemize}
\item[--] Gaussian priors on $P$ and $T_{\rm c}$ of Kepler-10\,b and Kepler-10\,c 
from the modeling of the \emph{Kepler} transits (D14); 
uniform priors on $P$ and $T_{\rm c}$ for the signals attributable to additional non-transiting planets.

\item[--] half-Gaussian priors centered on zero with $\sigma_{\rm e}=0.098$ for the orbital eccentricities 
of Kepler-10\,c and Kepler-10\,d, following the distribution of eccentricities in multiplanet systems \citep{2019AJ....157...61V}. 
This prior avoids spurious high eccentricities, and hence dynamical instabilities due to orbit crossings; spurious eccentricities may occur when the RV semiamplitudes are comparable to the noise level as in the present case (e.g., \citealt{2011MNRAS.410.1895Z, 2019MNRAS.489..738H}).

For the innermost planet, Kepler-10\,b, we used instead a circular orbit, because its orbit is expected to be well circularized given 
its extremely small (short) semimajor axis (orbital period; e.g., \citealt{2008ApJ...686L..29M, 2010ApJ...725.1995M}).

\item[--] uniform priors on all the other parameters ($K$, $\gamma$, $\sigma_{\rm j}$, and the GP hyper-parameters $h$, $\lambda$, $P_{\rm rot}$ and $w$).

\end{itemize}

We estimated the values and $1\sigma$ uncertainties of the model and derived parameters from the medians 
and the 15.86\%-84.14\% quantiles of their posterior distributions, as derived with the three different methods
described below. For distributions consistent with zero, we provided only $1\sigma$ upper limits.

\subsubsection{Differential evolution Markov chain Monte Carlo (DE-MCMC)}
\label{DE-MCMC_analysis}
The DE-MCMC method is the Markov chain Monte Carlo (MCMC) version of the differential evolution (DE) 
genetic algorithm \citep{TerBraak2006, 2013PASP..125...83E, 2019arXiv190709480E}. 
A number of chains equal to twice the number of free parameters are run simultaneously, 
and perform an optimized exploration of the parameter space through the automatic choice
of step scales and orientations. The sampling distribution for each step of a given chain consists in 
(i) randomly selecting two other chains; 
(ii) computing the difference of the model parameters of these two chains; 
(iii) determining the proposal step from (ii) and Eq.~2 in \citet{TerBraak2006}; and  
(iv) accepting or rejecting the proposal step according to the Metropolis-Hastings algorithm. 
For the identification of the burn-in steps and the convergence
and well mixing of the DE-MCMC chains, we followed the criteria proposed by \citet{2013PASP..125...83E}. 

We ran a DE-MCMC analysis first on the HARPS-N RVs only, and then on the combined 
HARPS-N and HIRES datasets. We accounted for uncorrelated (white) noise only, 
because the GP analysis with either QP or SE kernels did not achieve convergence of the chains, but led to very low acceptance rates and 
unconstrained hyperparameters, as expected from the low level of stellar magnetic activity. 

The results of the DE-MCMC for the various planet models are listed in Table~\ref{table_harpsn_param} for the HARPS-N RVs.
We used the Bayesian information criterion (BIC; e.g., \citealt{Kass_Raftery_1995, Jeffreys1998,  Mukherjee1998ApJ,   Burnham_Anderson_2004, 2007MNRAS.377L..74L})
to compute the relative probabilities of the models with two, three, and four planets, 
and found $BIC_{\rm 3pl}-BIC_{\rm 2pl}=-10.2$ and 
$BIC_{\rm 4pl}-BIC_{\rm 3pl}=+3.3$. This implies that the three-planet model is favored over both the 
two-planet and four-planet models, being $\sim165$ and 5 times more likely, respectively. 
Moreover, an additional planet with $P\sim83$~d in the four-planet model would not be compatible with the TTVs of Kepler-10\,c (see Sect.~\ref{sec:analysis_dynamical}).
A very important outcome is that the values of the RV semiamplitudes $K$ of planet b and, especially, c 
do not vary significantly with the adopted (two- or three-planet) model, being fully consistent within $1\sigma$ (see Table~\ref{table_harpsn_param}).

Adding the C2-decker HIRES data does not change significantly the planet parameters, as expected from their low number compared to the HARPS-N RV measurements (47 vs 291). The RV semiamplitudes agree
within $1\sigma$ with the values obtained with HARPS-N only (see Table~\ref{table_harpsn-hires_param}), though the RV semiamplitudes of both Kepler-10\,c and Kepler-10\,d 
get slightly lower by $1\sigma$ and have slightly larger relative errors; 
this might be due to some HIRES instrumental effects, the investigation of which goes beyond the scope of this work. 
In any case, we mainly rely on the solution obtained from the modeling of the HARPS-N data, given that the correction of systematics and/or low-level activity variations with YARARA-v2 could be applied to the HARPS-N spectra only.
 
\begin{figure}[h!]
\centering
\includegraphics[width=8.0 cm, angle=180]{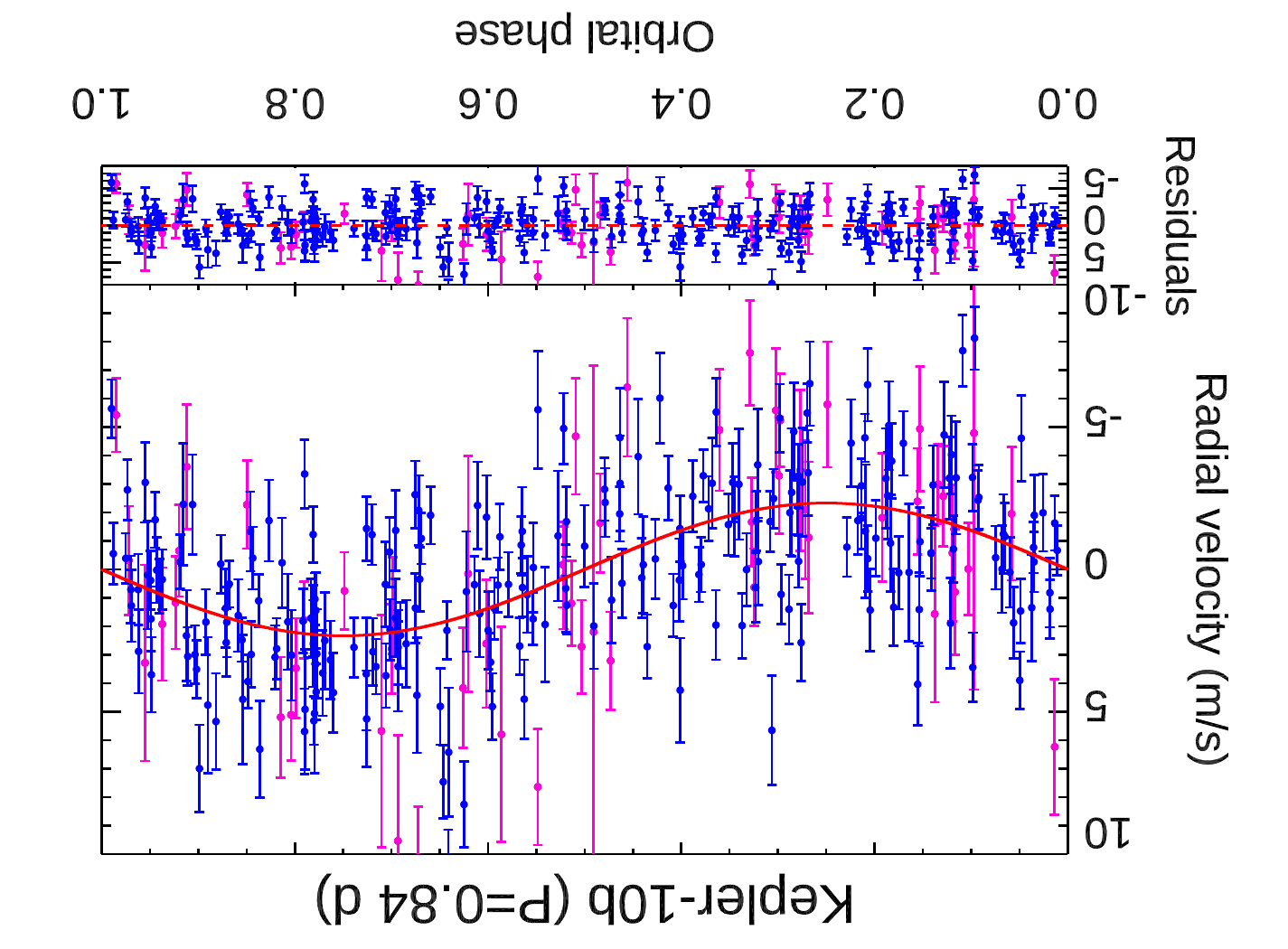}
\includegraphics[width=8.0 cm, angle=180]{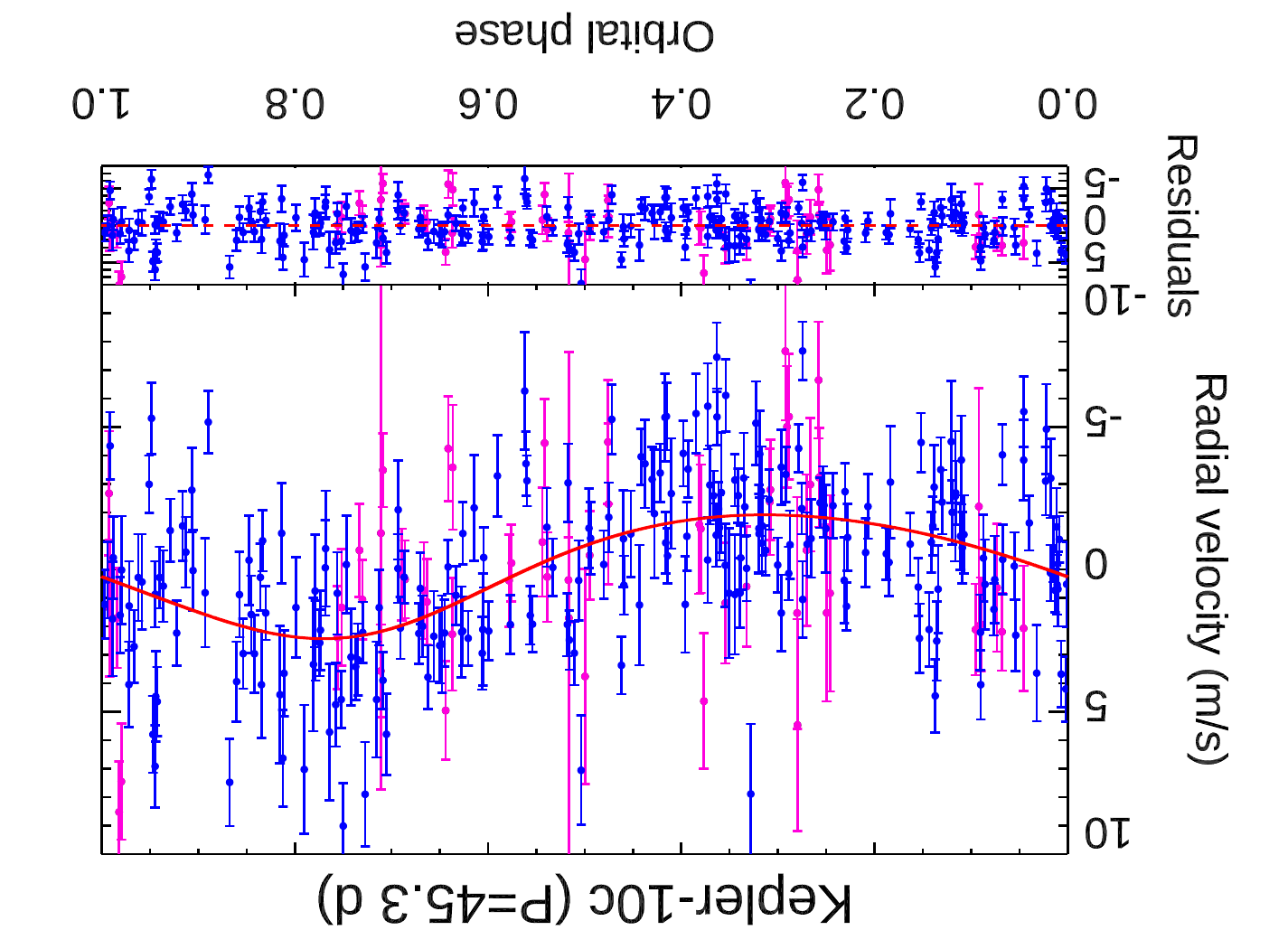}
\includegraphics[width=8.0 cm, angle=180]{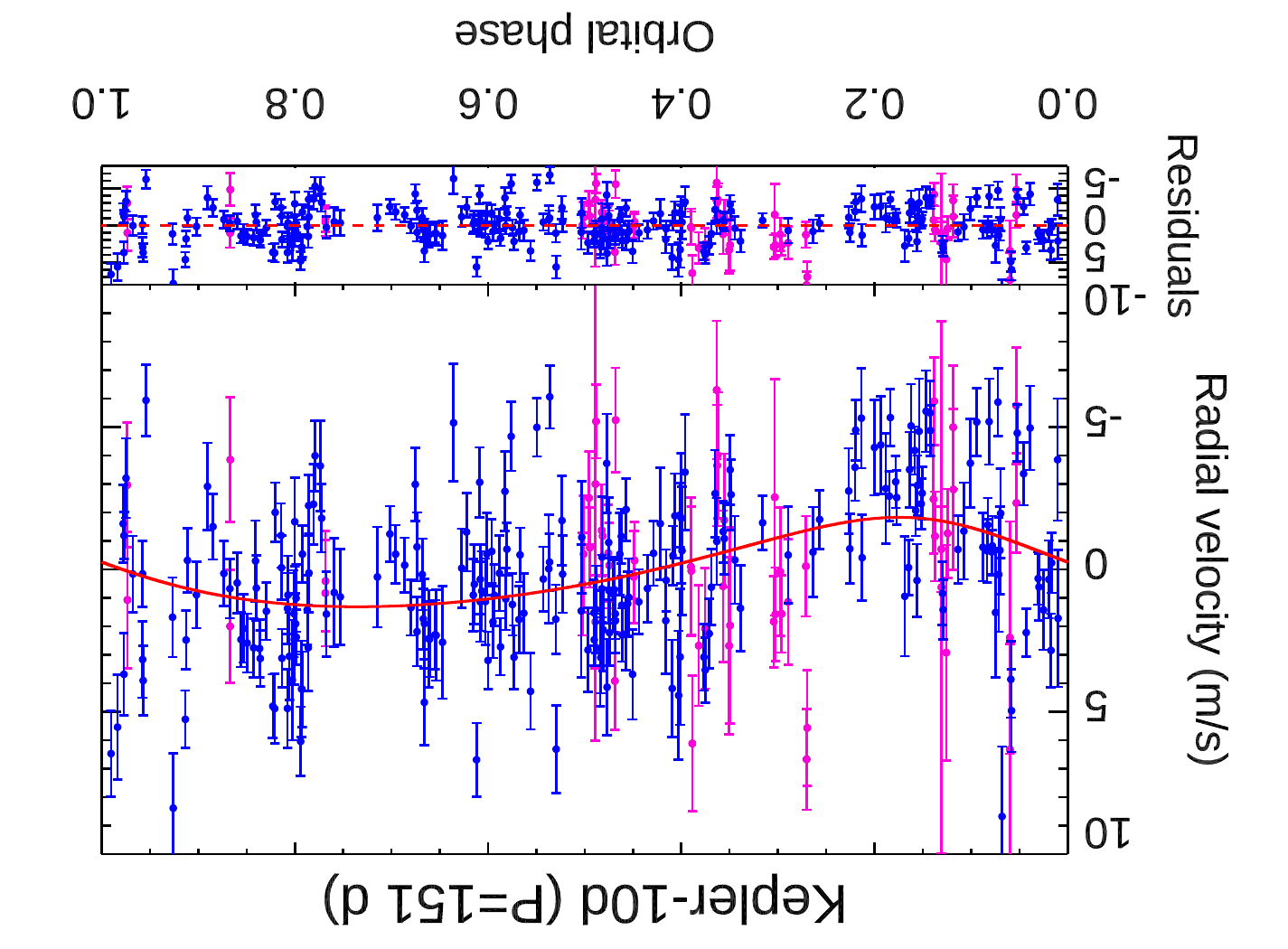}
\caption{Radial-velocity signals of Kepler-10\,b (top panel), c (middle panel),  and d (bottom panel), as a function of their orbital phase (phases 0 and 1 correspond to 
inferior conjunction times). Blue and magenta points refer to the HARPS-N data collected with the new and old CCD, respectively.}
\label{figure_3planet-fit_HARPSN}
\end{figure}

The best-fit with the three-planet model from the DE-MCMC analysis is shown in 
Fig.~\ref{figure_3planet-fit_HARPSN} for the HARPS-N RVs, and 
in Fig.~\ref{figure_3planet-fit_HARPSN-HIRES} for the HARPS-N and HIRES data.

\subsubsection{Nested sampling through \texttt{MultiNest}} 
\label{Nestedsampling_analysis}
The Kepler-10 RVs were also modeled using the publicly available Monte Carlo nested sampler and Bayesian inference tool \texttt{MultiNest v3.10} (e.g., \citealt{Feroz2019}), through the \texttt{pyMultiNest} wrapper \citep{Buchner2014}. The setup was characterized by 500 live points, a sampling efficiency of 0.3, and a Bayesian tolerance of 0.5. The Bayesian model selection was performed by comparing the values of the Bayesian evidence $\ln\mathcal{Z}$ calculated by \texttt{MultiNest} by assigning the same a-priori probability to each model.

We found that (i) the three-planet model is strongly favored over the two-planet model with $\ln\mathcal{Z}_{3\,pl.}-\ln\mathcal{Z}_{2\,pl.}$=8.4, and (ii) the four-planet model is weakly favored over the three-planet model, given $\ln\mathcal{Z}_{4\,pl.}-\ln\mathcal{Z}_{3\,pl.}$=1.1 and the empirical scale reported in \citet{feroz2011}; the posterior distribution of the fourth signal converged to a period of $P=82.95^{+0.20}_{-4.64}$~d, encompassing the 1-yr alias of the main mode at $83$~d.

\subsubsection{Nested sampling through \pchord ~with and without multidimensional Gaussian processes}
\label{Polychord_analysis}

The Kepler-10 RVs show only weak (linear) correlation with activity indicators, as expected for an old, quiet star. Specifically, the Pearson's coefficient between RV and BIS (\lrhk) measurements is $r=23.1\%, p=0.09$ ($r=-8.5\%, p=0.55$) before the CCD replacement, and $r=-15.0\%, p=0.021$ ($r=4.0\%, p=0.54$) afterwards.
Nevertheless, we used the multidimensional (MD) GP framework developed by \citet{Rajpaul2015} to model RVs simultaneously with \bis\ and \lrhk\ observations. This GP framework assumes that any observed stellar activity signals are generated by some underlying latent function and its derivatives; this function, not observed directly, is modeled with a GP \citep{rasmussen2006,roberts2013}. The framework includes components to account for convective blueshift suppression and active region evolution.

In addition to the GP component to describe correlated signals simultaneously present in RV, \lrhk\ and \bis\ time series, our model included possible non-interacting Keplerian terms in RVs only. Mutual orbital stability of all pairs of Keplerian orbits was checked using the criterion from \citet{stabilityGladman}.

For the computation of the posterior probabilities of the model parameters, 
we used \pchord, a nested sampling algorithm designed to work well even with parameter spaces with very large dimensionality \citep{polychord15}. A short discussion illuminating its favorable properties compared with \texttt{MultiNest}, its direct predecessor, can be found in \citet{hall2018}; a detailed study testing \pchord\ and validating its use for joint modeling of exoplanets and stellar activity in RVs is given by \citet{ahrer21}. We called \pchord\ via the \texttt{pypolychord} Python wrapper, leaving sampling parameters at their default values, except for the stopping (precision) criterion, which we changed from the default $10^{-3}$ to a (much) more stringent $10^{-9}$, to minimize the scatter in model evidence values from run to run \citep{ahrer21}. We ran all \pchord\ sampling on a high-performance computing platform, typically using several hundred CPU cores simultaneously. 

In addition to the MD GP analysis, we also used \pchord\ with models that describe RVs only, without any GP component: partly as a comparison with the MD GP analysis, and partly to be able to crosscheck results obtained with other approaches (e.g., \texttt{MultiNest}, Sect.~\ref{Nestedsampling_analysis}). We applied both classes of models (MD GP; no GP) to HARPS-N data only, and to HARPS-N plus HIRES data, for a total of four separate modeling `setups', (i)--(iv). For each of these four setups, we explored models with between $0$ and $4$ Keplerian terms (each model assumed equally likely a priori), and performed three separate \pchord\ runs for each model within a setup to obtain robust constraints on the evidences. The results from this analysis appear in Table~\ref{VMR-table}.

Across the four setups, planets b and c were reassuringly and resoundingly detected ($\Delta \ln \mathcal{Z}\sim30$, compared to models without either planet). A model incorporating three Keplerians (planets b and c, plus one non-transiting planet) was strongly favored over two-planet models across all four setups, with $\Delta\ln \mathcal{Z}\sim6$. As for the planet orbital parameters, $K_b$ was strongly consistent while $K_c$ was broadly consistent (within about $1\sigma$) across all setups and applicable models. For the third Keplerian, we obtained a period of $P_{\rm d}\sim151$~d (with a small secondary peak at $124.5$~d) and semiamplitude $K_{\rm d}\sim1.4$~\mps\ across all setups and applicable models -- this consistency lent additional weight to the conclusion that the third Keplerian term corresponds to a genuine planet. By contrast, there was insufficient statistical evidence for a fourth planet in any of our setups, with $\Delta\ln \mathcal{Z} \lesssim 1$ for the best $4$-four-planet versus the best three-planet models. For a fourth Keplerian, the period posterior distributions peaked at $P\sim$83~d (corresponding semiamplitude $K\sim 0.8$~\mps) with a prominent secondary peak at $114$~d.

Given the strong consistency between the results we obtained with the different modeling setups, and for direct comparison with results from other methods, we present in Table~\ref{table_harpsn_param} the results from the MD GP analysis applied to HARPS-N data only. We note that our planet parameter posteriors were generally broader when including HIRES data, with evidence in favor of the detection of the known planets b and c weakened slightly \citep[mirroring][]{2021MNRAS.507.1847R}. We also note that
the MD GP modeling with HARPS-N data only led us to a GP period of $54^{+10}_{-9}$~d, and a $1\sigma$ upper limit on the activity RV semiamplitude of $<1.32$~\mps. As already pointed out by \citet{2017MNRAS.471L.125R}, it is nevertheless unclear whether the GP period of $\sim54$~d corresponds to the stellar rotation period, given that neither the \emph{Kepler} light curve nor the time series of the CCF and CaII activity indicators show clear signals attributable to the stellar rotation.

\subsection{Search for Kepler-10\,d transits}\label{sec:TLS}
\begin{figure}[h!]
    \centering
    \includegraphics[width=\linewidth]{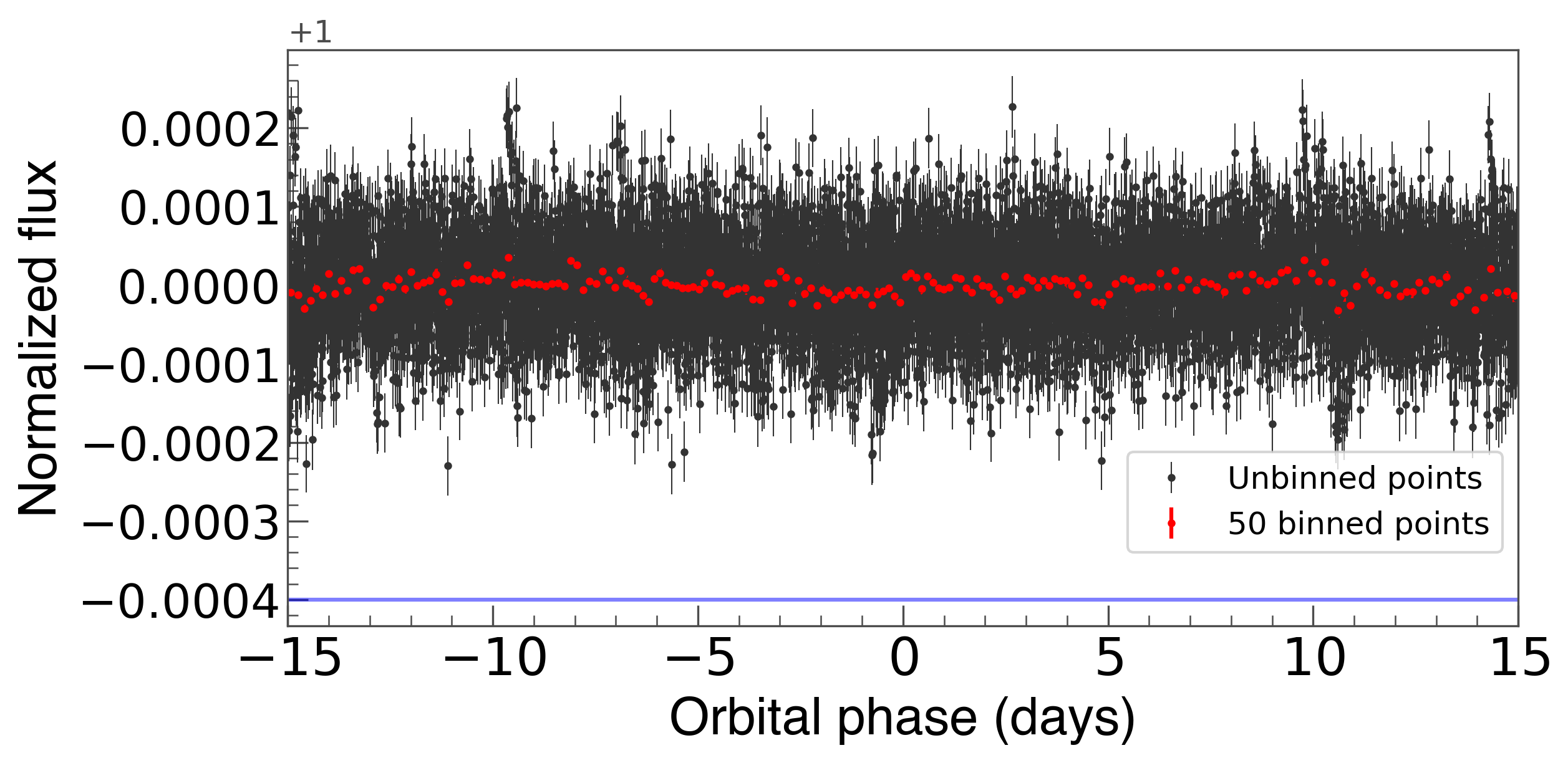}
    \caption{\emph{Kepler} light curve phase-folded to the period of planet~d (from Table\,\ref{table_harpsn_param}). No transit-like features with a depth comparable to planet~c (blue horizontal line) can be seen.}
    \label{fig:kepler_phased}
\end{figure}

Given the presence of a significant signal at $\sim151$~d, we analyzed the \emph{Kepler} light curve to check whether any transit could have been missed in previous studies of this system. In particular, we first modeled the light curves using the python package \texttt{wotan} \citep{Hippke2019b} and then performed a transit least squares (TLS) periodogram \citep{Hippke2019} on the residuals of the light curve, after modeling the transits of both Kepler-10\,b and Kepler-10\,c. We found no significant periodicity at $\sim151$~d or elsewhere, despite the Kepler curve being about $\sim1480$~d long (minus a few gaps in between). We also visually inspected the light curve around the eight predicted transits of Kepler-10\,d (from the ephemeris of Table~\ref{table_harpsn_param}) and found no significant dip. We point out that, based on the estimated RV semiamplitude of the outer planet, we would expect transits of a comparable depth to planet~c ($\sim0.5$~mmag), and thus easily detectable (see Fig.~\ref{fig:kepler_phased}). Planet~d would transit if it had the same inclination as planet~c, since $b_d\leq1$:

\begin{equation}
b_d = \frac{a_d}{R_\star}\cos{i_d}\left(\frac{1-e_d^2}{1+e_d\sin{\omega_d}}\right) = 0.64^{+0.07}_{-0.08}.
\end{equation}

\noindent
However, the transiting condition breaks with a small mutual inclination of only 0.2~deg between the two planets.

\subsection{Simultaneous modeling of radial velocities and transit timing variations}\label{sec:analysis_dynamical}
The TTVs of planet c have an amplitude, 
$A_\mathrm{TTV}$\footnote{$A_\mathrm{TTV}$ defined as
half of the difference between the maximum and the minimum of the O-C.},
of about 5.6 minutes and show a quasi-sinusoidal pattern (see Fig.~\ref{fig:K10c_OC}),
which suggests an interaction with an additional non-transiting
planet on an external orbit.
For this reason, we decided to run a dynamical analysis
with the latest version of TRADES\footnote{Publicly available at
\url{https://github.com/lucaborsato/trades}}
\citep{Borsato2014, Borsato2019, Nascimbeni2023}
and simultaneously fit the transit times ($T_{\rm c}$s) and the RVs during the
numerical integration of the planetary orbits.
The computational time required by the dynamical analysis scales with
the number of orbits of the inner planet with respect to the integration time (about 4222 days). This implies that the inclusion of planet b ($P=0.837$~d) causes the computational time to increase disproportionately. 
Therefore, we had to remove its signal from the RV dataset, according to the solution from Sect.~\ref{DE-MCMC_analysis}. In addition to the considerable reduction in computation time, this choice is justified by the fact that the USP planet b 
cannot induce a TTV on planet c, 
so we assumed a three-body system with the star, planet c, and the hypothetical planet d.

\begin{figure}[t!]
\centering
\includegraphics[width=0.90\columnwidth]
{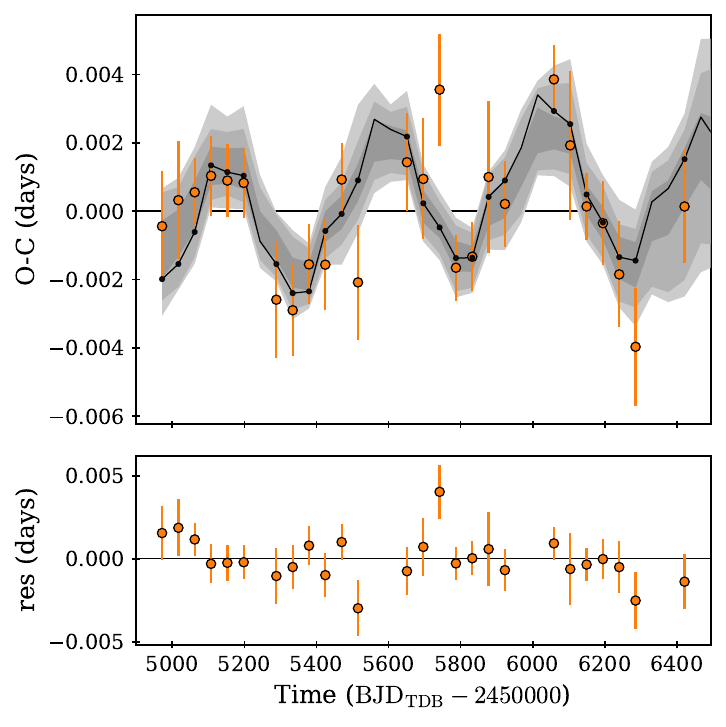}
\caption{Kepler-10\,c O-C diagram, defined as in Fig.~\ref{fig:K10c_OC}.
\textit{Top panel}: Observed (orange circles) TTVs, 
the best-fit model (black circles and line),
and the gray areas are the 1, 2, and $3\sigma$ uncertainty (from darker to lighter) of 100 samples drawn from the posterior distribution.
\textit{Bottom panel}: Residuals between the observed $T_{c}$ and simulated ones with \texttt{TRADES}.
}
\label{fig:K10c_OC_trades}
\end{figure}

\begin{figure}[t!]
\centering
\includegraphics[width=0.90\columnwidth]
{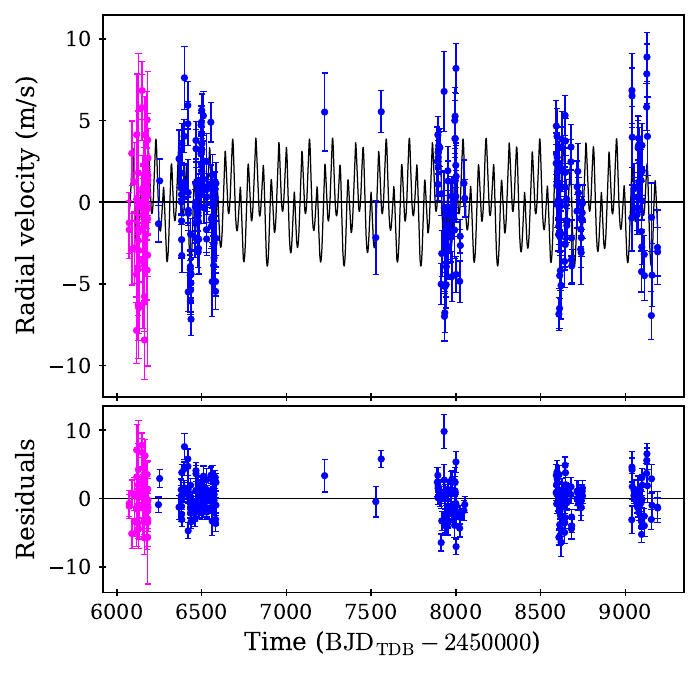}
\caption{Kepler-10 RVs from the \texttt{TRADES} analysis after subtracting the signal of Kepler-10\,b.
\textit{Top panel}: Observed RVs (removed RV offset for each dataset) as colored circles (magenta for HN-1 and blue for HN-2), 
and the \texttt{TRADES} best-fit model (black line).
\textit{Bottom panel}: Residuals between the observed RVs and simulated ones with \texttt{TRADES}.
}
\label{fig:K10_RV_trades}
\end{figure}

The TRADES code cannot manage complex stellar activity,
it can only fit an RV offset ($\gamma$) and
a jitter term ($\log_{2}{\sigma_{\rm j}}$) for each RV dataset; this is not an issue in our case, given the low level of stellar activity and the fact that the planet parameters do not critically depend on the noise model (Sect.~\ref{Polychord_analysis}).
We fixed the inclination of planet c ($i_\mathrm{c}$) to 89.6~deg (Table~\ref{Kepler10_system_table}), and assumed the external planet d to have an inclination of 90~deg. 
We also fixed the longitude of nodes ($\Omega$) of both planets to 0 deg.
We then varied the ratio of the planetary to stellar mass ($M_{\rm p}/M_\star$),
the period ($P$),
the eccentricity ($e$) and the argument of periastron ($\omega$) as
$\sqrt{e}\cos \omega$ and $\sqrt{e}\sin \omega$,
and the mean longitude ($\lambda$\footnote{
the mean longitude is defined as $\lambda = \mathcal{M} + \omega + \Omega$,
where $\mathcal{M}$ is the mean anomaly.}),
of both planets c and d.
All the orbital parameters are osculating astrocentric parameters
at the reference time $\mathrm{BJD_{TDB}} - 2450000.0 = 7081.0$
and have uniform priors with physical boundaries, which are:
$M_\mathrm{c,d} = \mathcal{U}(0.1, 20)\ M_\oplus$,
$P_\mathrm{c} = \mathcal{U}(42, 48)$~d,
$P_\mathrm{d} = \mathcal{U}(50, 250)$~d,
$e_\mathrm{c,d} = \mathcal{U}(0.0, 0.5)$,
$\omega_\mathrm{c,d} = \mathcal{U}(0.0, 360)$~deg,
$\lambda_\mathrm{c,d} = \mathcal{U}(0.0, 360)$~deg.
Firstly, we searched for an optimal configuration with a differential evolution algorithm \citep[DE,][]{Storn1997}
implemented as \texttt{PyDE} within \texttt{pytransit}\footnote{\url{https://github.com/hpparvi/PyTransit}} \citep{Parviainen2015}
with a population of 76 configurations that evolved for 76000 generations.
Then, we passed the optimal configuration to the python code \texttt{emcee} \citep{Foreman2013, Foreman2019}.
We used as walkers the same number of the DE population (76), 
and we mixed the sampler algorithm as in \citet{Nascimbeni2023}.
We ran emcee for 2.5 million steps, removed as burn-in the first two million, 
and used a thinning factor of 100.
We computed as parameter uncertainties the high density interval (HDI)
at $68.27\%$\footnote{This is the equivalent of the confidence intervals at the $16^\mathrm{th}$ and $84^\mathrm{th}$ percentile.}
of the posterior distribution, and 
our best-fit solution as the maximum likelihood estimator (MLE), that is,
the configuration of parameters that maximizes the log-likelihood ($\log\mathcal{L}$)
of the posterior distribution within the HDI.
Table~\ref{tab:trades_parameters} gives the results of the best-fit parameters, 
and Figs.~\ref{fig:K10c_OC_trades} and \ref{fig:K10_RV_trades} show the best-fit models of TTVs and RVs, respectively.
We checked the best-fit model assuming that planet d does not transit, 
and found an average difference on the transit times of less than one second. 
This is well below our sensitivity threshold given by the mean transit uncertainty of two minutes. 
Therefore, the effect of using an orbital inclination slightly different from 90~deg for planet d is negligible in our analysis.

The parameters are in agreement with the different solutions of Sect.\ref{data_analysis}. In particular, the eccentricities of both planets c and d are  more precise than those retrieved in the DE-MCMC and MultiNest analyses (Sects.~\ref{DE-MCMC_analysis} and \ref{Nestedsampling_analysis}).
We decided to test a model with the same number of bodies,
but fixing $e_\mathrm{d} = 0$,
and we ran \texttt{PyDE} and \texttt{emcee} with the same number of walkers (76),
but with a lower number of steps 
(76000 \texttt{PyDE} generations, 
one million \texttt{emcee} steps with 
400000 as burn-in, and a further thinning factor of 100).
We computed the BIC
for the eccentric (ecc) and for the circular (circ) best-fit model, and we found
that the eccentric model is strongly favored:
$\Delta \mathrm{BIC} = \mathrm{BIC(ecc)} - \mathrm{BIC(circ)} = -13$.
\citep[e.g.,][]{Kass_Raftery_1995}.

\begin{table}
    \centering\renewcommand{\arraystretch}{1.3}
    \caption{Best-fit MLE parameters for the Kepler-10 (c + d) system analyzed with TRADES.}
    {
    \small
    \begin{tabular}{lcc}
        \hline\hline
        \emph{Parameter} & \emph{MLE (HDI $68.27\%$)}\\
        \hline
        Kepler-10\,c & & \\
\hline
        $M_\mathrm{p}/M_\star\,[\rm M_\odot/M_\star \times10^{-3}]$ & $ 0.039 _{-0.005} ^{+0.003} $ \\
        $P$~[days]                                   & $ 45.29334 _{-0.00007} ^{+0.00097} $ \\
        $\sqrt{e}\cos\omega$                         & $ 0.347 _{-0.054} ^{+0.034} $ \\
        $\sqrt{e}\sin\omega$                         & $ 0.051 _{-0.037} ^{+0.133} $ \\
        $\lambda\,$[deg]                          & $ 282 _{-3} ^{+3} $ \\
        $M_\mathrm{p}\,[\Me]$                   & $11.70_{-1.64}^{+0.78}$ \\
        $e$                                          & $ 0.121 _{-0.018} ^{+0.035} $ \\
        $\omega\,$[deg]                           & $ 8 _{-8} ^{+21} $ \\
        $\mathcal{M}\,$[deg]                      & $ 273 _{-23} ^{+5} $ \\
        \hline       
        Kepler-10\,d & & \\
\hline
        $M_\mathrm{p}/M_\star\,[\rm M_\odot/M_\star \times10^{-3}]$ & $ 0.043 _{-0.007} ^{+0.003} $ \\
        $P$~[days]                                   & $ 151.09 _{-0.41} ^{+0.18} $ \\
        $\sqrt{e}\cos\omega$                         & $ -0.368 _{-0.057} ^{+0.276} $ \\
        $\sqrt{e}\sin\omega$                         & $ 0.233 _{-0.076} ^{+0.225} $ \\
        $\lambda\,$[deg]                          & $ 264 _{-6} ^{+8} $ \\
        $M_\mathrm{p}$ for $i=90$~deg [$\Me$]                   & $ 13.00_{-2.44} ^{+0.73} $ \\
        $e$                                          & $ 0.190 _{-0.070} ^{+0.027} $ \\
        $\omega\,$[deg]                           & $ 148 _{-55} ^{+13} $ \\
        $\mathcal{M}\,$[deg]                      & $ 117 _{-6}^{+50} $ \\
        \hline
        RV datasets & & \\
        \hline
        $\sigma_{\rm j,HN-1}\,[\ms]$    & $2.54_{-0.54}^{+0.22}$ \\
        $\sigma_{\rm j,HN-2}\,[\ms]$   & $ 2.26 _{-0.24} ^{+0.03} $ \\
        
        $\gamma_\mathrm{HN-1}\,[\ms]$    & $ 0.22 _{-0.52} ^{+0.28} $ \\
        $\gamma_\mathrm{HN-2}\,[\ms]$   & $ -0.01 _{-0.15} ^{+0.15} $ \\
    \end{tabular}
    }
    \tablefoot{The columns give the name of the parameter, its best-fit value with its associated $68.27\%$ HDI.
        Astrocentric orbital parameters are defined at $7081.0\, \mathrm{BJD_{TDB}}-2450000.0$.
         The mean anomaly $\mathcal{M}$ is computed as $\mathcal{M} = \lambda - \omega - \Omega$.
        We fixed the value of $i_\mathrm{d}=90$~deg, and so the $M_\mathrm{p}$ of planet d is the minimum mass.
        See Sect. \ref{sec:analysis_dynamical} for details.
    }
    \label{tab:trades_parameters}
\end{table}

\section{Planet parameters}
\label{planet_parameters}
We updated the orbital and physical parameters of the Kepler-10 planets in Table~\ref{Kepler10_system_table}, 
based on (i) the stellar parameters in D14; (ii) the results of the DE-MCMC transit and RV modeling (Sect.~\ref{data_analysis}), also for consistency with both D14 and \citet{2023A&A...677A..33B} (we recall that fully compatible results were obtained with the other two techniques employed; see Table~\ref{table_harpsn_param}); and (iii) the combined analysis of TTVs and RVs with TRADES for planets c and d. 

Thanks to the collection, refined extraction (Sect.~\ref{radial_velocities}), and analysis (Sect.~\ref{data_analysis}) of nearly 300 HARPS-N RVs, we determined the masses of Kepler-10\,b and Kepler-10\,c to be $M_{\rm p, b}=3.24 \pm 0.32$~\Me (10$\sigma$ precision) and $M_{\rm p, c}=11.29 \pm 1.24$~\Me (9$\sigma$ precision), respectively (see Table~\ref{Table_Kepler-10_system_parameters}, DE-MCMC solution), the latter being approximately in the middle between the previously determined values of $M_{\rm p, c}\sim6-7$~\Me (W16, \citealt{2017MNRAS.471L.125R}) and $M_{\rm p, c}\sim17$~\Me (D14).
The refined corresponding densities $\rho_{\rm p, b}=5.54 \pm 0.64~\rm g\;cm^{-3}$ and $\rho_{\rm p, c}=4.75 \pm 0.53~\rm g\;cm^{-3}$ 
allow us to better infer the planet compositions (Sect.~\ref{discussion_conclusions}), 
and place both Kepler-10\,b and Kepler-10\,c among the best characterized small planets.

For the outer non-transiting planet Kepler-10\,d, we found a more accurate orbital period of $P=151.06\pm0.48$~d from the HARPS-N RVs (Table~\ref{Table_Kepler-10_system_parameters}), which is consistent with both the analysis of TTVs (Sect.~\ref{sec:analysis_TTV}) and the combined analysis of TTVs and RVs (cf. Sect.~\ref{sec:analysis_dynamical} and Table~\ref{Table_Kepler-10_system_parameters}), and derived a minimum mass of $M_{\rm p, d}\sin{i}=12.00 \pm 2.15$~\Me.

\begin{table*}
\small
\centering
\caption{Kepler-10 system parameters.}     
\label{Kepler10_system_table}         
          
\begin{tabular}{l c c c}        
\hline\hline                 
\emph{\normalsize Stellar parameters}  & Value and $1\sigma$ error & & Reference \\
\hline
\hline
Star mass $M_\star$ [\Msun] &  $ 0.910 \pm 0.021 $ & & D14 \\
Star radius $R_\star$ [\Rsun] & $ 1.065 \pm 0.009$  & & D14 \\
Stellar density $\rho_{*}$ [$ \rm g\;cm^{-3}$] & $1.068 \pm 0.004$ & & D14 \\
Age $t$ [Gyr]  & $10.6^{+1.5}_{-1.3}$ & & D14 \\
Effective temperature $T_{\rm{eff}}$[K] & 5708 $\pm$ 28 & & D14 \\
Derived surface gravity log\,$g$ [cgs]&  4.344  $\pm$ 0.004 & & D14 \\
Metallicity $[\rm{Fe/H}]$ [dex] & -0.15  $\pm$ 0.04 & &  D14 \\
\hline
\hline
\emph{\normalsize Planetary parameters}  & Value and $1\sigma$ error & Value and $1\sigma$ error & Reference  \\
\hline
\hline
Kepler-10\,b  & DE-MCMC &  &  \\
\hline
Transit Time $T_{\rm c} \rm [BJD_{TDB}-2450000]$ & $5034.08687(18)$ & & D14 \\
Orbital period $P$ [days] & $0.83749070(20)$ & & D14 \\
Orbital semimajor axis $a$ [AU] & $0.01685 \pm 0.00013$ &  & D14 \\
Orbital eccentricity $e$  &  $\rm 0~(fixed)$ &  &  \\
Inclination $i$ [deg] & $84.8_{-3.9}^{+3.2}$ &  & D14  \\
Planet radius $R_{\rm p} ~[ \rm R_\oplus]$  &  $1.47^{+0.03}_{-0.02}$ &  & D14 \\
Planet mass $M_{\rm p} ~[\rm M_\oplus]$  &  $3.24 \pm 0.32$ &  & This work\\
Planet density $\rho_{\rm p}$ [$\rm g\;cm^{-3}$] &  $5.54^{+0.66}_{-0.62}$ &  & This work \\
Planet surface gravity log\,$g_{\rm p }$ [cgs] &  $3.163^{+0.046}_{-0.048}$ &  & This work\\
Equilibrium temperature$^1$ $T_{\rm eq}$ [K]  & $2188\pm16$ &  & This work \\
\hline
Kepler-10\,c  & DE-MCMC & TRADES &  \\
\hline
Transit Time $T_{\rm c} \rm [BJD_{TDB}-2450000]$ & $5062.26648(81)$ & & D14 \\
Orbital period $P$ [days] & 45.294301(48) & $45.29334^{+0.00097}_{-0.00007}$ & D14 and this work \\
Orbital semimajor axis $a$ [AU] & $0.2410 \pm 0.0019$ & & D14 \\
Orbital eccentricity $e$  &  $0.136\pm0.050$ & $0.121^{+0.035}_{-0.018}$ & This work\\
Argument of periastron $\omega$ & $330_{-23}^{+31}$ & $8_{-8}^{+21}$ & This work \\
Inclination $i$ [deg] & $89.623 \pm 0.011$ & & This work \\
Planet radius $R_{\rm p} ~[ \rm R_\oplus]$  &  $2.355 \pm 0.022$ & &  This work \\
Planet mass $M_{\rm p} ~[\rm M_\oplus]$  &  $11.29 \pm 1.24$ & $11.70_{-1.64}^{+0.78}$ & This work \\
Planet density $\rho_{\rm p}$ [$\rm g\;cm^{-3}$] &  $4.75\pm0.53$ & $4.82^{+0.40}_{-0.70}$ &  This work \\
Planet surface gravity log\,$g_{\rm p }$ [cgs] &  $3.300^{+0.044}_{-0.051}$ & $3.307^{+0.030}_{-0.070}$ & This work \\
Equilibrium temperature$^1$ $T_{\rm eq}$ [K]  & $579\pm4$ & & This work \\
\hline
Kepler-10\,d  & DE-MCMC & TRADES &  \\
\hline
Transit Time $T_{\rm c} \rm [BJD_{TDB}-2450000]$ & $7165.4^{+4.7}_{-5.3}$ & & This work \\
Orbital period $P$ [days] & $151.06\pm0.48$ & $151.09^{+0.18}_{-0.41}$ & This work \\
Orbital semimajor axis $a$ [AU] & $0.5379 \pm 0.0043$ & & This work \\
Orbital eccentricity $e$  &  $0.19\pm0.10~(<0.24)$ & $0.190^{+0.027}_{-0.070} $ & This work \\
Argument of periastron $\omega$ & $150_{-31}^{+28}$ & $148_{-55}^{+13}$ & This work \\
Planet minimum mass $M_{\rm p}\sin{i}~[\rm M_\oplus]$  &  $12.00\pm2.15$ & $13.00_{-2.44}^{+0.73}$ & This work \\
Equilibrium temperature$^1$ $T_{\rm eq}$ [K]  & $387\pm3$ & & This work \\
\hline
\hline
\end{tabular}
\label{Table_Kepler-10_system_parameters}
\begin{flushleft}
\footnotemark[1]{Blackbody equilibrium temperature assuming a null Bond albedo and uniform heat redistribution to the night side.} \\
\end{flushleft}
\end{table*}

\section{Discussion and conclusions}
\label{discussion_conclusions}
\begin{figure*}[t!]
\centering
\begin{minipage}{18.3 cm}
\hspace*{-0.75 cm}
\includegraphics[width=10.3cm, angle=180]{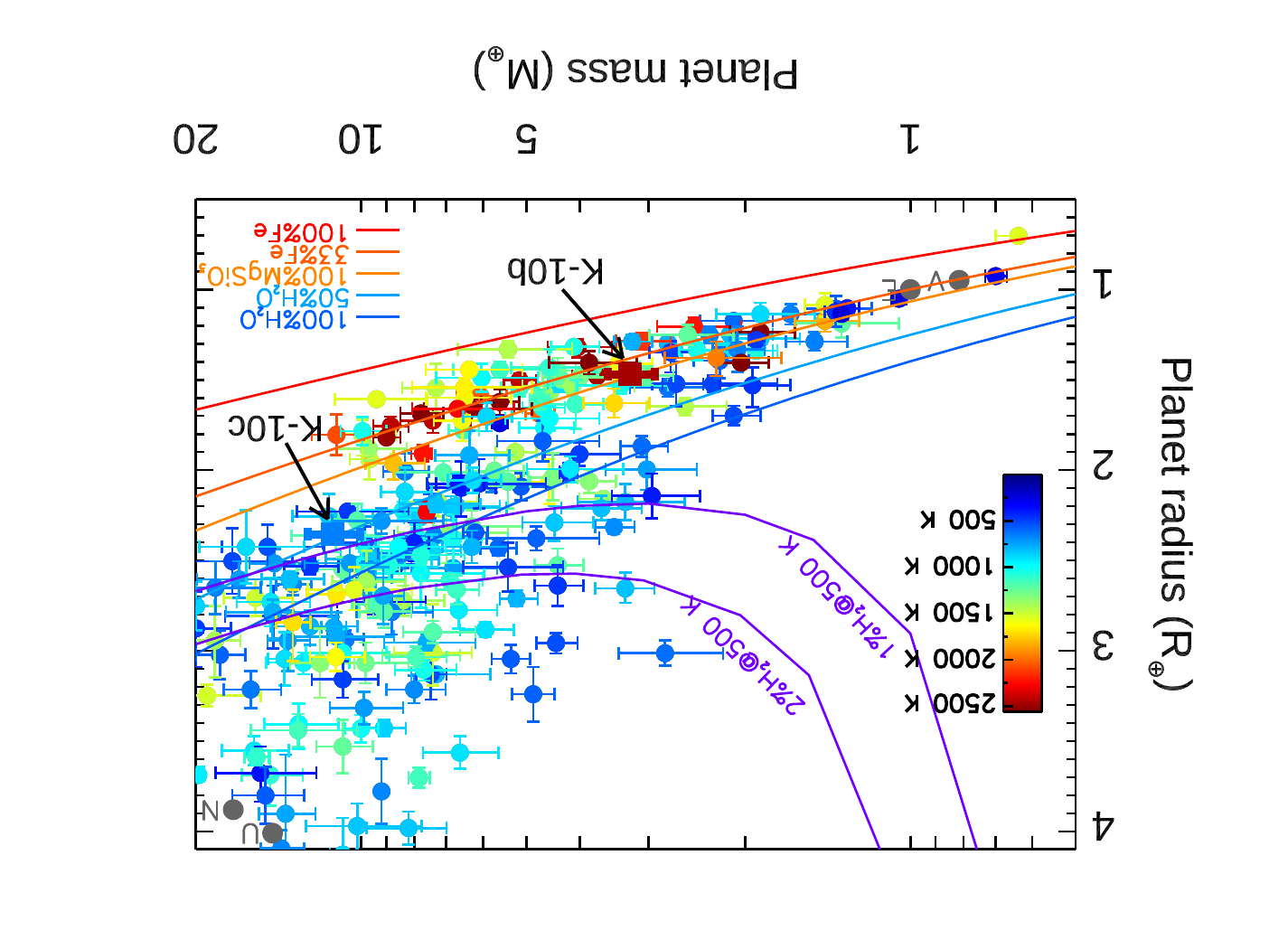}
\hspace*{-0.75 cm}
\includegraphics[width=10.3cm, angle=180]{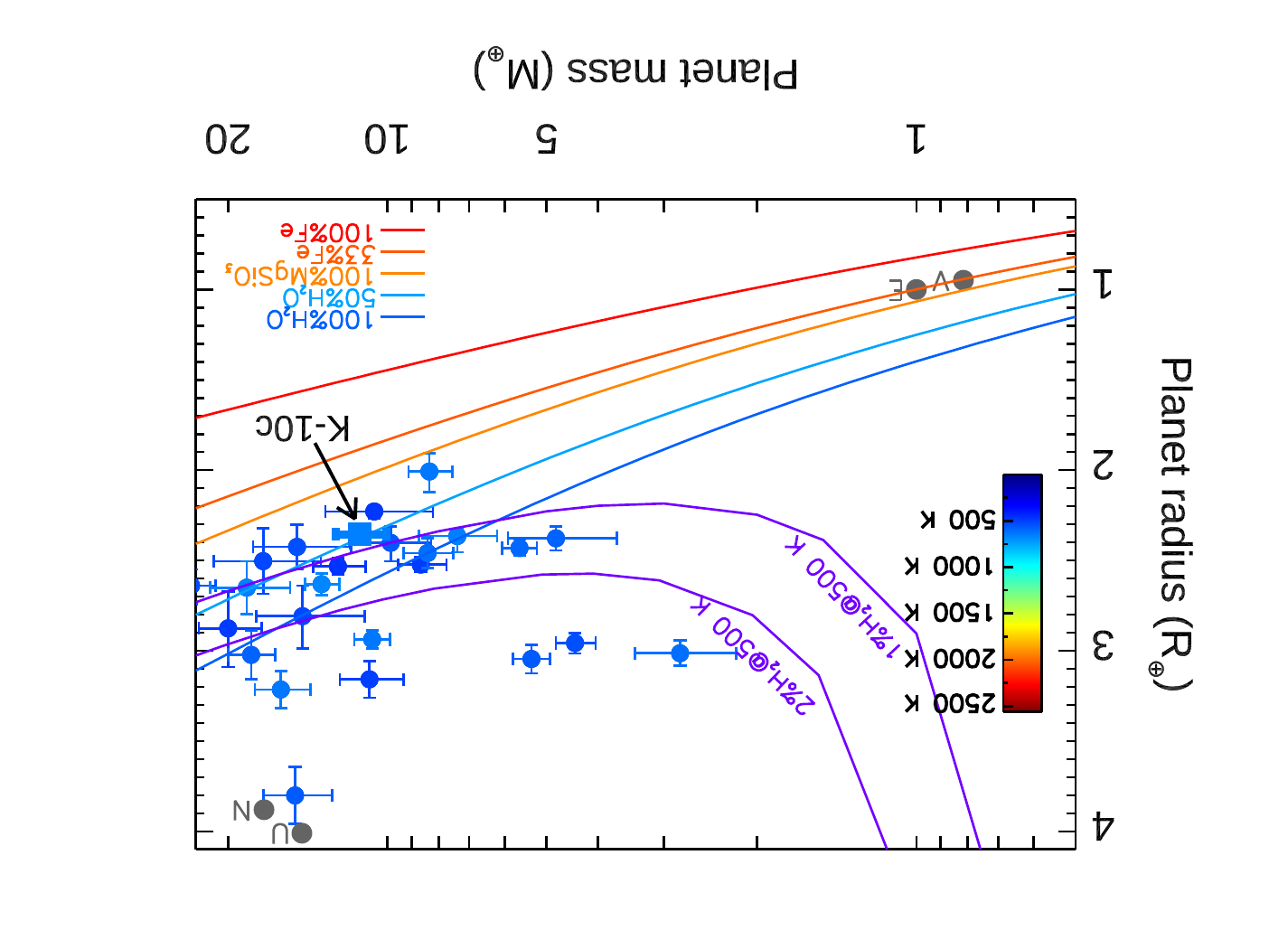}
\vspace{-0.5cm}
\caption{\emph{Left panel}: Mass-radius diagram of small ($R_{\rm p} \leq 4~\Re$) planets with mass and radius determinations better than $4\sigma$ and $10\sigma$, respectively, color-coded by planet equilibrium temperatures. 
The different solid curves, from bottom to top, correspond to planet compositions of $100\%$ iron, $33\%$ iron core and $67\%$ silicate mantle (Earth-like composition), $100\%$ silicates, $50\%$ rocky interior and $50\%$ water, $100\%$ water, rocky interiors and $1\%$ or $2\%$ hydrogen-dominated atmospheres \citep{Zeng2013}.
The gray dark circles indicate Venus (V), the Earth (E), Uranus (U), and Neptune (N). 
\emph{Right panel}: Same as left panel for planets orbiting FGK dwarfs, with color-coded equilibrium temperatures $T_{\rm eq} \leq 600$~K. 
Kepler-10\,b and c are indicated with squares.}
\label{massradius_Kepler10}
\end{minipage}
\end{figure*}

Figure~\ref{massradius_Kepler10}  shows the position of both Kepler-10\,b and Kepler-10\,c in the radius-mass diagram of small exoplanets with mass and radius determinations better than $4\sigma$ and $10\sigma$, respectively.

The innermost planet Kepler-10\,b is a rocky super-Earth that likely formed inside the water snowline at $\sim1-3$~AU, that is, in a possibly dry environment, and then migrated inward to the current semimajor axis where it is observed now. The planet may possibly have lost a primordial atmospheric H/He envelope because of the strong XUV stellar irradiation received in the first $\sim100$~Myr (e.g., \citealt{2013ApJ...775..105O}), but may have retained a thin, recently formed collisional secondary atmosphere \citep{2022A&A...658A.132S}. 
Since Kepler-10\,b is located on the pure-silicate isocomposition curve (Fig.~\ref{massradius_Kepler10}, left panel), its core mass fraction is broadly consistent with zero, though a small iron core may be present.

On the contrary, Kepler-10\,c, by its position in Fig.~\ref{massradius_Kepler10} and its low equilibrium temperature ($T_{\rm eq}\simeq580$~K), may be a water world, with a significant ($\sim40-70\%$) mass fraction of H$_2$O in a differentiated or miscible interior (e.g., \citealt{Luoetal2024}). 
Unlike many of the exoplanets in the figure, which are above the pure-H$_2$O mass-radius curve due to the presence of some atmospheric envelope, Kepler-10\,c lies below it, despite being on the right-hand side of the Cosmic Hydrogen and Ice Loss Line, where atmospheric loss due to stellar radiation is expected to be negligible (cf. \citealt{Zeng2024}).
The existence of water worlds is predicted by models of planet formation and migration (e.g., \citealt{Venturini2024, Burn2024}). Kepler-10\,c may thus join the fifteen or so possible water worlds that orbit relatively far from their FGK dwarf hosts, and hence have rather low equilibrium temperatures $T_{\rm eq} \leq 600$~K,  located in between the $100\%$ silicate and $100\%$ water isocomposition curves (Fig.~\ref{massradius_Kepler10}, right panel; see also \citealt{Luque_Palle_2022} for possible water worlds around M dwarfs).



Nevertheless, due to the well-known degeneracy in planet compositions from the measurement of the bulk density, a very thin H/He atmospheric envelope of less than $1\%$ of its total mass on top of a rocky interior (instead of a water-world composition) cannot be excluded. 
In this case, Kepler-10\,c might have accreted a more massive H/He envelope and lost a large fraction of it through ``boil-off'' \citep{2016ApJ...817..107O} and/or, more likely, core-powered mass loss \citep{2019MNRAS.487...24G}. 

The rather large density of Kepler-10\,c is in agreement with the trend that sub-Neptunes that are relatively far from MMR\footnote{according to the criterion given in \citet{Leleu2024}.} are denser \citep{Leleu2024}. 
This may provide an additional potential scenario to explain its density: the ejection of all or most of its atmosphere in post-disk dynamical instabilities \citep[e.g.,][]{Izidoro2017}.

Kepler-10\,d might be similar to Kepler-10\,c in terms of its composition, its minimum mass being consistent with the mass of Kepler-10\,c within $1\sigma$. Kepler-10\,d might thus potentially be a water world or have a composition more similar to Uranus and Neptune, that is, with a larger H/He envelope, given that it is colder and likely formed farther away. In any case, the lack of the measurement of both its radius and true mass makes any discussion quite speculative.

\begin{figure}[h!]
\centering
\includegraphics[width=0.90\linewidth]{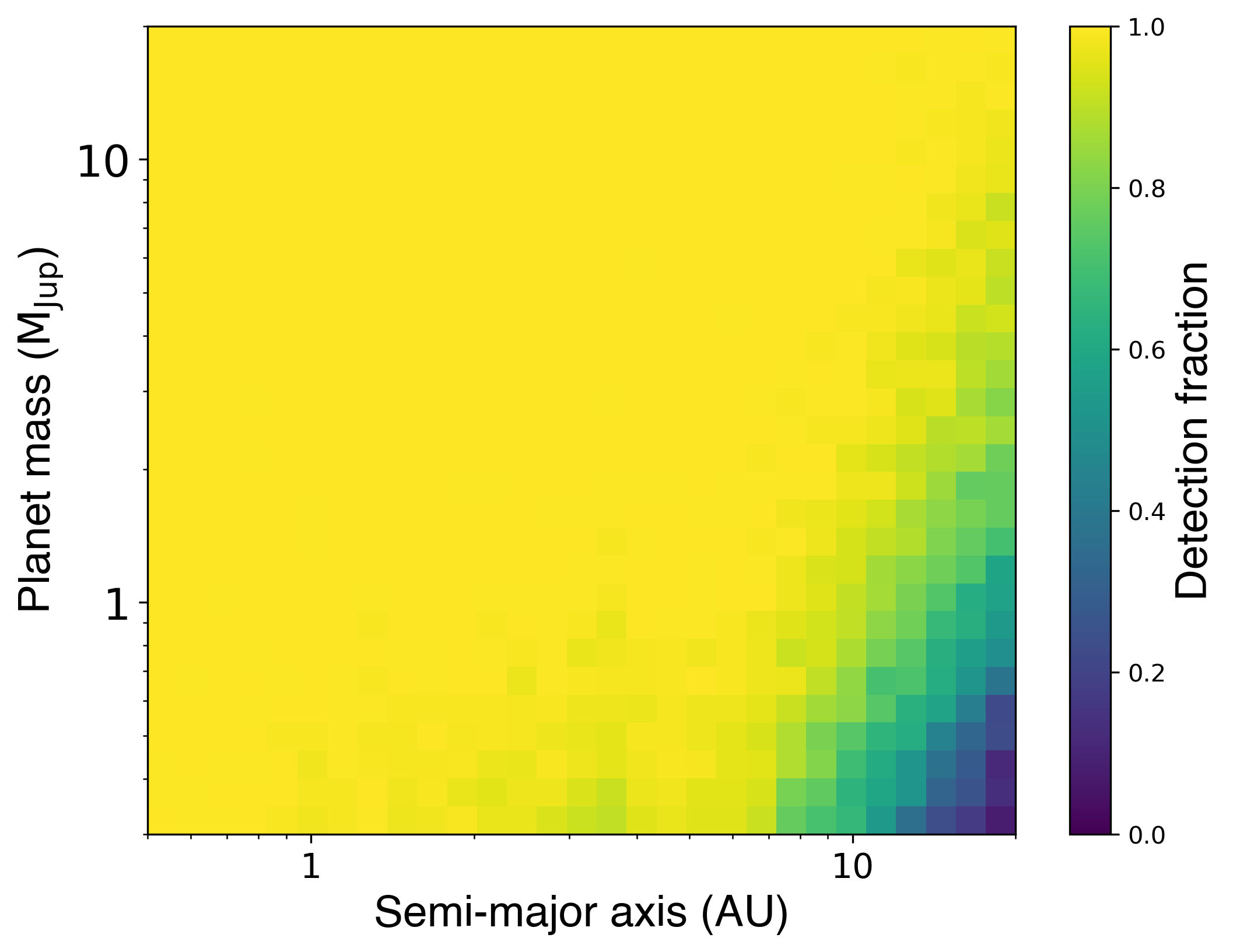}
\caption{Completeness map of Kepler-10, showing the sensitivity of the HARPS-N RV measurements to the presence
of possible cold Jupiters through experiments of injection and recovery of Doppler signals. The detectability inside the cell is indicated by its color, according to the colorbar on the right-hand side of the figure.}
\label{completeness_Kepler-10}
\end{figure}

Given the long RV time series of Kepler-10 spanning $\sim11$~years, we estimated the sensitivity of our data (completeness)
to the presence of possible outer cold Jupiters as in \citet[see their Sect.~3.3]{2023A&A...677A..33B}. 
The completeness is shown in Fig.~\ref{completeness_Kepler-10}, where yellow cells indicate
100\% detectability of the cold Jupiter, lighter to green cells show decreasing detection probability,
and the blue ones correspond to 0\% detectability. It is clear that there are no giant planets 
orbiting the Kepler-10 star within 10~AU, as is the case for the great majority of \emph{Kepler} and K2 small planet systems followed by HARPS-N \citep{2023A&A...677A..33B}; otherwise, we would have easily detected it.
This is consistent with the scenario that no giant planet could act as a dynamical barrier to the inward migration of the sub-Neptunes Kepler-10\,c and Kepler-10\,d (e.g., \citealt{2015ApJ...800L..22I}).

This work highlights the crucial importance of intensive long-term RV monitoring of planetary systems with relatively long-period ($P\gtrsim30$~d) transiting small planets, also in view of the forthcoming PLAnetary Transits and Oscillations of stars (PLATO) space mission \citep{Rauer2014, Rauer2024}. As in the case of Kepler-10, even several hundreds of RV measurements may be needed to (i) precisely determine the masses and hence bulk densities of the transiting planets, and (ii) properly characterize the system architecture by searching for additional non-transiting small and giant planets in outer regions. 
The increasing number of planetary systems better characterized in this way will lead to a more comprehensive understanding of the formation and evolution of sub-Neptunes and super-Earths.


\begin{acknowledgements}
The authors would like to thank the anonymous referee for her/his comments, which allowed them to improve the manuscript.
They also thank Lauren Weiss for helpful discussions about the HIRES RVs and the deckers used.
This work is based on observations made with the Italian Telescopio Nazionale Galileo (TNG)
operated on the island of La Palma by the Fundaci\'{o}n Galileo Galilei of the INAF at the Spanish Observatorio del Roque de los Muchachos
of the Instituto de Astrofisica de Canarias (GTO programme). 
The HARPS-N project was funded by the Prodex Program of the Swiss Space Office (SSO), the Harvard-University Origin of Life Initiative (HUOLI), the Scottish Universities Physics Alliance (SUPA), the University of Geneva, the Smithsonian Astrophysical Observatory (SAO), the Italian National Astrophysical Institute (INAF), the University of St. Andrews, Queen's University Belfast and the University of Edinburgh.
This paper is based on data collected by the \emph{Kepler} mission. Funding for the Kepler mission was provided by the NASA Science Mission directorate. 
This work has been carried out within the framework of the National Centre of Competence in Research PlanetS supported by the Swiss National Science Foundation under grants 51NF40\_182901 and 51NF40\_205606. The authors acknowledge financial contribution from the INAF Large Grant 2023 ``EXODEMO'' and from the SNSF. L.Bo. acknowledges support from CHEOPS ASI-INAF agreement n. 2019-29-HH.0. L.Z. acknowledges the support from the DOE-NNSA grant DE-NA0004084 awarded to Harvard University: Z Fundamental Science Program (PI: Stein B. Jacobsen). M.C. acknowledges the SNSF support under grant P500PT\_211024. 
A.L. acknowledges support of the Swiss National Science Foundation under grant number TMSGI2\_211697. X.D acknowledges the support from the European Research Council (ERC) under the European Union’s Horizon 2020 research and innovation programme (grant agreement SCORE No 851555) and from the Swiss National Science Foundation under the grant SPECTRE (No 200021\_215200). A.M. acknowledges funding from a UKRI Future Leader Fellowship, grant number MR/X033244/1 and a UK Science and Technology Facilities Council (STFC) small grant ST/Y002334/1. R.D.H. is funded by the UK Science and Technology Facilities Council (STFC)'s Ernest Rutherford Fellowship (grant number ST/V004735/1).
This publication makes use of The Data \& Analysis Center for Exoplanets (DACE), which is a facility based at the University of Geneva (CH) dedicated to extrasolar planets data visualisation, exchange and analysis. DACE is a platform of the Swiss National Centre of Competence in Research (NCCR) PlanetS, federating the Swiss expertise in Exoplanet research. The DACE platform is available at \url{https://dace.unige.ch}.
\end{acknowledgements}


\bibliographystyle{aa} 
\bibliography{aa53026-24} 

\begin{thebibliography}{108}
\expandafter\ifx\csname natexlab\endcsname\relax\def\natexlab#1{#1}\fi

\bibitem[{{Ahrer} {et~al.}(2021){Ahrer}, {Queloz}, {Rajpaul}, {S{\'e}gransan}, {Bouchy}, {Hall}, {Handley}, {Lovis}, {Mayor}, {Mortier}, {Pepe}, {Thompson}, {Udry}, \& {Unger}}]{ahrer21}
{Ahrer}, E., {Queloz}, D., {Rajpaul}, V.~M., {et~al.} 2021, \mnras, 503, 1248

\bibitem[{{Baluev}(2008)}]{baluev2008}
{Baluev}, R.~V. 2008, \mnras, 385, 1279

\bibitem[{{Batalha} {et~al.}(2011){Batalha}, {Borucki}, {Bryson}, {Buchhave}, {Caldwell}, {Christensen-Dalsgaard}, {Ciardi}, {Dunham}, {Fressin}, {Gautier}, {Gilliland}, {Haas}, {Howell}, {Jenkins}, {Kjeldsen}, {Koch}, {Latham}, {Lissauer}, {Marcy}, {Rowe}, {Sasselov}, {Seager}, {Steffen}, {Torres}, {Basri}, {Brown}, {Charbonneau}, {Christiansen}, {Clarke}, {Cochran}, {Dupree}, {Fabrycky}, {Fischer}, {Ford}, {Fortney}, {Girouard}, {Holman}, {Johnson}, {Isaacson}, {Klaus}, {Machalek}, {Moorehead}, {Morehead}, {Ragozzine}, {Tenenbaum}, {Twicken}, {Quinn}, {VanCleve}, {Walkowicz}, {Welsh}, {Devore}, \& {Gould}}]{2011ApJ...729...27B}
{Batalha}, N.~M., {Borucki}, W.~J., {Bryson}, S.~T., {et~al.} 2011, \apj, 729, 27

\bibitem[{{Bonomo} {et~al.}(2023){Bonomo}, {Dumusque}, {Massa}, {Mortier}, {Bongiolatti}, {Malavolta}, {Sozzetti}, {Buchhave}, {Damasso}, {Haywood}, {Morbidelli}, {Latham}, {Molinari}, {Pepe}, {Poretti}, {Udry}, {Affer}, {Boschin}, {Charbonneau}, {Cosentino}, {Cretignier}, {Ghedina}, {Lega}, {L{\'o}pez-Morales}, {Margini}, {Mart{\'\i}nez Fiorenzano}, {Mayor}, {Micela}, {Pedani}, {Pinamonti}, {Rice}, {Sasselov}, {Tronsgaard}, \& {Vanderburg}}]{2023A&A...677A..33B}
{Bonomo}, A.~S., {Dumusque}, X., {Massa}, A., {et~al.} 2023, \aap, 677, A33

\bibitem[{{Bonomo} {et~al.}(2014){Bonomo}, {Sozzetti}, {Lovis}, {Malavolta}, {Rice}, {Buchhave}, {Sasselov}, {Cameron}, {Latham}, {Molinari}, {Pepe}, {Udry}, {Affer}, {Charbonneau}, {Cosentino}, {Dressing}, {Dumusque}, {Figueira}, {Fiorenzano}, {Gettel}, {Harutyunyan}, {Haywood}, {Horne}, {Lopez-Morales}, {Mayor}, {Micela}, {Motalebi}, {Nascimbeni}, {Phillips}, {Piotto}, {Pollacco}, {Queloz}, {S{\'e}gransan}, {Szentgyorgyi}, \& {Watson}}]{2014A&A...572A...2B}
{Bonomo}, A.~S., {Sozzetti}, A., {Lovis}, C., {et~al.} 2014, \aap, 572, A2

\bibitem[{{Bonomo} {et~al.}(2019){Bonomo}, {Zeng}, {Damasso}, {Leinhardt}, {Justesen}, {Lopez}, {Lund}, {Malavolta}, {Silva Aguirre}, {Buchhave}, {Corsaro}, {Denman}, {Lopez-Morales}, {Mills}, {Mortier}, {Rice}, {Sozzetti}, {Vanderburg}, {Affer}, {Arentoft}, {Benbakoura}, {Bouchy}, {Christensen-Dalsgaard}, {Collier Cameron}, {Cosentino}, {Dressing}, {Dumusque}, {Figueira}, {Fiorenzano}, {Garc{\'\i}a}, {Handberg}, {Harutyunyan}, {Johnson}, {Kjeldsen}, {Latham}, {Lovis}, {Lundkvist}, {Mathur}, {Mayor}, {Micela}, {Molinari}, {Motalebi}, {Nascimbeni}, {Nava}, {Pepe}, {Phillips}, {Piotto}, {Poretti}, {Sasselov}, {S{\'e}gransan}, {Udry}, \& {Watson}}]{2019NatAs...3..416B}
{Bonomo}, A.~S., {Zeng}, L., {Damasso}, M., {et~al.} 2019, Nature Astronomy, 3, 416

\bibitem[{{Borsato} {et~al.}(2019){Borsato}, {Malavolta}, {Piotto}, {Buchhave}, {Mortier}, {Rice}, {Collier Cameron}, {Coffinet}, {Sozzetti}, {Charbonneau}, {Cosentino}, {Dumusque}, {Figueira}, {Latham}, {Lopez-Morales}, {Mayor}, {Micela}, {Molinari}, {Pepe}, {Phillips}, {Poretti}, {Udry}, \& {Watson}}]{Borsato2019}
{Borsato}, L., {Malavolta}, L., {Piotto}, G., {et~al.} 2019, \mnras, 484, 3233

\bibitem[{{Borsato} {et~al.}(2014){Borsato}, {Marzari}, {Nascimbeni}, {Piotto}, {Granata}, {Bedin}, \& {Malavolta}}]{Borsato2014}
{Borsato}, L., {Marzari}, F., {Nascimbeni}, V., {et~al.} 2014, \aap, 571, A38

\bibitem[{{Brewer}(2014)}]{brewer2014}
{Brewer}, B.~J. 2014, ArXiv e-prints [\eprint[arXiv]{1411.3921}]

\bibitem[{{Buchner} {et~al.}(2014){Buchner}, {Georgakakis}, {Nandra}, {Hsu}, {Rangel}, {Brightman}, {Merloni}, {Salvato}, {Donley}, \& {Kocevski}}]{Buchner2014}
{Buchner}, J., {Georgakakis}, A., {Nandra}, K., {et~al.} 2014, \aap, 564, A125

\bibitem[{{Burn} {et~al.}(2024){Burn}, {Mordasini}, {Mishra}, {Haldemann}, {Venturini}, {Emsenhuber}, \& {Henning}}]{Burn2024}
{Burn}, R., {Mordasini}, C., {Mishra}, L., {et~al.} 2024, Nature Astronomy, 8, 463

\bibitem[{Burnham \& Anderson(2004)}]{Burnham_Anderson_2004}
Burnham, K.~P. \& Anderson, D.~R. 2004, Sociological Methods \& Research, 33, 261

\bibitem[{{Cosentino} {et~al.}(2012){Cosentino}, {Lovis}, {Pepe}, {Collier Cameron}, {Latham}, {Molinari}, {Udry}, {Bezawada}, {Black}, {Born}, {Buchschacher}, {Charbonneau}, {Figueira}, {Fleury}, {Galli}, {Gallie}, {Gao}, {Ghedina}, {Gonzalez}, {Gonzalez}, {Guerra}, {Henry}, {Horne}, {Hughes}, {Kelly}, {Lodi}, {Lunney}, {Maire}, {Mayor}, {Micela}, {Ordway}, {Peacock}, {Phillips}, {Piotto}, {Pollacco}, {Queloz}, {Rice}, {Riverol}, {Riverol}, {San Juan}, {Sasselov}, {Segransan}, {Sozzetti}, {Sosnowska}, {Stobie}, {Szentgyorgyi}, {Vick}, \& {Weber}}]{2012SPIE.8446E..1VC}
{Cosentino}, R., {Lovis}, C., {Pepe}, F., {et~al.} 2012, in Society of Photo-Optical Instrumentation Engineers (SPIE) Conference Series, Vol. 8446, Ground-based and Airborne Instrumentation for Astronomy IV, ed. I.~S. {McLean}, S.~K. {Ramsay}, \& H.~{Takami}, 84461V

\bibitem[{{Cosentino} {et~al.}(2014){Cosentino}, {Lovis}, {Pepe}, {Collier Cameron}, {Latham}, {Molinari}, {Udry}, {Bezawada}, {Buchschacher}, {Figueira}, {Fleury}, {Ghedina}, {Glenday}, {Gonzalez}, {Guerra}, {Henry}, {Hughes}, {Maire}, {Motalebi}, \& {Phillips}}]{2014SPIE.9147E..8CC}
{Cosentino}, R., {Lovis}, C., {Pepe}, F., {et~al.} 2014, in Society of Photo-Optical Instrumentation Engineers (SPIE) Conference Series, Vol. 9147, Ground-based and Airborne Instrumentation for Astronomy V, ed. S.~K. {Ramsay}, I.~S. {McLean}, \& H.~{Takami}, 91478C

\bibitem[{{Cretignier} {et~al.}(2021){Cretignier}, {Dumusque}, {Hara}, \& {Pepe}}]{cretignier2021}
{Cretignier}, M., {Dumusque}, X., {Hara}, N.~C., \& {Pepe}, F. 2021, \aap, 653, A43

\bibitem[{{Cretignier} {et~al.}(2022){Cretignier}, {Dumusque}, \& {Pepe}}]{2022A&A...659A..68C}
{Cretignier}, M., {Dumusque}, X., \& {Pepe}, F. 2022, \aap, 659, A68

\bibitem[{{Crossfield} {et~al.}(2015){Crossfield}, {Petigura}, {Schlieder}, {Howard}, {Fulton}, {Aller}, {Ciardi}, {L{\'e}pine}, {Barclay}, {de Pater}, {de Kleer}, {Quintana}, {Christiansen}, {Schlafly}, {Kaltenegger}, {Crepp}, {Henning}, {Obermeier}, {Deacon}, {Weiss}, {Isaacson}, {Hansen}, {Liu}, {Greene}, {Howell}, {Barman}, \& {Mordasini}}]{2015ApJ...804...10C}
{Crossfield}, I. J.~M., {Petigura}, E., {Schlieder}, J.~E., {et~al.} 2015, \apj, 804, 10

\bibitem[{{Dai} {et~al.}(2019){Dai}, {Masuda}, {Winn}, \& {Zeng}}]{2019ApJ...883...79D}
{Dai}, F., {Masuda}, K., {Winn}, J.~N., \& {Zeng}, L. 2019, \apj, 883, 79

\bibitem[{{Damasso} {et~al.}(2018){Damasso}, {Bonomo}, {Astudillo-Defru}, {Bonfils}, {Malavolta}, {Sozzetti}, {Lopez}, {Zeng}, {Haywood}, {Irwin}, {Mortier}, {Vanderburg}, {Maldonado}, {Lanza}, {Affer}, {Almenara}, {Benatti}, {Biazzo}, {Bignamini}, {Borsa}, {Bouchy}, {Buchhave}, {Cameron}, {Carleo}, {Charbonneau}, {Claudi}, {Cosentino}, {Covino}, {Delfosse}, {Desidera}, {Di Fabrizio}, {Dressing}, {Esposito}, {Fares}, {Figueira}, {Fiorenzano}, {Forveille}, {Giacobbe}, {Gonz{\'a}lez-{\'A}lvarez}, {Gratton}, {Harutyunyan}, {Johnson}, {Latham}, {Leto}, {Lopez-Morales}, {Lovis}, {Maggio}, {Mancini}, {Masiero}, {Mayor}, {Micela}, {Molinari}, {Motalebi}, {Murgas}, {Nascimbeni}, {Pagano}, {Pepe}, {Phillips}, {Piotto}, {Poretti}, {Rainer}, {Rice}, {Santos}, {Sasselov}, {Scandariato}, {S{\'e}gransan}, {Smareglia}, {Udry}, {Watson}, \& {W{\"u}nsche}}]{2018A&A...615A..69D}
{Damasso}, M., {Bonomo}, A.~S., {Astudillo-Defru}, N., {et~al.} 2018, \aap, 615, A69

\bibitem[{{Deck} \& {Agol}(2015)}]{Deck2015}
{Deck}, K.~M. \& {Agol}, E. 2015, \apj, 802, 116

\bibitem[{{Deck} {et~al.}(2013){Deck}, {Payne}, \& {Holman}}]{DePaHo2013}
{Deck}, K.~M., {Payne}, M., \& {Holman}, M.~J. 2013, \apj, 774, 129

\bibitem[{{Delisle} {et~al.}(2020){Delisle}, {Hara}, \& {S{\'e}gransan}}]{delisle2020}
{Delisle}, J.~B., {Hara}, N., \& {S{\'e}gransan}, D. 2020, \aap, 635, A83

\bibitem[{{Dorn} {et~al.}(2017){Dorn}, {Venturini}, {Khan}, {Heng}, {Alibert}, {Helled}, {Rivoldini}, \& {Benz}}]{Dorn2017}
{Dorn}, C., {Venturini}, J., {Khan}, A., {et~al.} 2017, \aap, 597, A37

\bibitem[{{Dressing} {et~al.}(2015){Dressing}, {Charbonneau}, {Dumusque}, {Gettel}, {Pepe}, {Collier Cameron}, {Latham}, {Molinari}, {Udry}, {Affer}, {Bonomo}, {Buchhave}, {Cosentino}, {Figueira}, {Fiorenzano}, {Harutyunyan}, {Haywood}, {Johnson}, {Lopez-Morales}, {Lovis}, {Malavolta}, {Mayor}, {Micela}, {Motalebi}, {Nascimbeni}, {Phillips}, {Piotto}, {Pollacco}, {Queloz}, {Rice}, {Sasselov}, {S{\'e}gransan}, {Sozzetti}, {Szentgyorgyi}, \& {Watson}}]{2015ApJ...800..135D}
{Dressing}, C.~D., {Charbonneau}, D., {Dumusque}, X., {et~al.} 2015, \apj, 800, 135

\bibitem[{{Dumusque}(2021)}]{2021plat.confE.106D}
{Dumusque}, X. 2021, in Plato Mission Conference 2021. Presentations and posters of the online PLATO Mission Conference 2021, 106

\bibitem[{{Dumusque} {et~al.}(2014){Dumusque}, {Bonomo}, {Haywood}, {Malavolta}, {S{\'e}gransan}, {Buchhave}, {Collier Cameron}, {Latham}, {Molinari}, {Pepe}, {Udry}, {Charbonneau}, {Cosentino}, {Dressing}, {Figueira}, {Fiorenzano}, {Gettel}, {Harutyunyan}, {Horne}, {Lopez-Morales}, {Lovis}, {Mayor}, {Micela}, {Motalebi}, {Nascimbeni}, {Phillips}, {Piotto}, {Pollacco}, {Queloz}, {Rice}, {Sasselov}, {Sozzetti}, {Szentgyorgyi}, \& {Watson}}]{2014ApJ...789..154D}
{Dumusque}, X., {Bonomo}, A.~S., {Haywood}, R.~D., {et~al.} 2014, \apj, 789, 154

\bibitem[{{Eastman} {et~al.}(2013){Eastman}, {Gaudi}, \& {Agol}}]{2013PASP..125...83E}
{Eastman}, J., {Gaudi}, B.~S., \& {Agol}, E. 2013, \pasp, 125, 83

\bibitem[{{Eastman} {et~al.}(2019){Eastman}, {Rodriguez}, {Agol}, {Stassun}, {Beatty}, {Vanderburg}, {Gaudi}, {Collins}, \& {Luger}}]{2019arXiv190709480E}
{Eastman}, J.~D., {Rodriguez}, J.~E., {Agol}, E., {et~al.} 2019, arXiv e-prints, arXiv:1907.09480

\bibitem[{{Feroz} {et~al.}(2011){Feroz}, {Balan}, \& {Hobson}}]{feroz2011}
{Feroz}, F., {Balan}, S.~T., \& {Hobson}, M.~P. 2011, \mnras, 415, 3462

\bibitem[{{Feroz} {et~al.}(2019){Feroz}, {Hobson}, {Cameron}, \& {Pettitt}}]{Feroz2019}
{Feroz}, F., {Hobson}, M.~P., {Cameron}, E., \& {Pettitt}, A.~N. 2019, The Open Journal of Astrophysics, 2, 10

\bibitem[{{Fogtmann-Schulz} {et~al.}(2014){Fogtmann-Schulz}, {Hinrup}, {Van Eylen}, {Christensen-Dalsgaard}, {Kjeldsen}, {Silva Aguirre}, \& {Tingley}}]{2014ApJ...781...67F}
{Fogtmann-Schulz}, A., {Hinrup}, B., {Van Eylen}, V., {et~al.} 2014, \apj, 781, 67

\bibitem[{{Foreman-Mackey} {et~al.}(2019){Foreman-Mackey}, {Farr}, {Sinha}, {Archibald}, {Hogg}, {Sanders}, {Zuntz}, {Williams}, {Nelson}, {de Val-Borro}, {Erhardt}, {Pashchenko}, \& {Pla}}]{Foreman2019}
{Foreman-Mackey}, D., {Farr}, W., {Sinha}, M., {et~al.} 2019, The Journal of Open Source Software, 4, 1864

\bibitem[{{Foreman-Mackey} {et~al.}(2013){Foreman-Mackey}, {Hogg}, {Lang}, \& {Goodman}}]{Foreman2013}
{Foreman-Mackey}, D., {Hogg}, D.~W., {Lang}, D., \& {Goodman}, J. 2013, \pasp, 125, 306

\bibitem[{{Fressin} {et~al.}(2011){Fressin}, {Torres}, {D{\'e}sert}, {Charbonneau}, {Batalha}, {Fortney}, {Rowe}, {Allen}, {Borucki}, {Brown}, {Bryson}, {Ciardi}, {Cochran}, {Deming}, {Dunham}, {Fabrycky}, {Gautier}, {Gilliland}, {Henze}, {Holman}, {Howell}, {Jenkins}, {Kinemuchi}, {Knutson}, {Koch}, {Latham}, {Lissauer}, {Marcy}, {Ragozzine}, {Sasselov}, {Still}, {Tenenbaum}, \& {Uddin}}]{2011ApJS..197....5F}
{Fressin}, F., {Torres}, G., {D{\'e}sert}, J.-M., {et~al.} 2011, \apjs, 197, 5

\bibitem[{{Fulton} {et~al.}(2017){Fulton}, {Petigura}, {Howard}, {Isaacson}, {Marcy}, {Cargile}, {Hebb}, {Weiss}, {Johnson}, {Morton}, {Sinukoff}, {Crossfield}, \& {Hirsch}}]{2017AJ....154..109F}
{Fulton}, B.~J., {Petigura}, E.~A., {Howard}, A.~W., {et~al.} 2017, \aj, 154, 109

\bibitem[{{Ginzburg} {et~al.}(2018){Ginzburg}, {Schlichting}, \& {Sari}}]{2018MNRAS.476..759G}
{Ginzburg}, S., {Schlichting}, H.~E., \& {Sari}, R. 2018, \mnras, 476, 759

\bibitem[{Gladman(1993)}]{stabilityGladman}
Gladman, B. 1993, Icarus, 106, 247

\bibitem[{{Gregory}(2005)}]{2005ApJ...631.1198G}
{Gregory}, P.~C. 2005, \apj, 631, 1198

\bibitem[{{Gregory}(2016)}]{gregory2016}
{Gregory}, P.~C. 2016, \mnras, 458, 2604

\bibitem[{{Grunblatt} {et~al.}(2015){Grunblatt}, {Howard}, \& {Haywood}}]{2015ApJ...808..127G}
{Grunblatt}, S.~K., {Howard}, A.~W., \& {Haywood}, R.~D. 2015, \apj, 808, 127

\bibitem[{{Gupta} \& {Schlichting}(2019)}]{2019MNRAS.487...24G}
{Gupta}, A. \& {Schlichting}, H.~E. 2019, \mnras, 487, 24

\bibitem[{{Hadden} \& {Lithwick}(2016)}]{HaLi2016}
{Hadden}, S. \& {Lithwick}, Y. 2016, \apj, 828, 44

\bibitem[{{Hall} {et~al.}(2018){Hall}, {Thompson}, {Handley}, \& {Queloz}}]{hall2018}
{Hall}, R.~D., {Thompson}, S.~J., {Handley}, W., \& {Queloz}, D. 2018, \mnras, 479, 2968

\bibitem[{{Handley} {et~al.}(2015){Handley}, {Hobson}, \& {Lasenby}}]{polychord15}
{Handley}, W.~J., {Hobson}, M.~P., \& {Lasenby}, A.~N. 2015, \mnras, 453, 4384

\bibitem[{{Hara} {et~al.}(2017){Hara}, {Bou{\'e}}, {Laskar}, \& {Correia}}]{hara2017}
{Hara}, N.~C., {Bou{\'e}}, G., {Laskar}, J., \& {Correia}, A.~C.~M. 2017, \mnras, 464, 1220

\bibitem[{{Hara} {et~al.}(2019){Hara}, {Bou{\'e}}, {Laskar}, {Delisle}, \& {Unger}}]{2019MNRAS.489..738H}
{Hara}, N.~C., {Bou{\'e}}, G., {Laskar}, J., {Delisle}, J.~B., \& {Unger}, N. 2019, \mnras, 489, 738

\bibitem[{{Hara} {et~al.}(2024){Hara}, {de Poyferr{\'e}}, {Delisle}, \& {Hoffmann}}]{hara2024}
{Hara}, N.~C., {de Poyferr{\'e}}, T., {Delisle}, J.-B., \& {Hoffmann}, M. 2024, Annals of Applied Statistics, 18, 749

\bibitem[{{Hara} {et~al.}(2022{\natexlab{a}}){Hara}, {Delisle}, {Unger}, \& {Dumusque}}]{hara2022b}
{Hara}, N.~C., {Delisle}, J.-B., {Unger}, N., \& {Dumusque}, X. 2022{\natexlab{a}}, \aap, 658, A177

\bibitem[{{Hara} {et~al.}(2022{\natexlab{b}}){Hara}, {Unger}, {Delisle}, {D{\'\i}az}, \& {S{\'e}gransan}}]{hara2022a}
{Hara}, N.~C., {Unger}, N., {Delisle}, J.-B., {D{\'\i}az}, R.~F., \& {S{\'e}gransan}, D. 2022{\natexlab{b}}, \aap, 663, A14

\bibitem[{{Haywood} {et~al.}(2014){Haywood}, {Collier Cameron}, {Queloz}, {Barros}, {Deleuil}, {Fares}, {Gillon}, {Lanza}, {Lovis}, {Moutou}, {Pepe}, {Pollacco}, {Santerne}, {S{\'e}gransan}, \& {Unruh}}]{2014MNRAS.443.2517H}
{Haywood}, R.~D., {Collier Cameron}, A., {Queloz}, D., {et~al.} 2014, \mnras, 443, 2517

\bibitem[{{Hippke} {et~al.}(2019){Hippke}, {David}, {Mulders}, \& {Heller}}]{Hippke2019b}
{Hippke}, M., {David}, T.~J., {Mulders}, G.~D., \& {Heller}, R. 2019, \aj, 158, 143

\bibitem[{{Hippke} \& {Heller}(2019)}]{Hippke2019}
{Hippke}, M. \& {Heller}, R. 2019, \aap, 623, A39

\bibitem[{{Howard} {et~al.}(2010){Howard}, {Johnson}, {Marcy}, {Fischer}, {Wright}, {Bernat}, {Henry}, {Peek}, {Isaacson}, {Apps}, {Endl}, {Cochran}, {Valenti}, {Anderson}, \& {Piskunov}}]{2010ApJ...721.1467H}
{Howard}, A.~W., {Johnson}, J.~A., {Marcy}, G.~W., {et~al.} 2010, \apj, 721, 1467

\bibitem[{{Izidoro} {et~al.}(2017){Izidoro}, {Ogihara}, {Raymond}, {Morbidelli}, {Pierens}, {Bitsch}, {Cossou}, \& {Hersant}}]{Izidoro2017}
{Izidoro}, A., {Ogihara}, M., {Raymond}, S.~N., {et~al.} 2017, \mnras, 470, 1750

\bibitem[{{Izidoro} {et~al.}(2015){Izidoro}, {Raymond}, {Morbidelli}, {Hersant}, \& {Pierens}}]{2015ApJ...800L..22I}
{Izidoro}, A., {Raymond}, S.~N., {Morbidelli}, A., {Hersant}, F., \& {Pierens}, A. 2015, \apjl, 800, L22

\bibitem[{Jeffreys(1998)}]{Jeffreys1998}
Jeffreys, H. 1998, Theory of probability, Oxford Classic Texts in the Physical Sciences (The Clarendon Press, Oxford University Press, New York), xii+459, reprint of the 1983 edition

\bibitem[{Kass \& Raftery(1995)}]{Kass_Raftery_1995}
Kass, R.~E. \& Raftery, A.~E. 1995, Journal of the American Statistical Association, 90, 773

\bibitem[{{Kipping}(2010)}]{2010MNRAS.408.1758K}
{Kipping}, D.~M. 2010, \mnras, 408, 1758

\bibitem[{{Kipping}(2014)}]{kipping2014}
{Kipping}, D.~M. 2014, \mnras, 444, 2263

\bibitem[{{Kipping} {et~al.}(2015){Kipping}, {Schmitt}, {Huang}, {Torres}, {Nesvorn{\'y}}, {Buchhave}, {Hartman}, \& {Bakos}}]{2015ApJ...813...14K}
{Kipping}, D.~M., {Schmitt}, A.~R., {Huang}, X., {et~al.} 2015, \apj, 813, 14

\bibitem[{{Kosiarek} {et~al.}(2019){Kosiarek}, {Crossfield}, {Hardegree-Ullman}, {Livingston}, {Benneke}, {Henry}, {Howard}, {Berardo}, {Blunt}, {Fulton}, {Hirsch}, {Howard}, {Isaacson}, {Petigura}, {Sinukoff}, {Weiss}, {Bonfils}, {Dressing}, {Knutson}, {Schlieder}, {Werner}, {Gorjian}, {Krick}, {Morales}, {Astudillo-Defru}, {Almenara}, {Delfosse}, {Forveille}, {Lovis}, {Mayor}, {Murgas}, {Pepe}, {Santos}, {Udry}, {Corbett}, {Fors}, {Law}, {Ratzloff}, \& {del Ser}}]{2019AJ....157...97K}
{Kosiarek}, M.~R., {Crossfield}, I. J.~M., {Hardegree-Ullman}, K.~K., {et~al.} 2019, \aj, 157, 97

\bibitem[{{Leleu} {et~al.}(2024){Leleu}, {Delisle}, {Burn}, {Izidoro}, {Udry}, {Dumusque}, {Lovis}, {Millholland}, {Parc}, {Bouchy}, {Bourrier}, {Alibert}, {Faria}, {Mordasini}, \& {S{\'e}gransan}}]{Leleu2024}
{Leleu}, A., {Delisle}, J.-B., {Burn}, R., {et~al.} 2024, \aap, 687, L1

\bibitem[{{Liddle}(2007)}]{2007MNRAS.377L..74L}
{Liddle}, A.~R. 2007, \mnras, 377, L74

\bibitem[{{Lissauer} {et~al.}(2011){Lissauer}, {Fabrycky}, {Ford}, {Borucki}, {Fressin}, {Marcy}, {Orosz}, {Rowe}, {Torres}, {Welsh}, {Batalha}, {Bryson}, {Buchhave}, {Caldwell}, {Carter}, {Charbonneau}, {Christiansen}, {Cochran}, {Desert}, {Dunham}, {Fanelli}, {Fortney}, {Gautier}, {Geary}, {Gilliland}, {Haas}, {Hall}, {Holman}, {Koch}, {Latham}, {Lopez}, {McCauliff}, {Miller}, {Morehead}, {Quintana}, {Ragozzine}, {Sasselov}, {Short}, \& {Steffen}}]{2011Natur.470...53L}
{Lissauer}, J.~J., {Fabrycky}, D.~C., {Ford}, E.~B., {et~al.} 2011, \nat, 470, 53

\bibitem[{{Lithwick} {et~al.}(2012){Lithwick}, {Xie}, \& {Wu}}]{Lithwick2012}
{Lithwick}, Y., {Xie}, J., \& {Wu}, Y. 2012, \apj, 761, 122

\bibitem[{{Liu} {et~al.}(2015){Liu}, {Hori}, {Lin}, \& {Asphaug}}]{2015ApJ...812..164L}
{Liu}, S.-F., {Hori}, Y., {Lin}, D.~N.~C., \& {Asphaug}, E. 2015, \apj, 812, 164

\bibitem[{{Lopez}(2017)}]{2017MNRAS.472..245L}
{Lopez}, E.~D. 2017, \mnras, 472, 245

\bibitem[{{Lopez} \& {Fortney}(2014)}]{2014ApJ...792....1L}
{Lopez}, E.~D. \& {Fortney}, J.~J. 2014, \apj, 792, 1

\bibitem[{{Lopez} \& {Rice}(2018)}]{2018MNRAS.479.5303L}
{Lopez}, E.~D. \& {Rice}, K. 2018, \mnras, 479, 5303

\bibitem[{{Luo} {et~al.}(2024){Luo}, {Dorn}, \& {Deng}}]{Luoetal2024}
{Luo}, H., {Dorn}, C., \& {Deng}, J. 2024, Nature Astronomy, 8, 1399

\bibitem[{{Luque} \& {Pall{\'e}}(2022)}]{Luque_Palle_2022}
{Luque}, R. \& {Pall{\'e}}, E. 2022, Science, 377, 1211

\bibitem[{{Malavolta}(2016)}]{2016ascl.soft12008M}
{Malavolta}, L. 2016, {PyORBIT: Exoplanet orbital parameters and stellar activity}, Astrophysics Source Code Library, record ascl:1612.008

\bibitem[{{Martinez} {et~al.}(2019){Martinez}, {Cunha}, {Ghezzi}, \& {Smith}}]{2019ApJ...875...29M}
{Martinez}, C.~F., {Cunha}, K., {Ghezzi}, L., \& {Smith}, V.~V. 2019, \apj, 875, 29

\bibitem[{{Matsumura} {et~al.}(2010){Matsumura}, {Peale}, \& {Rasio}}]{2010ApJ...725.1995M}
{Matsumura}, S., {Peale}, S.~J., \& {Rasio}, F.~A. 2010, \apj, 725, 1995

\bibitem[{{Matsumura} {et~al.}(2008){Matsumura}, {Takeda}, \& {Rasio}}]{2008ApJ...686L..29M}
{Matsumura}, S., {Takeda}, G., \& {Rasio}, F.~A. 2008, \apjl, 686, L29

\bibitem[{{Mortier} {et~al.}(2018){Mortier}, {Bonomo}, {Rajpaul}, {Buchhave}, {Vanderburg}, {Zeng}, {L{\'o}pez-Morales}, {Malavolta}, {Collier Cameron}, {Dressing}, {Figueira}, {Nascimbeni}, {Rice}, {Sozzetti}, {Watson}, {Affer}, {Bouchy}, {Charbonneau}, {Harutyunyan}, {Haywood}, {Johnson}, {Latham}, {Lovis}, {Martinez Fiorenzano}, {Mayor}, {Micela}, {Molinari}, {Motalebi}, {Pepe}, {Piotto}, {Phillips}, {Poretti}, {Sasselov}, {S{\'e}gransan}, \& {Udry}}]{2018MNRAS.481.1839M}
{Mortier}, A., {Bonomo}, A.~S., {Rajpaul}, V.~M., {et~al.} 2018, \mnras, 481, 1839

\bibitem[{{Mukherjee} {et~al.}(1998){Mukherjee}, {Feigelson}, {Jogesh Babu}, {Murtagh}, {Fraley}, \& {Raftery}}]{Mukherjee1998ApJ}
{Mukherjee}, S., {Feigelson}, E.~D., {Jogesh Babu}, G., {et~al.} 1998, \apj, 508, 314

\bibitem[{{Nascimbeni} {et~al.}(2023){Nascimbeni}, {Borsato}, {Zingales}, {Piotto}, {Pagano}, {Beck}, {Broeg}, {Ehrenreich}, {Hoyer}, {Majidi}, {Granata}, {Sousa}, {Wilson}, {Van Grootel}, {Bonfanti}, {Salmon}, {Mustill}, {Delrez}, {Alibert}, {Alonso}, {Anglada}, {B{\'a}rczy}, {Barrado}, {Barros}, {Baumjohann}, {Beck}, {Benz}, {Bergomi}, {Billot}, {Bonfils}, {Brandeker}, {Cabrera}, {Charnoz}, {Collier Cameron}, {Csizmadia}, {Cubillos}, {Davies}, {Deleuil}, {Deline}, {Demangeon}, {Demory}, {Erikson}, {Fortier}, {Fossati}, {Fridlund}, {Gandolfi}, {Gillon}, {G{\"u}del}, {Isaak}, {Kiss}, {Laskar}, {Lecavelier des Etangs}, {Lendl}, {Lovis}, {Luque}, {Magrin}, {Maxted}, {Mordasini}, {Olofsson}, {Ottensamer}, {Pall{\'e}}, {Peter}, {Piazza}, {Pollacco}, {Queloz}, {Ragazzoni}, {Rando}, {Ratti}, {Rauer}, {Ribas}, {Santos}, {Scandariato}, {S{\'e}gransan}, {Simon}, {Smith}, {Steinberger}, {Steller}, {Szab{\'o}}, {Thomas}, {Udry}, {Venturini}, {Walton}, \& {Wolter}}]{Nascimbeni2023}
{Nascimbeni}, V., {Borsato}, L., {Zingales}, T., {et~al.} 2023, \aap, 673, A42

\bibitem[{{Nava} {et~al.}(2020){Nava}, {L{\'o}pez-Morales}, {Haywood}, \& {Giles}}]{nava2020}
{Nava}, C., {L{\'o}pez-Morales}, M., {Haywood}, R.~D., \& {Giles}, H. A.~C. 2020, \aj, 159, 23

\bibitem[{{Nesvorn{\'y}} \& {Vokrouhlick{\'y}}(2016)}]{NeVo2016}
{Nesvorn{\'y}}, D. \& {Vokrouhlick{\'y}}, D. 2016, \apj, 823, 72

\bibitem[{{Owen} \& {Wu}(2013)}]{2013ApJ...775..105O}
{Owen}, J.~E. \& {Wu}, Y. 2013, \apj, 775, 105

\bibitem[{{Owen} \& {Wu}(2016)}]{2016ApJ...817..107O}
{Owen}, J.~E. \& {Wu}, Y. 2016, \apj, 817, 107

\bibitem[{{Owen} \& {Wu}(2017)}]{2017ApJ...847...29O}
{Owen}, J.~E. \& {Wu}, Y. 2017, \apj, 847, 29

\bibitem[{Parviainen(2015)}]{Parviainen2015}
Parviainen, H. 2015, MNRAS, 450, 3233

\bibitem[{{Pepe} {et~al.}(2002){Pepe}, {Mayor}, {Galland}, {Naef}, {Queloz}, {Santos}, {Udry}, \& {Burnet}}]{2002A&A...388..632P}
{Pepe}, F., {Mayor}, M., {Galland}, F., {et~al.} 2002, \aap, 388, 632

\bibitem[{Rajpaul {et~al.}(2015)Rajpaul, Aigrain, Osborne, Reece, \& Roberts}]{Rajpaul2015}
Rajpaul, V., Aigrain, S., Osborne, M.~A., Reece, S., \& Roberts, S. 2015, \mnras, 452, 2269

\bibitem[{{Rajpaul} {et~al.}(2017){Rajpaul}, {Buchhave}, \& {Aigrain}}]{2017MNRAS.471L.125R}
{Rajpaul}, V., {Buchhave}, L.~A., \& {Aigrain}, S. 2017, \mnras, 471, L125

\bibitem[{{Rajpaul} {et~al.}(2021){Rajpaul}, {Buchhave}, {Lacedelli}, {Rice}, {Mortier}, {Malavolta}, {Aigrain}, {Borsato}, {Mayo}, {Charbonneau}, {Damasso}, {Dumusque}, {Ghedina}, {Latham}, {L{\'o}pez-Morales}, {Magazz{\`u}}, {Micela}, {Molinari}, {Pepe}, {Piotto}, {Poretti}, {Rowther}, {Sozzetti}, {Udry}, \& {Watson}}]{2021MNRAS.507.1847R}
{Rajpaul}, V.~M., {Buchhave}, L.~A., {Lacedelli}, G., {et~al.} 2021, \mnras, 507, 1847

\bibitem[{Rasmussen \& Williams(2006)}]{rasmussen2006}
Rasmussen, C.~E. \& Williams, C. K.~I. 2006, {Gaussian Processes for Machine Learning} (MIT Press, Cambridge MA)

\bibitem[{{Rauer} {et~al.}(2024){Rauer}, {Aerts}, {Cabrera}, {Deleuil}, {Erikson}, {Gizon}, {Goupil}, {Heras}, {Lorenzo-Alvarez}, {Marliani}, {Martin-Garcia}, {Mas-Hesse}, {O'Rourke}, {Osborn}, {Pagano}, {Piotto}, {Pollacco}, {Ragazzoni}, {Ramsay}, {Udry}, {Appourchaux}, {Benz}, {Brandeker}, {G{\"u}del}, {Janot-Pacheco}, {Kabath}, {Kjeldsen}, {Min}, {Santos}, {Smith}, {Suarez}, {Werner}, {Aboudan}, {Abreu}, {Acu a}, {Adams}, {Adibekyan}, {Affer}, {Agneray}, {Agnor}, {Aguirre B{\o}rsen-Koch}, {Ahmed}, {Aigrain}, {Al-Bahlawan}, {Alcacera Gil}, {Alei}, {Alencar}, {Alexander}, {Alfonso-Garz{\'o}n}, {Alibert}, {Allende Prieto}, {Almeida}, {Alonso Sobrino}, {Altavilla}, {Althaus}, {Alonso Alvarez Trujillo}, {Amarsi}, {Ammler-von Eiff}, {Am{\^o}res}, {Andrade}, {Antoniadis-Karnavas}, {Ant{\'o}nio}, {Aparicio del Moral}, {Appolloni}, {Arena}, {Armstrong}, {Aroca Aliaga}, {Asplund}, {Audenaert}, {Auricchio}, {Avelino}, {Baeke}, {Bailli{\'e}}, {Balado}, {Ballber Balaguer{\'o}}, {Balestra}, {Ball}, {Ballans}, {Ballot},
  {Barban}, {Barbary}, {Barbieri}, {Barcel{\'o} Forteza}, {Barker}, {Barklem}, {Barnes}, {Barrado Navascues}, {Barragan}, {Baruteau}, {Basu}, {Baudin}, {Baumeister}, {Bayliss}, {Bazot}, {Beck}, {Bedding}, {Belkacem}, {Bellinger}, {Benatti}, {Benomar}, {B{\'e}rard}, {Bergemann}, {Bergomi}, {Bernardo}, {Biazzo}, {Bignamini}, {Bigot}, {Billot}, {Binet}, {Biondi}, {Biondi}, {Birch}, {Bitsch}, {Bluhm Ceballos}, {B{\'o}di}, {Bogn{\'a}r}, {Boisse}, {Bolmont}, {Bonanno}, {Bonavita}, {Bonfanti}, {Bonfils}, {Bonito}, {Bonomo}, {B{\"o}rner}, {Boro Saikia}, {Borreguero Mart{\'\i}n}, {Borsa}, {Borsato}, {Bossini}, {Bouchy}, {Bou{\'e}}, {Boufleur}, {Boumier}, {Bourrier}, {Bowman}, {Bozzo}, {Bradley}, {Bray}, {Bressan}, {Breton}, {Brienza}, {Brito}, {Brogi}, {Brown}, {Brown}, {Brun}, {Bruno}, {Bruns}, {Buchhave}, {Bugnet}, {Buldgen}, {Burgess}, {Busatta}, {Busso}, {Buzasi}, {Caballero}, {Cabral}, {Cabrero Gomez}, {Calderone}, {Cameron}, {Cameron}, {Campante}, {Campos Gestal}, {Canto Martins}, {Cara}, {Carone}, {Carrasco},
  {Casagrande}, {Casewell}, {Cassisi}, {Castellani}, {Castro}, {Catala}, {Catal{\'a}n Fern{\'a}ndez}, {Catelan}, {Cegla}, {Cerruti}, {Cessa}, {Chadid}, {Chaplin}, {Charpinet}, {Chiappini}, {Chiarucci}, {Chiavassa}, {Chinellato}, {Chirulli}, {Christensen-Dalsgaard}, {Church}, {Claret}, {Clarke}, {Claudi}, {Clermont}, {Coelho}, {Coelho}, {Cogato}, {Colom{\'e}}, {Condamin}, {Conde Garc{\'\i}a}, \& {Conseil}}]{Rauer2024}
{Rauer}, H., {Aerts}, C., {Cabrera}, J., {et~al.} 2024, arXiv e-prints, arXiv:2406.05447

\bibitem[{{Rauer} {et~al.}(2014){Rauer}, {Catala}, {Aerts}, {Appourchaux}, {Benz}, {Brandeker}, {Christensen-Dalsgaard}, {Deleuil}, {Gizon}, {Goupil}, {G{\"u}del}, {Janot-Pacheco}, {Mas-Hesse}, {Pagano}, {Piotto}, {Pollacco}, {Santos}, {Smith}, {Su{\'a}rez}, {Szab{\'o}}, {Udry}, {Adibekyan}, {Alibert}, {Almenara}, {Amaro-Seoane}, {Eiff}, {Asplund}, {Antonello}, {Barnes}, {Baudin}, {Belkacem}, {Bergemann}, {Bihain}, {Birch}, {Bonfils}, {Boisse}, {Bonomo}, {Borsa}, {Brand{\~a}o}, {Brocato}, {Brun}, {Burleigh}, {Burston}, {Cabrera}, {Cassisi}, {Chaplin}, {Charpinet}, {Chiappini}, {Church}, {Csizmadia}, {Cunha}, {Damasso}, {Davies}, {Deeg}, {D{\'\i}az}, {Dreizler}, {Dreyer}, {Eggenberger}, {Ehrenreich}, {Eigm{\"u}ller}, {Erikson}, {Farmer}, {Feltzing}, {de Oliveira Fialho}, {Figueira}, {Forveille}, {Fridlund}, {Garc{\'\i}a}, {Giommi}, {Giuffrida}, {Godolt}, {Gomes da Silva}, {Granzer}, {Grenfell}, {Grotsch-Noels}, {G{\"u}nther}, {Haswell}, {Hatzes}, {H{\'e}brard}, {Hekker}, {Helled}, {Heng}, {Jenkins},
  {Johansen}, {Khodachenko}, {Kislyakova}, {Kley}, {Kolb}, {Krivova}, {Kupka}, {Lammer}, {Lanza}, {Lebreton}, {Magrin}, {Marcos-Arenal}, {Marrese}, {Marques}, {Martins}, {Mathis}, {Mathur}, {Messina}, {Miglio}, {Montalban}, {Montalto}, {Monteiro}, {Moradi}, {Moravveji}, {Mordasini}, {Morel}, {Mortier}, {Nascimbeni}, {Nelson}, {Nielsen}, {Noack}, {Norton}, {Ofir}, {Oshagh}, {Ouazzani}, {P{\'a}pics}, {Parro}, {Petit}, {Plez}, {Poretti}, {Quirrenbach}, {Ragazzoni}, {Raimondo}, {Rainer}, {Reese}, {Redmer}, {Reffert}, {Rojas-Ayala}, {Roxburgh}, {Salmon}, {Santerne}, {Schneider}, {Schou}, {Schuh}, {Schunker}, {Silva-Valio}, {Silvotti}, {Skillen}, {Snellen}, {Sohl}, {Sousa}, {Sozzetti}, {Stello}, {Strassmeier}, {{\v{S}}vanda}, {Szab{\'o}}, {Tkachenko}, {Valencia}, {Van Grootel}, {Vauclair}, {Ventura}, {Wagner}, {Walton}, {Weingrill}, {Werner}, {Wheatley}, \& {Zwintz}}]{Rauer2014}
{Rauer}, H., {Catala}, C., {Aerts}, C., {et~al.} 2014, Experimental Astronomy, 38, 249

\bibitem[{{Reinhardt} {et~al.}(2022){Reinhardt}, {Meier}, {Stadel}, {Otegi}, \& {Helled}}]{Reinhardt2022}
{Reinhardt}, C., {Meier}, T., {Stadel}, J.~G., {Otegi}, J.~F., \& {Helled}, R. 2022, \mnras, 517, 3132

\bibitem[{Roberts {et~al.}(2013)Roberts, Osborne, Ebden, Reece, Gibson, \& Aigrain}]{roberts2013}
Roberts, S., Osborne, M., Ebden, M., {et~al.} 2013, Philosophical Transactions of the Royal Society A: Mathematical, Physical and Engineering Sciences, 371, 20110550

\bibitem[{Rogers \& Seager(2010)}]{Rogers_2010}
Rogers, L.~A. \& Seager, S. 2010, The Astrophysical Journal, 712, 974

\bibitem[{{Singh} {et~al.}(2022){Singh}, {Bonomo}, {Scandariato}, {Cibrario}, {Barbato}, {Fossati}, {Pagano}, \& {Sozzetti}}]{2022A&A...658A.132S}
{Singh}, V., {Bonomo}, A.~S., {Scandariato}, G., {et~al.} 2022, \aap, 658, A132

\bibitem[{{Spiegel} {et~al.}(2014){Spiegel}, {Fortney}, \& {Sotin}}]{2014PNAS..11112622S}
{Spiegel}, D.~S., {Fortney}, J.~J., \& {Sotin}, C. 2014, Proceedings of the National Academy of Science, 111, 12622

\bibitem[{Storn \& Price(1997)}]{Storn1997}
Storn, R. \& Price, K. 1997, Journal of Global Optimization, 11, 341

\bibitem[{{Ter Braak}(2006)}]{TerBraak2006}
{Ter Braak}, C. J.~F. 2006, Statistics and Computing, 16, 239

\bibitem[{{Van Eylen} {et~al.}(2018){Van Eylen}, {Agentoft}, {Lundkvist}, {Kjeldsen}, {Owen}, {Fulton}, {Petigura}, \& {Snellen}}]{2018MNRAS.479.4786V}
{Van Eylen}, V., {Agentoft}, C., {Lundkvist}, M.~S., {et~al.} 2018, \mnras, 479, 4786

\bibitem[{{Van Eylen} {et~al.}(2019){Van Eylen}, {Albrecht}, {Huang}, {MacDonald}, {Dawson}, {Cai}, {Foreman-Mackey}, {Lundkvist}, {Silva Aguirre}, {Snellen}, \& {Winn}}]{2019AJ....157...61V}
{Van Eylen}, V., {Albrecht}, S., {Huang}, X., {et~al.} 2019, \aj, 157, 61

\bibitem[{{Venturini} {et~al.}(2024){Venturini}, {Ronco}, {Guilera}, {Haldemann}, {Mordasini}, \& {Miller Bertolami}}]{Venturini2024}
{Venturini}, J., {Ronco}, M.~P., {Guilera}, O.~M., {et~al.} 2024, \aap, 686, L9

\bibitem[{{Weiss} {et~al.}(2016){Weiss}, {Rogers}, {Isaacson}, {Agol}, {Marcy}, {Rowe}, {Kipping}, {Fulton}, {Lissauer}, {Howard}, \& {Fabrycky}}]{2016ApJ...819...83W}
{Weiss}, L.~M., {Rogers}, L.~A., {Isaacson}, H.~T., {et~al.} 2016, \apj, 819, 83

\bibitem[{{Zakamska} {et~al.}(2011){Zakamska}, {Pan}, \& {Ford}}]{2011MNRAS.410.1895Z}
{Zakamska}, N.~L., {Pan}, M., \& {Ford}, E.~B. 2011, \mnras, 410, 1895

\bibitem[{{Zechmeister} \& {K{\"u}rster}(2009)}]{2009A&A...496..577Z}
{Zechmeister}, M. \& {K{\"u}rster}, M. 2009, \aap, 496, 577

\bibitem[{Zeng \& Jacobsen(2024)}]{Zeng2024}
Zeng, L. \& Jacobsen, S.~B. 2024, Icarus

\bibitem[{{Zeng} {et~al.}(2019){Zeng}, {Jacobsen}, {Sasselov}, {Petaev}, {Vanderburg}, {Lopez-Morales}, {Perez-Mercader}, {Mattsson}, {Li}, {Heising}, {Bonomo}, {Damasso}, {Berger}, {Cao}, {Levi}, \& {Wordsworth}}]{2019PNAS..116.9723Z}
{Zeng}, L., {Jacobsen}, S.~B., {Sasselov}, D.~D., {et~al.} 2019, Proceedings of the National Academy of Science, 116, 9723

\bibitem[{{Zeng} \& {Sasselov}(2013)}]{Zeng2013}
{Zeng}, L. \& {Sasselov}, D. 2013, \pasp, 125, 227

\bibitem[{{Zeng} \& {Seager}(2008)}]{2008PASP..120..983Z}
{Zeng}, L. \& {Seager}, S. 2008, \pasp, 120, 983

\end{thebibliography}

\appendix

\onecolumn

\section{Generalized Lomb-Scargle periodograms of the HARPS-N radial velocities}

\begin{figure*}[h!]
\centering
\includegraphics[width=16 cm]{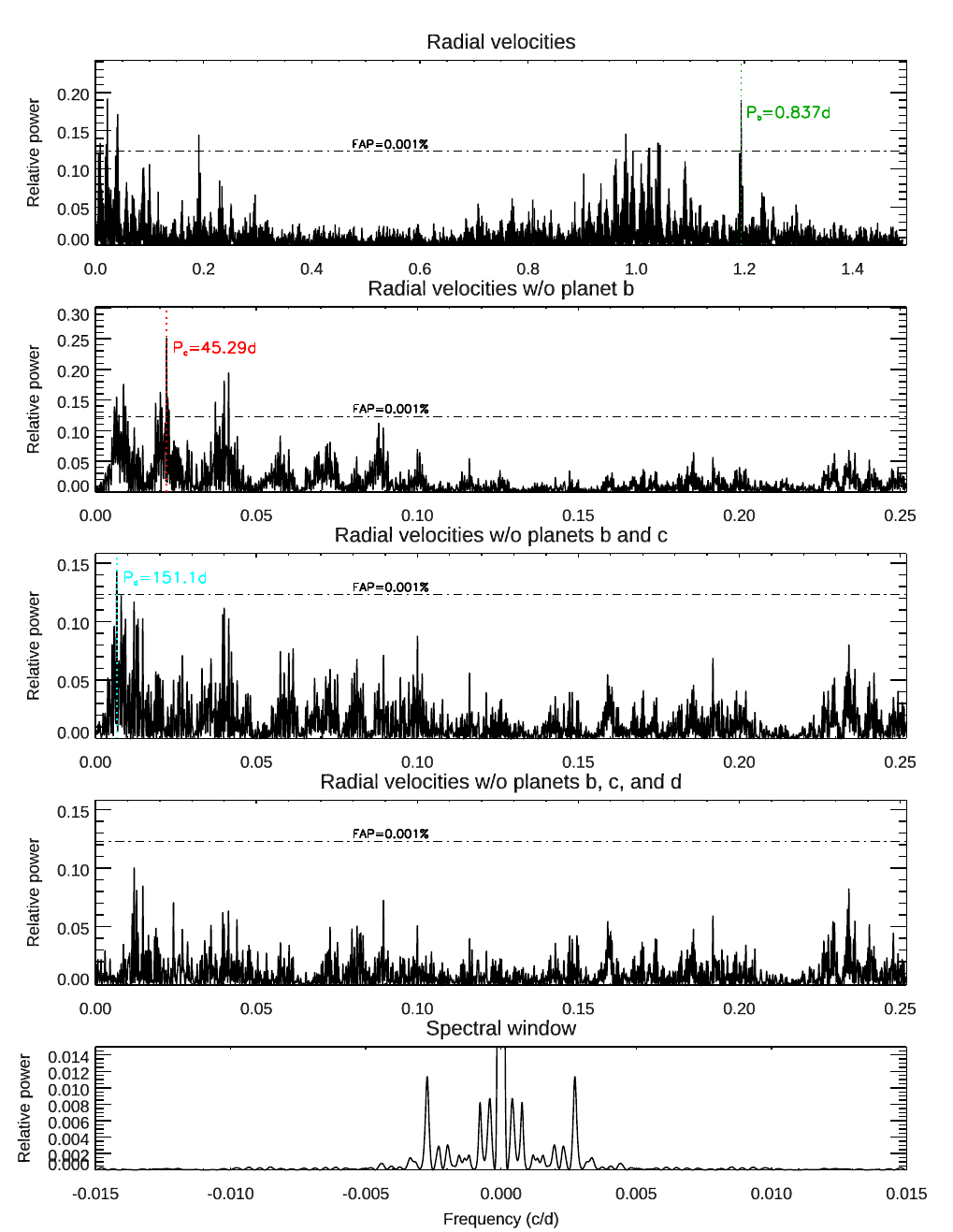}
\caption{Generalized Lomb-Scargle periodograms of the 291 HARPS-N radial velocities and residuals as a function of frequency in cycles/day (or day$^{-1}$). From top to bottom are displayed the periodograms of radial velocities; residuals after subtracting the signal of Kepler-10\,b; residuals after subtracting the signals of both Kepler-10\,b and c; residuals after subtracting the signals of Kepler-10\,b, c, and d; and the spectral window. The dotted vertical lines indicate the orbital periods of planets b (green), c (red), and d (light blue). The horizontal dash-dotted lines show the false alarm probability level of $10^{-5}$.}
\label{figure_GLS_periodograms}
\end{figure*}

\clearpage

\twocolumn

\section{Exploring the parameter space}
\label{app:fip_asp}
\subsection{False inclusion probability}
\label{app:fip}

This appendix gives the details of the analysis of the HARPS-N data on Kepler-10 processed with YARARA-v2 \citep{cretignier2021, 2022A&A...659A..68C}.
We define the true inclusion probability (TIP) as  the posterior probability of the following event: for a given range of periods $I$, there is at least one planet with period $P \in I$. By analogy with the FAP, we also define the false inclusion probability (FIP), which is the probability that there is no planet with period $P$ in interval $I$. Formally, denoting the data by $y$, the TIP and FIP are 
\begin{align}
    \label{eq:criterion}
        \mathrm{TIP}_I &= %
    \mathrm{Pr}\{ \exists P, P \in I| y  \}. 
        \\
        \label{eq:fip}
    \mathrm{FIP}_I &= 1 - \mathrm{TIP}_I.
\end{align}
In practice, we can evaluate Eq.~\eqref{eq:fip}  as follows.
We suppose that there is a maximum number of Keplerian signals in the data $k_{\mathrm{max}}$. We denote by $p(\vec \theta|k)$ the prior probability of model parameters knowing there are $k$ planets, and $p(y|\vec \theta,k)$ the likelihood of the data $y$ knowing the number of planets $k$ and the orbital parameters. We suppose a prior probability for the model with $k$ planets $\mathrm{Pr}\{k\}$. Then, as defined in Eq.~\eqref{eq:criterion}, the TIP is
\begin{align}
        \label{eq:calc0}
        \mathrm{TIP}_I  = \sum\limits_{k=0}^{k_{\mathrm{max}}}    \mathrm{Pr}\{ \exists i \in [1..k] , P_i \in I | y, k  \} \mathrm{Pr}\{k | y \}, 
\end{align}
where the $\mathrm{Pr}\{k | y \}$ term is computed from the Bayesian evidences of $k$ planet models, and is called PNP for posterior number of planets, as defined in \cite{brewer2014}. It is the probability to have $k$ planets knowing the data. 
Denoting by $\mathcal{Z}_k$ the Bayesian evidence of the model with $k$ planets, assuming there are at most $k_{\mathrm{max}}$ planets, and denoting by $\mathrm{Pr}\{i\} $ the prior probability to have $i$ planets,
         \begin{align}
                \mathrm{Pr}\{k | y \}  =  \frac{  \mathcal{Z}_k\mathrm{Pr}\{k\}  }{    \sum\limits_{i=0}^{k_{\mathrm{max}} }   \mathcal{Z}_i \mathrm{Pr}\{i\}    }.
                \label{eq:jointpost}
         \end{align}

The main advantage of the TIP/FIP is that it is very easy to interpret. If the model is correct, among detections made with TIP = $\alpha$ = 1- FIP, on average a fraction $\alpha$ are correct detections and a fraction $1-\alpha$ are spurious ones. 

We perform the FIP calculation up to four planets with two different types of priors for the semiamplitude. The rationale is that, as we have shown in \cite{hara2022a}, the significance of low-amplitude signals highly depends on the chosen prior.  In the two sets of priors, we use a $\log$-uniform prior on period, we constrain the periods and phase of two of the planets according to the constraints from the transit. The eccentricity has a Beta prior as in \cite{kipping2014}, the prior on the argument of periastron is uniform.

We then compute the FIP/TIP as the probability to have at least one planet with frequency in the interval $[f-1/T_{\mathrm{obs}}, f+1/T_{\mathrm{obs}}]$ where $T_{\mathrm{obs}}$ is the total time of observation, for an array of equispaced frequencies.  In the first set of priors, we use a log-uniform prior from 0.1 to 20 \ms on the RV semiamplitudes, and obtain the FIP periodogram shown in Fig.~\ref{fig:fip_perio} (a). 

In the second set of priors, the prior is not defined on semiamplitude but on the mass. This means that, for each set of period, eccentricity and time of passage at periastron (or equivalently, mean anomaly at a reference time), the prior on semiamplitude is scaled to correspond to the same prior on mass. This choice implicitly supposes that the probability of occurrence of a given type of planet (super-Earth, Neptune) does not depend on semimajor axis. Since, for a given mass, the semiamplitude decreases with separation, this will also have the effect to boost the significance of low amplitude signals. We use a Gaussian mixture model for the prior on mass, which follows a Rayleigh prior with $\sigma = 5 M_\oplus$ or $\sigma = 15 M_\oplus$ with a probability 1/2. With this choice, we can marginalize the likelihood analytically over the linear parameters as shown in Appendix C of \cite{hara2022a}. In that case, a 68~d signal reaches a FIP of 4\% (see Fig.~\ref{fig:fip_perio}, b).

\begin{figure*}
    \centering
    \vspace{1 cm}    \includegraphics[width=\linewidth]{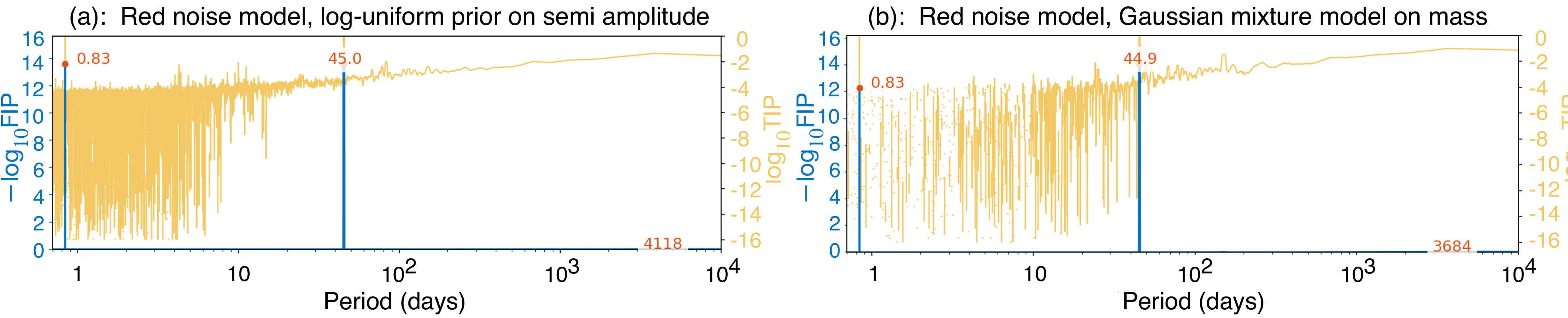}
    \caption{False inclusion probability periodograms of the Kepler-10 HARPS-N data processed with YARARA-v2 for a red noise model, for different priors on semiamplitude. (a) is obtained with a log-uniform prior on semiamplitude, (b) with a Gaussian mixture model on mass. In yellow we represent the true inclusion probability (TIP), the probability of having a planet in the range $[\omega - \Delta \omega,  \omega + \Delta \omega]$ as a function of $\omega$. In blue, we represent $-\log_{10} FIP$, where FIP = 1 - TIP is the false inclusion probability. }
    \label{fig:fip_perio_appendix}
    \vspace{0.75 cm}
\end{figure*}

We did the same analyses as before but including a red noise term with an exponential kernel (free jitter and time-scale). The corresponding FIP periodograms are shown in Fig. \ref{fig:fip_perio_appendix}, where the 150.7~d signal is much less significant and completely disappears, respectively. However, we caution that the red noise model is expected to damp the amplitude of long-period signals, as shown in~\cite{delisle2020}. It is common for this type of noise model to suppress viable planets. If what happens at low frequency was truly red noise, then we would expect the white noise model to find several unclear planet candidates at long periods, due to a diffuse power at low frequencies. However, in Kepler-10 the white noise model shows a clear candidate with a well defined period. As a consequence, the interpretation of the FIP periodogram with a red noise model with exponential kernel is subject to caution.

\begin{table}
\small
        \centering

\caption{YARARA-v2 reduction, white noise model. }
                \label{tab:wn}
\begin{tabular}{ccccc}
Number of planets & $\ln  \mathcal{Z}$ & $\sigma_\mathcal{Z}$ &$ \Delta \ln  \mathcal{Z}$ &PNP \\
\hline
0 &-799.83      &0.24 & 0.0     & 7.009-30\\
1 &-778.50      &0.19   &21.33 & 1.286-20\\
2 &-741.08      &0.10&  37.41 & 2.294-04\\
3 &-732.70      &0.36&  8.38 &  9.998e-01\\
\end{tabular}

\vspace{0.5 cm}

        \centering
        \caption{YARARA-v2 reduction, red noise model}
        \label{tab:rn}
        \begin{tabular}{ccccc}
                Number of planets & $\ln  \mathcal{Z}$ & $\sigma_\mathcal{Z}$ &$ \Delta \ln  \mathcal{Z}$ &PNP \\
        \hline
        0&      -777.55 &0.12 & 0.0     &1.79e-19 \\
        1&      -771.34 &0.17&  6.20  & 8.89e-17  \\
        2&      -734.77&        0.17    &36.57& 6.81e-01  \\
        3&      -735.53 &0.13&  -0.76 & 3.19e-01 
        \end{tabular}
\tablefoot{$\mathcal{Z}$  is the Bayesian evidence and PNP is the posterior probability of the number of planets. }
\end{table}

In Table~\ref{tab:wn} and Table~\ref{tab:rn} we show the evidences $\mathcal{Z}$ and uncertainty on them (standard deviation of three runs) for the log-uniform prior on K, white and red noise assumptions, respectively. We can now compute the TIP of the 150.7~d planet averaged over the noise model.  Denoting by $I$ the frequency interval $[\frac{2\pi}{150.7} -\Delta\omega, \frac{2\pi}{150.7} +\Delta\omega ] $, we have
\begin{align}
p( \omega \in I \mid y) = p( \omega \in I \mid W, y) p(W \mid y )  + p( \omega \in I \mid R,y) p(R \mid y) ,
\end{align}
where $W$ and $R$ represent the white and red noise signal. The two terms $p( \omega \in I \mid W, y) $ and $p( \omega \in I \mid R,y)$ are equal to 1-FIP of the 150.7~d signal for the white and red noise models, respectively. We have $p( \omega \in I \mid R,y)=0$. Assuming white and red noise have both a prior probability of 1/2, we have 
\begin{align}
p(  W \mid y) =  \frac{p(  y \mid W) } {p(  y \mid W)  + p(  y \mid R) }
\end{align}
and 
\begin{align}
p(  y \mid W) = \sum_{k=0}^3 p(  y \mid W, k) p(k)
\end{align}
where $k$ is the number of planets. With the values of Table~\ref{tab:wn} and Table~\ref{tab:rn}, assuming $p(k) = 1/4$, we have 
\begin{align}
        p(  y \mid W) &= \e^{-732.70}/4  \\
        p(  y \mid R) &= ( \e^{-734.77} + \e^{-735.53}   )/4  
\end{align}
which means that  $p(  W \mid y) = 0.84$. As a result, the probability to have a signal at 150.7~d is $(1- 10^{-2.16}) \times 0.84 \sim 0.84$. The probability that there is a signal in $I$ marginalized on the noise models considered here is 84\%.  Conversely, we have a FIP for the 150.7~d signal of 1 - 0.84 = 0.16.


\subsection{Apodized sine periodograms, white noise}
\label{app:asp}

The result from the two previous analyses show that there is a significant signal at 150.7~d, and hints of signals at 24 and 68~d, the latter being the most promising, but still not significant enough to be a clear detection. The question is then mainly to determine if the 150.7~d signal, which is clearly significant,  can be attributed to a planet. 

To address this question, we analyze the data in a different framework: the apodized sine periodogram \citep{hara2022b}. This framework, proposed by \cite{gregory2016}, is designed to determine if signals are present throughout the whole dataset, and have a constant phase, amplitude and frequency. This approach is particularly useful when there is no clear model for the data.
In practice, we fit the apodized sine function 
        \begin{align}
        \mu_K(t, \omega, \tau,t_0, A,B ) = w(\tau, t_0) (A \cos \omega t + B \sin \omega t ) ,
        \label{eq:mu1}
\end{align}
where  $w(\tau, t_0) $ is the apodization function, $t_0$ is its center and $\tau$ its temporal width. The resulting signal $\mu$ is then wavelet-like, quasi periodic from $\sim t_0 - \tau $ to $\sim t_0 + \tau $ and 0 elsewhere. 
We here use Gaussian and box-shaped functions, that is
\begin{align}
        w_G(\tau, t_0) &:= \e^{-\frac{(t-t_0)^2}{2\tau^2}}   \label{eq:gauss}\\
        w_B(\tau, t_0) &:= \mathbf{1}_{\left[t_0 - \frac{\tau}{2}, t_0 + \frac{\tau}{2}\right]} (\tau, t_0) \label{eq:box}
\end{align}

We then compute a periodogram, using the same defintion as \cite{baluev2008},
        \begin{align}
        z(\omega, t_0,\tau) &= \chi^2_H - \chi^2_K(\omega, t_0,\tau), \label{eq:z}
\end{align}
where $\chi^2_H $ is the $\chi^2$ of the residuals after fitting the model $H$ (the base model, without sinusoidal function) and the model $K(t, \omega, \tau,t_0)$, that is, linearly fitting $A$ and $B$ for model Eq.~\ref{eq:mu1} simultaneously with the base model. Then, for a given value of $\tau$, we represent $z$ as a function of $\omega$ maximized on $t_0$, that is, for a given frequency and timescale, the best fitting model when the center of the window varies. 

\begin{figure*}
\centering 
\vspace{2.5cm}
\hspace*{-1.2 cm}
\includegraphics[width=1.15\linewidth]{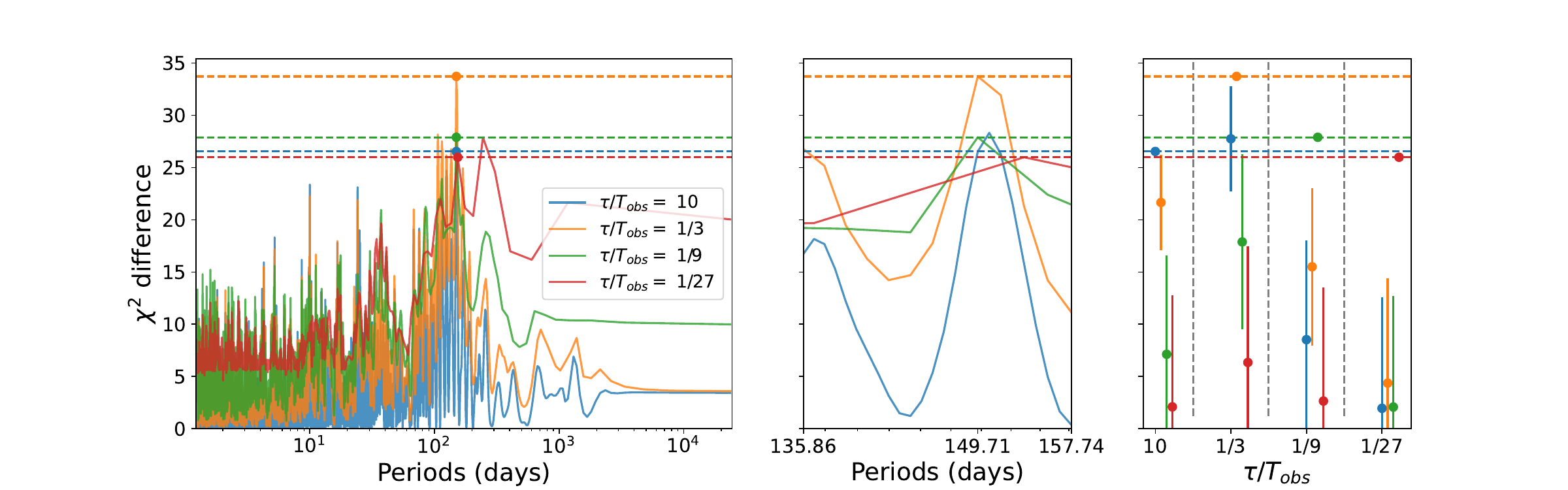}
        \caption{Apodized sine periodogram of the Kepler-10 HARPS-N data with YARARA-v2 reduction.}
        \label{fig:asprv2}
\end{figure*}

\begin{figure*}
\vspace{3cm}
\hspace*{0.25 cm}
\centering
\includegraphics[width=\linewidth]{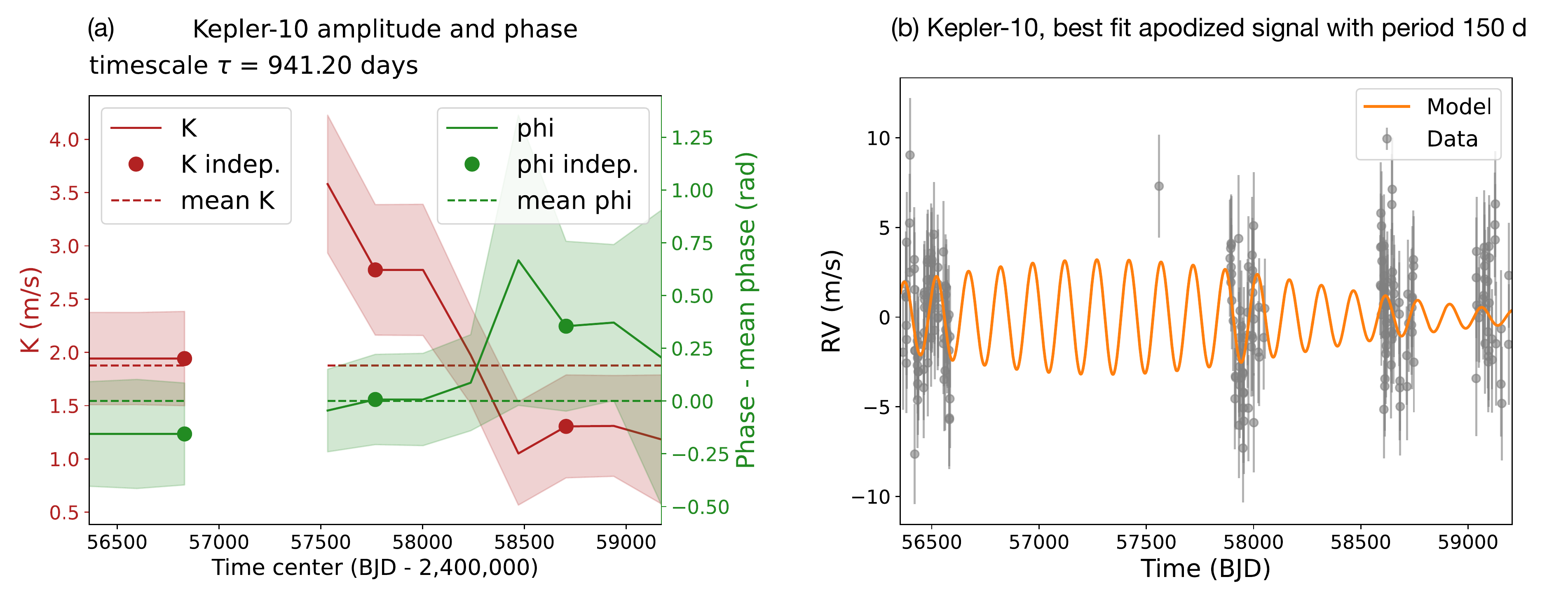}
    \caption{(a) Semiamplitude (red) and phase (green) of the 150~d signal averaged over a window of size $ [t_0 - \tau, t_0 + \tau ]$ where $\tau =T_\mathrm{obs}/3$, and $T_\mathrm{obs}$ is the total observation timespan. The solid lines represent the least square fit, and the shaded areas the $\pm1\sigma$ intervals. Red and green dotted lines represent the mean value of the amplitude and phase, respectively.  (b) In grey, HARPS-N data processed with YARARA-v2; in orange, a model of the form $ \mathrm{e}^{-\frac{(t-t_0)^2}{2\tau^2}}  (A \cos \omega t + B \sin \omega t )$, where $\omega, A, B, \tau$, and $t_0$ are adjusted starting from $ \omega = 2\pi/150$ rad/day.}
    \label{fig:fip_asp}
\end{figure*}

The transiting planets are added to the base model (they are assumed to be there in the null hypothesis).  We take four values of $\tau$ in the grid, such that  $\tau/T_{\mathrm{obs}} = 10, 1/3, 1/9, 1/27$ where $T_{\mathrm{obs}}$ is the total observation time-span. The apodized sine periodogram for Kepler-10 (with the transiting planets in the null hypothesis model) is shown in Fig.~\ref{fig:asprv2}. The best fit occurs at $\tau =T_\mathrm{obs}/3$, and the corresponding model is shown in Fig.~\ref{fig:fip_asp} (b). Given the gap in the data, the best fitting timescale might be spuriously shorter than the true ones due to random fluctuations. The right panel of Fig.~\ref{fig:asprv2} shows a statistical test to see if some values of $\tau$ can be excluded (see \citealt{hara2022b}). It shows that the highest value of $\tau$ is still 2$\sigma$ compatible with the data. 

In Fig.~\ref{fig:fip_asp} (a) we show the evolution of the phase and amplitude of that signal over the observations. While the phase is constant, the amplitude tends to vanish towards the end of the observations. We then multiplied the sinusoidal model of the signal by a so called apodization factor \citep{gregory2016, hara2022b}, $\e^{(t-t_0)^2/(2\tau^2)}$ where $t$ is the time, and $t_0$ and $\tau$ are free parameters. The best fit is represented in Fig.~\ref{fig:fip_asp} (b), where it seems that the amplitude of the signal vanishes at the end of the dataset. As said above, this is simply the best fitting timescale, but a purely sinusoidal signal ($\tau = 10\,T_{\mathrm{obs}}$) is still compatible with the data. 
We performed the same analysis on the ancillary indicators and did not find trace of signals at $\sim151$~d.

\clearpage


\twocolumn
~\\
\section{Radial-velocity analyses}

\begin{table}[h!]
\small
\vspace{1.0 cm}
\centering 
\caption{Priors imposed in the radial-velocity analyses.} 
\begin{tabular}{l c}
\hline
$\gamma_{\rm HN-1}$ [\ms] & $U]-\infty, +\infty[$ \\
$\gamma_{\rm HN-2}$ [\ms] & $U]-\infty, +\infty[$ \\
$\gamma_{\rm HIRES}$ [\ms] & $U]-\infty, +\infty[$ \\
$\sigma_{\rm j, HN-1}$ [\ms] & $U[0, +\infty[$ \\
$\sigma_{\rm j, HN-2}$ [\ms] & $U[0, +\infty[$ \\
$\sigma_{\rm j, HIRES}$ [\ms] & $U[0, +\infty[$ \\
$h$ [\ms] & $U[0, +\infty[$ \\
$\lambda$ [days] & $U[0, 300]$ \\
$P_{\rm rot}$ [days] & $U[10, 80]$ \\
$w$ & $U[0.1, 5]$ \\
\hline
\multicolumn{2}{c}{Kepler-10\,b} \\
\hline
$T_{\rm c} \rm [BJD_{TDB}-2450000]$ & $\mathcal{N}(5034.08687, 1.8~\mbox{\sc{e}-04}) $ \\
$P$ [days] & $ \mathcal{N} (0.83749070, 2.0~\mbox{\sc{e}-07})$ \\
$e$  &  $\rm 0~(fixed)$\\
$K$ [\ms] & $U[0, +\infty[$  \\
\hline
\multicolumn{2}{c}{Kepler-10\,c} \\
\hline
$T_{\rm c} \rm [BJD_{TDB}-2450000]$ & $\mathcal{N}(5062.26648, 8.1~\mbox{\sc{e}-04})$ \\
$P$ [days] & $\mathcal{N}(45.294301, 4.8~\mbox{\sc{e}-05})$ \\
$e$  &  $\mathcal{N}(0.0, 0.098)~\rm AND ~ e < 1$ \\
$K$ [\ms] & $U[0, +\infty[$  \\
\hline
\multicolumn{2}{c}{Kepler-10\,d}\\
\hline
$T_{\rm c} \rm [BJD_{TDB}-2450000]$ & $U[7100, 7220]$ \\
$P$ [days] & $U[0, +\infty[$  \\
$e$  &  $ \mathcal{N}(0.0, 0.098)~\rm AND ~ e < 1 $ \\
$K$ [\ms] & $U[0, +\infty[$ \\
\hline
\label{table_priors_RV_parameters}
\end{tabular}
\end{table}
\tablefoot{$\gamma$ and $\sigma_{\rm j}$ are the RV zero points and uncorrelated jitter terms (HN-1 and HN-2 refer to the old and new HARPS-N CCD, respectively); $h$, $\lambda$, $P_ {\rm rot}$, and $w$ are the Gaussian process hyper-parameters -- namely the radial-velocity semiamplitude, the exponential decay time, the rotation period, and the inverse harmonic complexity term;
$P$ and $T_{\rm c}$ are the orbital period and inferior conjunction times; 
$e$ is the orbital eccentricity; 
and $K$ is the RV semiamplitude. $\mathcal{N}$ and $U$ stand for normal (Gaussian) and uniform priors, respectively.}


\begin{figure}[h!]
\centering
\vspace{2.5 cm}
\includegraphics[width=8.0 cm, angle=180]
{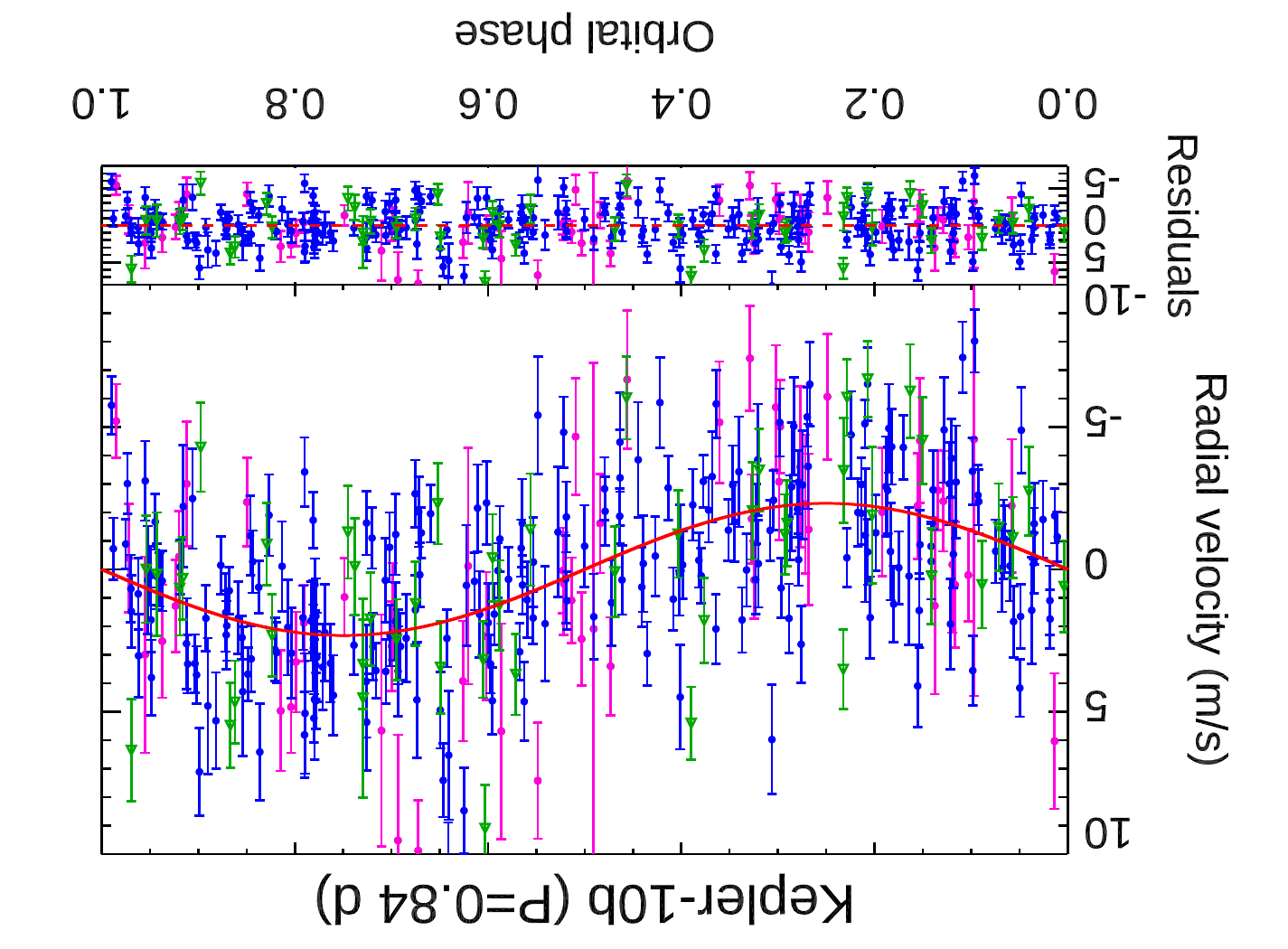}
\includegraphics[width=8.0 cm, angle=180]{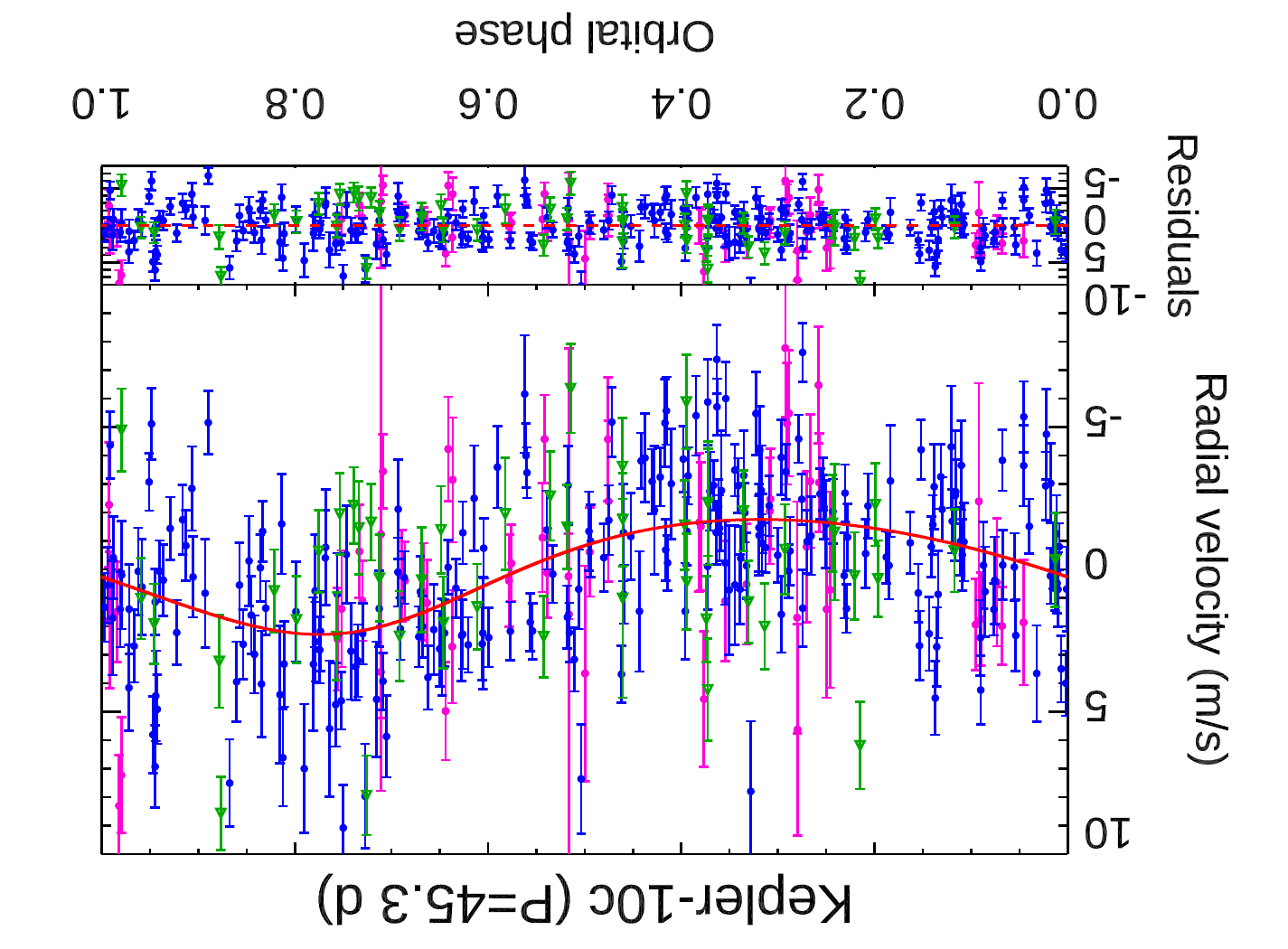}
\includegraphics[width=8.0 cm, angle=180]
{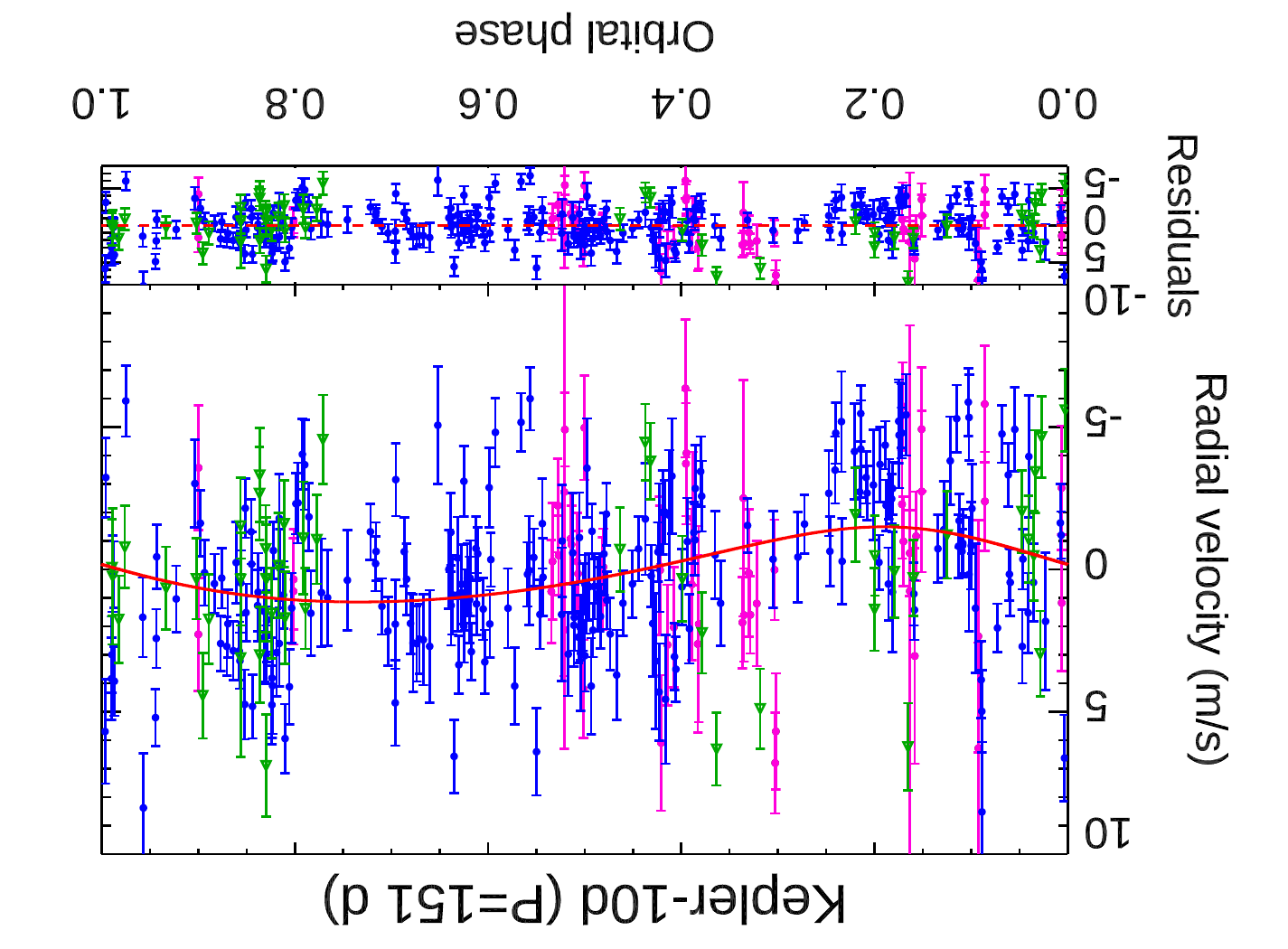}
\caption{Same as Fig.~\ref{figure_3planet-fit_HARPSN} with the addition of HIRES RVs (green triangles).}
\label{figure_3planet-fit_HARPSN-HIRES}
\end{figure}


\newpage

\onecolumn


\begin{landscape}
\small
\centering
\begin{longtable}{|c|c|c|c|l|c|c|c|c|c|}
\caption{Kepler-10 radial-velocity parameters obtained with the HARPS-N data.} 
\label{table_harpsn_param} \\
\hline
\multicolumn{4}{|c|}{\textbf{\large RV zero points and jitters}} & 
\multicolumn{6}{c|}{\textbf{\large Planet parameters}} \\
\hline
\multicolumn{1}{|c|}{$\gamma_{\rm HN-1}$} & $\gamma_{\rm HN-2}$ & $\sigma_{\rm j, HN-1}$ & $\sigma_{\rm j, HN-2}$ & \multicolumn{1}{l|}{Planet} & \multicolumn{1}{|c|}{$P$} & $T_{\rm c}$ & $e$ & $\omega$ & \multicolumn{1}{c|}{$K$} \\
\multicolumn{1}{|c|}{[\ms]} & [\ms] & [\ms] & [\ms] & \multicolumn{1}{l|}{ }& \multicolumn{1}{|c|}{[d]} & [$\rm BJD_{\rm TDB}-2\,450\,000$] & & [deg] & \multicolumn{1}{c|}{[\ms]} \\
\hline  
\endfirsthead
\hline
\endhead
\hline
\multicolumn{10}{|c|}{}\\
\multicolumn{10}{|c|}{\textbf{\large Differential evolution Markov chain Monte Carlo (DE-MCMC)}}\\
\multicolumn{10}{|c|}{}\\
\hline
\multicolumn{10}{|c|}{two-planet model}\\
\hline
$-98737.58\pm0.43$ & $0.26\pm0.19$ & $2.31_{-0.39}^{+0.41}$ & $2.45\pm0.15$ & Kepler-10\,b & $0.83749070(20)$ & $5034.08687(18)$ & 0 & - & $2.33\pm0.23$ \\ 
\cline{5-10} 
& & & & Kepler-10\,c & $45.294301(48)$ & $5062.26648(81)$ & $0.145_{-0.054}^{+0.056}$ & $318_{-18}^{+27}$ & $2.28\pm0.24$ \\
\hline
\multicolumn{10}{|c|}{three-planet model}\\
\hline
$-98737.13\pm0.48$ & $0.12\pm0.17$ & $2.51_{-0.40}^{+0.45}$ & $2.15\pm0.15$ & Kepler-10\,b & $0.83749070(20)$ & $5034.08687(18)$ & 0 & - & $2.33\pm0.22$ \\ 
\cline{5-10} 
& & & & Kepler-10\,c & $45.294301(48)$ & $5062.26648(81)$ & $0.136\pm0.050$ & $330_{-23}^{+31}$ & $2.17\pm0.23$ \\ 
\cline{5-10} 
& & & & Kepler-10\,d & $151.06\pm0.48$ & $7165.1_{-5.6}^{+4.9}$ & $0.19\pm0.10~(<0.24)$ & $150^{+28}_{-31}$ & $1.57\pm0.28$ \\
\hline
\multicolumn{10}{|c|}{}\\
\multicolumn{10}{|c|}{\textbf{\large \texttt{MultiNest} }}\\
\multicolumn{10}{|c|}{}\\
\hline
\multicolumn{10}{|c|}{two-planet model}\\
\hline
$-98737.66\pm0.50$ & $0.27\pm0.19$ & $2.35^{+0.43}_{-0.39}$ & $2.45\pm0.16$ & Kepler-10\,b & $0.83749070(19)$ & $5034.08687(18)$ & $0$ & - & $2.35\pm0.24$ \\ 
\cline{5-10} 
& & & & Kepler-10\,c & $45.294300(48)$ & $5062.26650(81)$ & $0.146\pm0.056$ & $313.4^{+21.2}_{-18.3}$ & $2.28\pm0.25$ \\
\hline
\multicolumn{10}{|c|}{three-planet model}\\
\hline
$-98737.52^{+0.54}_{-0.51}$ & $0.12\pm0.17$ & $2.53^{+0.43}_{-0.39}$ & $2.12^{+0.15}_{-0.14}$ & Kepler-10\,b & $0.83749070(20)$ & $5034.08687(18)$ & 0 & - & $2.35^{+0.21}_{-0.22}$ \\ 
\cline{5-10} 
& & & & Kepler-10\,c & $45.294301(46)$ & $5062.26647(81)$ & $0.134^{+0.048}_{-0.046}$ & $321.4^{+22.3}_{-39.0}$ & $2.19^{+0.23}_{-0.22}$ \\ 
\cline{5-10} 
& & & & Kepler-10\,d & $151.03\pm0.43$ & $7165.8^{+5.2}_{-5.3}$ & $0.216^{+0.091}_{-0.099}$ & $152.0^{+24.1}_{-26.9}$ & $1.68\pm0.27$ \\
\hline
\multicolumn{10}{|c|}{}\\
\multicolumn{10}{|c|}{\textbf{\large Multi-dimensional GP plus \pchord\ }}\\
\multicolumn{10}{|c|}{}\\
\hline
\multicolumn{10}{|c|}{two-planet model}\\
\hline
$-98737.53\pm 0.41$ & $0.26\pm 0.19  $ & $2.19^{+0.23}_{-0.14}$ & $2.41\pm 0.17 $ & Kepler-10\,b & $0.83749070(20)$ & $5034.08687(18)$ & 0 & - & $2.31\pm 0.23$ \\ 
\cline{5-10} 
& & & & Kepler-10\,c & $45.294301(48)$ & $5062.26648(81)$ & $0.138\pm 0.054$ & $314^{+30}_{-50}$ & $2.29\pm 0.26$ \\
\hline
\multicolumn{10}{|c|}{three-planet model}\\
\hline
$-98737.23\pm 0.42$ & $0.164^{-0.029}_{-0.14}   $ & $2.24^{+0.15}_{-0.20}$ & $2.113^{+0.042}_{-0.092} $ & Kepler-10\,b & $0.83749070(20)$ & $5034.08687(18)$ & 0 & - & $2.30^{+0.16}_{-0.20}      $ \\ 
\cline{5-10} 
& & & & Kepler-10\,c & $45.294301(48)$ & $5062.26648(81)$ & $0.131^{+0.039}_{-0.047}   $ & $317^{+22}_{-16}$ & $2.22^{+0.20}_{-0.17}      $ \\ 
\cline{5-10} 
& & & & Kepler-10\,d & $150.58^{+0.80}_{+0.31}$ & $7167.2^{+3.5}_{-5.7}$ &  $0.179^{+0.077}_{-0.096}   $ & $154\pm40$ & $1.42\pm 0.23$ \\
\hline
\end{longtable}
\tablefoot{$\gamma$ and $\sigma_{\rm j}$ are the RV zero points and uncorrelated jitter terms (HN-1 and HN-2 refer to the old and new CCD, respectively); $P$ and $T_{\rm c}$ are the orbital period and inferior conjunction times; 
$e$ and $\omega$ are the orbital eccentricity and argument of periastron; 
and $K$ is the RV semiamplitude.}

\small
\centering
\begin{longtable}{|c|c|c|c|c|c|l|c|c|c|c|c|}
\caption{Kepler-10 radial-velocity parameters from the DE-MCMC analysis of the HARPS-N and HIRES data.}
\label{table_harpsn-hires_param}
\\ \hline
\multicolumn{6}{|c|}{\textbf{\large RV zero points and jitters}} & \multicolumn{6}{c|}{\textbf{\large Planet parameters}} \\
\hline
\multicolumn{1}{|c|}{$\gamma_{\rm HN-1}$} & $\gamma_{\rm HN-2}$ & $\gamma_{\rm HIRES}$ & $\sigma_{\rm j, HN-1}$ & $\sigma_{\rm j, HN-2}$ & 
$\sigma_{\rm j, HIRES}$ & \multicolumn{1}{l|}{Planet} & \multicolumn{1}{|c|}{$P$} & $T_{\rm c}$ & $e$ & $\omega$ & \multicolumn{1}{c|}{$K$} \\
\multicolumn{1}{|c|}{[\ms]} & [\ms] & [\ms] & [\ms] & [\ms] & [\ms] & \multicolumn{1}{l|}{ }& \multicolumn{1}{|c|}{[d]} & [$\rm BJD_{\rm TDB}-2\,450\,000$] & & [deg] & \multicolumn{1}{c|}{[\ms]} \\ 
\hline  
\multicolumn{12}{|c|}{two-planet model}\\
\hline
$-98737.58\pm0.43$ & $0.25\pm0.18$ & $-0.02\pm0.43$ & $2.32_{-0.39}^{+0.43}$ & $2.46\pm0.15$ & $2.53_{-0.33}^{+0.41}$ & Kepler-10\,b & $0.83749070(20)$ & $5034.08687(18)$ & 0 & - & $2.35\pm0.21$ \\ 
\cline{7-12} 
& & & & & & Kepler-10\,c & $45.294301(48)$ & $5062.26648(81)$ & $0.144\pm0.056$ & $319_{-19}^{+30}$ & $2.04\pm0.23$ \\
\hline
\multicolumn{12}{|c|}{three-planet model}\\
\hline
$-98737.24\pm0.46$ & $0.13\pm0.17$ & $-0.07\pm0.49$ & $2.40_{-0.40}^{+0.44}$ & $2.19\pm0.15$ & $2.75_{-0.37}^{+0.42}$ &  Kepler-10\,b & $0.83749070(20)$ & $5034.08687(18)$ & 0 & - & $2.32\pm0.20$ \\ 
\cline{7-12} 
& & & & & & Kepler-10\,c & $45.294301(48)$ & $5062.26648(81)$ & $0.149\pm0.049$ & $332_{-22}^{+29}$ & $2.02\pm0.23$ \\ 
\cline{7-12} 
& & & & & & Kepler-10\,d & $151.21\pm0.60$ & $7161.2_{-6.0}^{+5.5}$ & $0.14\pm0.10~(<0.19)$ & $152_{-43}^{+41}$ & $1.32\pm0.25$ \\
\hline
\end{longtable}
\tablefoot{$\gamma$ and $\sigma_{\rm j}$ are the RV zero points and uncorrelated jitter terms (HN-1 and HN-2 refer to the old and new HARPS-N CCD, respectively); $P$ and $T_{\rm c}$ are the orbital period and inferior conjunction times; 
$e$ and $\omega$ are the orbital eccentricity and argument of periastron; 
and $K$ is the RV semiamplitude.
Very similar results were obtained with the \texttt{MultiNest} and multi-dimensional GP plus \pchord\ analyses.}
        

\begin{table}[b]
\small
\centering
\caption{Relative log Bayesian evidences, $\Delta \ln \mathcal{Z}$, as computed by \pchord\ for non-GP and multidimensional (MD) GP models including different numbers of Keplerian terms, both with and without the inclusion of HIRES data. }    
\label{VMR-table}       \begin{tabular}{lcccc}
\hline
& \multicolumn{2}{c}{Non-GP modelling of RVs only} & \multicolumn{2}{c}{MD-GP modelling of RVs, BIS, \lrhk} \\
Keplerians & (i) HARPS-N only     & (ii) HARPS-N + HIRES     & (iii) HARPS-N only       & (iv) HARPS-N + HIRES      \\ \hline
--          & $-75.98\pm0.16$      & $-77.13\pm0.21$          & $-76.24\pm0.14$          & $-73.91\pm0.14$           \\
$b$          & $-42.2\pm0.13$       & $-38.08\pm0.14$          & $-38.19\pm0.21$          & $-33.51\pm0.18$           \\
$b, c$       & $-7.32\pm0.12$       & $-6.14\pm0.16$           & $-9.12\pm0.12$           & $-6.5\pm0.13$             \\
$b, c, d$    & $-0.94\pm0.41$       & $-1.55\pm0.46$           & $-1.6\pm0.53$            & $-1.07\pm0.18$            \\
$b, c, d, e$ & $0\pm0.32$           & $0\pm0.13$               & $0\pm0.38$               & $0\pm0.13$                \\
$c$          & $-49.79\pm0.12$      & $-56.51\pm0.13$          & $-51.92\pm0.12$          & $-55.64\pm0.13$           \\
$d$          & $-59.24\pm0.18$      & $-63.01\pm1.49$          & $-60.15\pm0.95$          & $-63.25\pm0.30$            \\
$d, e$       & $-54.87\pm1.50$       & $-60.61\pm1.33$          & $-58.82\pm0.21$          & $-51.23\pm4.81$           \\ \hline
\end{tabular}
\tablefoot{Uncertainties correspond to the standard error on the mean evidence across three \pchord\ runs, or to the highest individual run error provided by \pchord\ (the greater of the two). For ease of interpretation, the highest evidence in a given column is normalized to zero; all other evidences in that column are defined relative to that zero point. Note that evidences cannot be compared meaningfully across setups (i)--(iv), as the data being modeled are different in each case. In each setup, a three-planet model is strongly favored, with weak but insufficient statistical evidence in support of a four-planet model.}
\end{table}
\end{landscape}

\end{document}